# Complex Networks and the Drugs Repositioning Problem

## Felipe Bivort Haiek

### Licenciatura's Thesis in Physical Sciences

### Faculty of Exact and Natural Sciences

### University of Buenos Aires

### October 2017

**TOPIC:** Network Physics / Biophysics

**STUDENT:** Felipe Bivort Haiek

**LU N °:** 609/09

**LOCATION OF WORK:** FCEyN

**DIRECTOR OF WORK:** Ariel Chernomoretz

## FINAL REPORT APPROVED BY

______________________________    ______________________________

**Author**                                                **Jury**

______________________________    ______________________________

**Director**                                              **Jury**

______________________________    ______________________________

**Professor of Licenciatura's Thesis**                    **Jury**

# Chapter 1

Introduction

Cells, the basic morphological and functional units of every living organism must interpret and respond to a wide variety of physical and chemical stimuli, both external and internal, in order to guarantee their survival [1]. For this, they need to maintain a high degree of organization that allows them to carry out their basic vital functions such as nourishing, growing, multiplying, differentiating, registering and transporting signals, controlling and coordinating a huge amount of biochemical reactions that take place inside them. The cell's capacity for coordination and control over these processes and functions, despite the number of variables involved, makes it an extremely complex system of study that has aroused interest in different scientific disciplines, including Physics.

Although in previous centuries Biology was for the most part dominated by reductionism, which managed to identify simple cellular structures and their functions, cells present multiple phenotypes and emergent properties. These are complex systems, which depend on intricate mechanisms of interaction and coupling between genetic material and proteins, and present various methods of sensing, signaling and control. In particular, the mere problem of analyzing the coupling of a protein and a small molecule at a given binding site, within the cellular environment, taking into account the possible universe of alternative competing interactions is too computationally costly.

One of the methods used to describe biological systems, without losing complete vision, but retaining characterization power, is that of complex networks [2]. This approach has had a strong boom in the field of Physics since the publications of Watts and Strogatz on small-world networks [3] and that of Barabasi and Albert on the study of scale-free networks [4] . In this type of approach, molecular components within a cell are usually considered as nodes, and the possible interactions (physical, chemical, direct or indirect) as edges or connections between them.

The enormous advances in experimental technologies in recent decades have made it possible to develop techniques that collect experimental data on a massive scale and facilitate the construction of this type of network at previously unthinkable scales. Current technologies, such as next-generation sequencing, allow complete genomes or proteomes to be surveyed with minimal effort, and techniques such as Y2H (yeast-2hybrid) can determine whether two proteins interact within the yeast cellular environment. As a direct consequence, current biomolecular networks have reached the scales of complete organisms, where statistical, computational and physics approaches are extremely enriching to address questions underlying the area of cell biology.

In recent years, attention has increased on drug-protein networks. These networks generally have two types of nodes: proteins, and drugs, where the activity, that is, the experimentally reported interaction of a drug on a protein is represented by an edge. For certain concentrations of the drug, this interaction affects the protein and results in an obvious change in its properties.

To understand the boom in the use of this type of network, it is necessary to consider the socio-economic context that underlies the development of drugs today. On average, the approval and inclusion of a new drug on the market, for use in humans, takes between 12 and 15 years (depending on the therapeutic area) and the costs to do so amount to the trillion dollars. If we also consider the fact that 1 in 24 drugs that enter the preclinical phase is approved, it becomes clear that the area requires a high capital investment and involves a high degree of risk. In fact, one of the most fruitful and efficient methods to identify a new drug-target relationship of interest to health is to start with an old, existing drug, which has passed some of the research phases that require more time and money. . This resource for finding therapeutic targets is known as drug repositioning (or reuse).

The use of biomolecular networks such as drug-protein networks, chemical similarity networks between drugs or even protein interaction networks have been very useful in the area. In particular, these networks have shown efficacy in proposing new drug targets that do not seem obvious to guide the repositioning of existing drugs or predict possible side effects of them, and provide, in general, different tools for conceptual synthesis and integrative analysis for the design of new ones. drugs [2] .

A case of special interest is the neglected tropical diseases (NTD) that include diseases such as Malaria, Sleeping Sickness, Chagas Disease, Yellow Fever, etc. and mainly affect poor people in developing countries. The limited commercial interest in the development and therapeutic improvements mainly underlies the high investment costs and the low expected return when dealing with low-income patients [5]. For this reason, repositioning strategies, particularly those based on the use of molecular networks and integration of chemogenomic data, have become a fundamental tool to address the problem of identifying therapeutic targets for this type of disease [6,7 , 8]. Thus, in the area of NTD, the repositioning of existing drugs plays a fundamental role. In particular, through research and development efforts that come from the academic area and government agencies. The strategy here is to make use of drugs already designed and tested in model organisms and to test their efficacy in the treatment of NTD. A clear example of success is *eflornithine,* which was developed as an anticancer compound and is being used to treat African trypanosomiasis (sleeping sickness)[9,10].

Following this line of work, this thesis presents a work based on chemogenomic information from the TDR targets database, dedicated to NTD. This database was introduced by Aguero, Crowther et al [11,12,13] in order to guide the prioritization process of putative targets in drug development in NTDs. Initially, the targeting prioritizations were based only on protein characteristics with limited use of the available information on bioactive compounds in guiding these prioritizations. The authors have integrated information into this database (TDR targets [11] ) on a large number of bioactive compounds, from public domain sources and a series of high-throughput assays, at an unusual scale for NTDs [ 12,14] . These works have now brought the integration of chemogenomic data associated with NTDs to a stage in which large-scale data mining exercises are both feasible and promising.

It is particularly relevant for the present work that the similarity relationships between pairs of compounds and proteins can be efficiently described using complex network concepts. Under this paradigm, non-trivial interconnectivity patterns can be explored to discover underlying organizing principles, identify relevant entities, and novel drug-target associations ([15],[16],[2]).

In accordance with this approach, in this work we set out to analyze the topological properties of a chemogenomic network conceived to carry out a drug prioritization and repositioning program. We conceptualize an organization of the data accumulated in TDR based on a network structure composed of three main layers: one of drugs, linked by relationships of structural similarity and substructure (see below); a protein layer, which is linked to the previous layer by edges of experimentally reported bioactivity; and finally a layer with three different types of annotations, which represent molecular characteristics and / or cellular functions that the proteins adjacent to them have.

We begin by analyzing the structure of the network in the drug layer. This allowed us to characterize and understand how the chemical, genomic and annotation spaces of the network are related. In particular, we were able to quantify the extent to which similarities reported in the drug space spoke to us of similarities reported in the protein space, a fundamental property on which the development of any prioritization algorithm that seeks to propose novel drug-target relationships is based. Subsequently, we use information on the temporal evolution of the structure of the network to recognize patterns that reflect biases in the development and discovery of drugs and bioactive molecules. Finally, we carried out a prioritization task for proteins, in order to extend the probability of *druggability* (ie the probability that some drug has activity on it) from one protein to another, and we validated it.

It is important to verify that it is possible to use the idea that similarities between elements in one space are consistent with similarities reported in another to develop prioritization algorithms that allow proposing new drug-target relationships. In our case this means that similar drugs target similar proteins and vice versa: similar proteins are targeted by similar drugs. To carry out this analysis, we check to what extent similarity between drugs implies that they share proteins. In addition, it was established that there is a relationship between the number of shared domains per protein pair and the association of this same pair with common drugs. For the same purpose, the layer of targets and annotations was taken and quantities related to the passing of information about drug-ability between proteins were analyzed, and annotations of particular interest were found. A projection was made on the annotations layer and densely connected sets were used to determine related drugs.Later we use the structure of the network, with "timestamps" to recognize patterns that reflect biases in the development and discovery of drugs and bioactive molecules, and finally hicimos prioritization process for proteins neighbors to a planned network, to extend drogabilidad of one protein to another, and validate.

The structure of this thesis is organized in 8 chapters. Chapter 1 gives the introduction to the work. Subsequently, basic foundations of network theory are described that will be developed in the rest of the work. Finally, prioritization methods are explained and the way to validate them is introduced. Then projection methods in multiparty networks and corresponding notation are established.

Chapter 3 introduces the basic characteristics of the network in its three layers and between them. Chapter 4 discusses the topology of the drug layer. First, the relationship between drug similarities and the protein layer is discussed. Chapter 5 discusses the topological characteristics of the annotation layer in relation to the protein layer. Chapter 6 establishes the relationship between groups. densely related annotations and sets of drugs with similar spectrum of action.

In chapter 7 studies the temporal evolution of the network, as motivation for prioritizing proteins in the network, with the methods specified in Chapter 2.

# Chapter 2

Fundamentals.

## *2.1. Introduction*

This chapter introduces and discusses fundamental concepts of complex networks that will be used extensively throughout this work, (a brief introduction to basic ideas and concepts is included in `appendix A`). Two techniques are described to predict missing links in complex networks and the usual criteria to evaluate their performance in the context of classification problems. Finally, the classic concept of networks is extended to other types of graphs, such as multiparty networks and multilayer networks. In both cases, their usefulness is motivated and the formal notation necessary to work with these types of networks is presented.

## *2.2.General definitions*

Formally, a graph $G(\mathfrak{N}, \mathfrak{E})$ consists of a set of entities $\mathfrak{N}$ and a set of interactions or edges $\mathfrak{E}$ established between pairs of them. The elements of $\mathfrak{N} := \{n_1, n_2, ..., n_N\}$ are called nodes or vertices of the graph and the elements of $\mathfrak{E} = \{e_1, e_2, ..., e_m\}$ are called edges, arcs or connections of the same. The number of elements of the set $\mathfrak{N}$ and $\mathfrak{E}$ we will denote them by $N$ and $m$ respectively, the first determines the number of objects in the graph, usually referred to as the size of the graph or mass, and the second the number of edges or connections of the same. i-th node $n_i$ through the letter i. Each edge is associated with a pair of two numbers identifying the nodes that it connects. Namely, if the k-th edge $e_k \in E$ connects the nodes $i$ and $j \in N$ we will denote it $e_k := e_{ij} = (i, j)$. In that case the nodes $i$ and $j$ are adjacent nodes or first neighbors. An edge that connects a node with itself ( $e_i$ ) is called a loop. On the other hand, if there is more than one edge $e_k$ , $e_k$ both connecting the nodes $i$ and $j$, those nodes are said to have multiple edges. Both loops and multiple edges are not included in the standard definition of a graph, which we will use here.

There are two well differentiated classes of networks depending on whether the connections between nodes have a defined preferential direction or not. We say that a graph is undirected if the connections lack a preferential direction. This implies that the elements $e_{ij} \in E$ can be described by unordered pairs $i(,)j$ and is $e_{ij} = e_{ji}$ therefore,. $\forall e_{ij} \in E$ On the contrary, a graph is said to be directed if the nature of the connections it represents has some preferential direction or orientation. In general, for directed graphs, we have a $e_{ij} = (i, j) \neq e_{ji}$ way that $e_{i,j} = (i, j)$ implies the existence of a connection with a well-defined sense, which goes from node $i$ to node $j$.

## *2.3 Prioritization*

Prioritization algorithms in complex networks are essentially predictive models. Suppose we have a graph G = G (N, E), where each node of the set $N = \{n_1, n_2, ...n_N\}$ represents an individual and the edges $e_{ij} \in E$ represent relationships between pairs of these, be it work, family, friendship, etc. Suppose further that we have concrete information that a subset $N_a = \{n_1, n_2, ...n_k\}$ has bought a product P, and none of the remaining individuals $N_b = \{n_{k+1}, n_{k+2}, ...n_j...n_N\}$ you have yet purchased this product. The basic idea of the prioritization algorithm in complex networks is to use information from the set $N_a$ called seeds, and the information embedded in patterns of network connectivity to infer potential buyers of the product *P*. The typical result of a prioritization algorithm of this nature is a list of nodes L ordered according to the degree of confidence given to each node $n_j \in N_b$ as a potential buyer of P.

There is a wide variety of prioritization algorithms in complex networks, each based on very varied ideas and principles (for a comprehensive review see [17]). In this work, we will present and use a prioritization algorithm. This is an analogy of a voting scheme *VS* where each node in the set $N_a$ can transmit information to its first neighbors.

### 2.3.1 Voting Scheme

Let it be $G = G(N, E, W)$ a weighted graph and $N_a = \{n_1, n_2, ...n_k\}$ a subset of its nodes that is known from information external to the network, associated with a category or class P. For the remaining nodes of the graph $N_b = \{n_{k+1}, n_{k+2}, ...n_j...n_N\}$ we want to infer those with the greatest potential to belong to the same category P of the set $N_a$.

The simplest strategy that can be implemented is a voting scheme (VS), which in recommendation systems theory is known as KNN with K = 1, that is, a weighted sum over first neighbors of the set of nodes used as seeds $N_a$. It is a simple method, but it has a good success rate, comparable to more complex prioritization algorithms [18,19] with the added benefit of being extremely efficient. In a *VS* scheme, given a weighted graph $G = G(N, E, W)$ represented by its weight matrix $M_P(w_{ij})$, and the subset of seeds $N_a = n_1, n_2, ...n_k$, the VS algorithm prioritizes the remaining nodes of the network $N_b = \{n_{k+1}, n_{k+2}, ...n_j...n_N\}$ from the score assignment function

$$f_P(nj) = \sum_{l \in Nei(n_j), l \in N_a} W_{jl} \forall n_j \in N_b \qquad (2.1)$$

where the sum traverses only nodes that are simultaneously in the set of first neighbors of $n_j$, that is $Nei(n_j)$, and the set of seeds used $N_a$. The sum $w_{jl}$ weights are the weights of the connections in the matrix $M_P$. As a result, we get an ordered list L of the nodes that obtain the highest score in equation 2.1, where the highests will be inferred as potential candidates to belong to the category P. Note that every node $n_j \in N_b$ that is not a direct neighbor of some node in the set of seeds $N_a$ will get a null score in equation 2.1.

### 2.3.2 Performance metrics: ROC curves

As mentioned above, prioritization algorithms are essentially predictive models. There is a graph G and a functional class P that involves at least a subset $N_a$ of nodes in G. The problem at hand here is to predict which of the remaining nodes of the network $N_b$ belong to P and which do not. That is, we are in the presence of a binary classification problem (belonging or not, to class P). The algorithms presented in the preceding sections result in a list of nodes $L = \{n_i, / n_i \in N_b\}$ ordered according to a scalar magnitude $f_P(ni)$ (see equations 2.15 and 2.19). That is, the first nodes in the list will be those with the highest value of $f_P$. Furthermore, it is expected that these are associated with the functional class P with a higher level of confidence than nodes with lower values of $f_P$. This list can respond to the classification problem posed by defining a threshold $f_P(n_i) = u$, so that every node in $N_b$ will be classified as I verified

$$L(n_i, u) = \begin{cases} n_i \in P & if \quad f_P(n_i) \geq u \\ n_i \notin P & if \quad f_P(n_i) < u \end{cases} \tag{2.2}$$

The function $L(n_i, u)$ represents a binary classifier. This classifier depends on the cutoff threshold u chosen in the list L provided by each prioritization algorithm. To evaluate the predictive capacity of a classifier, it is necessary to have among the elements of L some subset that is known a priori belongs to P. Having this reference set, it is possible to count the number of hits and misses that the classifier $L(n_i, u)$ commits, which allow in turn define different performance metrics.

In practice, to perform the evaluation of a binary classifier it is usual to divide the set $N_a$ into two subsets $N_\alpha^T, N_\alpha^E$, so that it is verified $N_\alpha^T \cup N_\alpha^E = N_\alpha \quad N_\alpha^T \cap N_\alpha^E = \emptyset$. The largest $N_\alpha^T$ (suppose 90% of the nodes in $N_\alpha$) is called the training set while $N_\alpha^E$ (the remaining 10%) is known as the reference evaluation set. The nodes in $N_\alpha^T$ will be used as seeds of the prioritization algorithm, while the $N_\alpha^E$ Nodes in will be added to the list whose class you want to infer and will allow you to evaluate the predictive capacity of the classifier. Note that now the result of a prioritization algorithm is a list *L* that contains elements of $N_\alpha^E$ and of $N_b$ assigning each element an observable $f_P(n_i)$. For a cutoff or discrimination threshold *u*, we can calculate the hit rate of the classifier $L(n_i, u)$ in the prediction of elements of the set $N_\alpha^E$, that is, the fraction of true positives (TPR) or sensitivity of the node predictor in the evaluation set.

$$TPR(u) = \frac{1}{|N_\alpha^E|} \sum_{n_i \epsilon L} \delta_i^{TP} \text{ with } \delta_i^{TP} = \begin{cases} 1 & if \quad f_P(n_i) \geq u \quad \wedge \quad n_i \in N_\alpha^E \\ 0 & in \quad other \quad case \end{cases}$$

**(2.2)**

On the other hand, the failure rate, that is, the *fraction of false positives* (FPR) is given by

$$FPR(u) = \frac{1}{|L - N_\alpha^E|} \sum_{n_i \epsilon L} \delta_i^{FP} \text{ with } \delta_i^{FP} = \begin{cases} 1 & if \quad f_P(n_i) \geq u \quad \wedge \quad n_i \notin N_\alpha^E \\ 0 & in \quad other \quad case \end{cases}$$

**(2.3)**

Note that both quantities $FPR(u)$ and $TPR(u)$ are normalized in the interval [0, 1]. The FPR rate can alternatively be expressed in terms of the specificity of the predictor by $FPR = 1 - especificidad$. Also, note that both FPR and TPR are monotonic increasing functions of u and it is verified that for $u_{min} = min(f_P(u))$ is had $TPR(u_{min}) = FPR(u_{min}) = 1$. Intuitively, it is expected that a "good classifier" can infer nodes of functional class P with a high true positive rate (TPR) at the expense of a low false-positive rate (FPR). That is, the predictor is expected to have simultaneously high specificity and sensitivity. **Figure 2.1** setpoint observable $TPR(u)$ and $FPR(u)$ varying discrimination threshold or from the maximum to the minimum $f_P$ value. This type of graph is called ROC (Receiver Operating Characteristic) curve and is very useful for comparing the performance of different binary classifiers. In particular, the area under this curve denoted by AUC (Area Under Curve) is used as a performance measure of the classifier under study. Let's analyze two extreme examples to gain insight into the interpretation of AUC values. On the one hand, an ideal classifier should be able to assign for every node $N_\alpha^E$ a higher value $f_P$ than for any other node in the list L, that is

$$f_P(n_i) > f_P(n_j), \forall n_i \in N_a^E, n_j \in N_b \quad \textbf{(2.4)}$$

from which the existence of ais deduced $u$ verifying ( $TPR(U*) = 1$ and $FPR(u*) = 0$ ). Therefore, the ROC curve associated with an ideal classifier contains a unit area AUC = 1 (see **Figure 2.2**). In contrast, in a random classifier, the magnitude of $f_P(n_i)$ is completely uncorrelated with the class to which each node does not belong $(n_i \in P, on_i \notin P)$

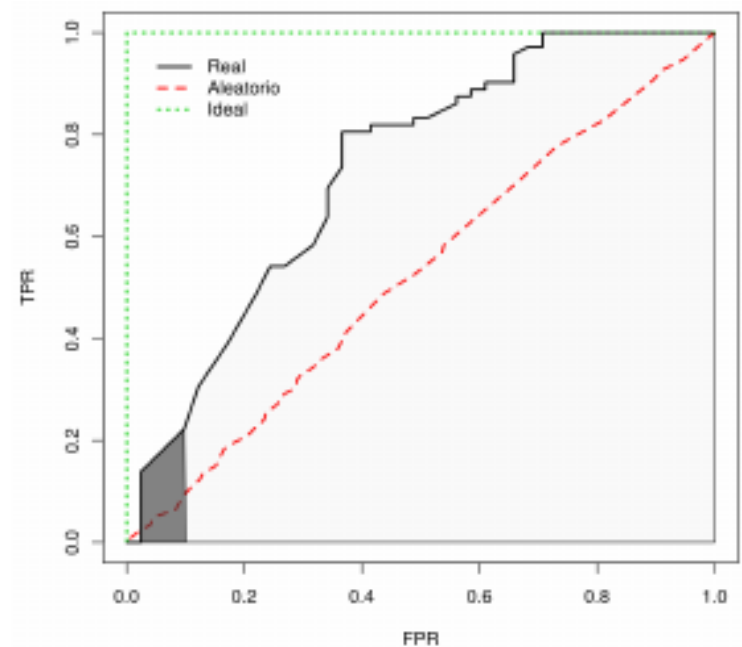

**FIGURE 2.1**: ROC curves: rate of true positives TPR of a predictor (or sensitivity) as a function of the rate of false positives FPR (1-predictor specificity). The green dotted curve corresponds to an ideal predictor, whose enclosed area is unitary (AUC = 1). The red dashed curve represents a random predictor whose graph oscillates on the identity line and therefore presents an AUC~ 0.5. The solid black line illustrates the case of a real predictor. In the illustrated case, the total area AUC = 0.73 and is shaded faint gray for illustration. The dark gray shaded surface corresponds to the area under the curve limited to 10% FPR, AUC0.1 = 0.0137. Under McClish's correction (see Eq.2.6), we have AUC0c.1 = 0.546.

In such a case, regardless of the discrimination threshold *u* selected, it is expected that the rate of hits and misses in the classifier will be of the same order. Therefore, the ROC curve associated with a random classifier must approximate a line with unit slope and the associated area is AUC ~1 / 2. In general, a predictive algorithm obtains

190a94fc7003d5cd18-19112AUC values in the range (1/2, 1). Within this range, the higher AUC, the better the performance of the algorithm under study.

In practice, however, it is not very useful to compare two algorithms based on the entire list L. In contrast, it is more appropriate to compare algorithms considering the elements with the best scores in their respective lists (that is, with a higher value of $f_P$). With this idea in mind, a very useful measure to compare predictive algorithms can be defined, the AUC-01, defined as the area under the ROC curve in the interval $FP \in [0, 0.1]$. This is, limiting the analysis, to what occurs for a false positive rate equal to 10%. If it is available $|N_\alpha^E| << |L|$, the AUC-01 is equivalent to considering approximately 10% of the nodes with the highest L. Note that in the case of AUC-01 a random predictor presents a $AUC - 01 = 0.005$, while an ideal predictor presents an area $AUC - 01 = 0.1$. It is therefore useful to consider some type of normalization of the AUC-01 to bring it to the interval [0.5,1]. In this work, McClish's correction was used for this purpose, [20] which is expressed as

$$AUC_\alpha^c = \tfrac{1}{2}(1 + \tfrac{AUC_\alpha - AUC_\alpha^{aleat}}{AUC_\alpha^{mx} - AUC_\alpha^{aleat}}) \qquad (2.5)$$

$$AUC_\alpha^c = \tfrac{1}{2}(1 + \tfrac{AUC_{0.1} - 0.005}{0.1 - 0.005}) \qquad (2.6)$$

where $\alpha$ is the maximum FPR value considered, $AUC_c$ the renormalized area, $AUC_{aleat}^\alpha$ the area corresponding to a random predictor, and $AUC_{max}^\alpha$ the area corresponding to an ideal predictor. In our case of interest ec.2.24 considers $\alpha = 0.1$.

## 2.4 Structure of bipartite networks and projection methods

### Bipartite networks

A network G (N, E) is said to be bipartite if there is a partition of N, ($N_1$, $N_2$) that verifies $N_1$ ∪ $N_2$ = N, $N_1 \cap N_2$ = ø, and also no arc $e_i$ ∈ E joins nodes of a same set $N_1$ or $N_2$. There are many concrete cases of bipartite networks. For example, collaborative or co-authorship networks have two well-distinguished types of nodes: authors and their publications. A typical biological example is metabolic networks [21] where nodes can be classified into chemical compounds and chemical reactions. Another case of particular interest for this thesis is affiliation networks, where objects are identified as a class of nodes and characteristics common to them as another class. An example of an affiliation network would be to consider actors as objects and the films in which they participated as characteristics. Another possible case that will be addressed in this work consists of taking proteins from different species as objects and functional or structural groups of them as a set of characteristics.

A natural extension of the concept of two-party networks is to introduce the idea of multi-party networks. In the latter, there is a partition $N_1, N_2, ....N_m$ that verifies the conditions $\cup_i N_i = N$, and for any pair it $i,j \in \{1...m\}$ is fulfilled $N_i \cap N_j = \emptyset$. Furthermore, no pair of nodes of the same class $N_i$ is connected. An example of a multipartite network with three types of nodes (tripartite) is the so-called folksonomies, a term that refers to social indexing methods [22,23], where users, labels, and online resources are the three types of nodes in the network. For example, flick.com is a website where users can assign tags to different photos, or CiteUlike.com is another site where users can assign tags to publication references.

### 2.4.1. Projection of bipartite networks in single-party networks

A common mechanism for calculating the structural similarity between nodes of the same class in a bipartite network, is projecting it on a single-party network that contains only one of the two types of nodes. In this way, the existence of the other class of nodes will be implicit in the new links. In this type of projected network, two nodes share a connection only if they are both connected to at least one common node in the original two-part network.

It is a bipartite graph with two sets of nodes $X = \{x_1, x_2, ...x_n\}$ $Y = \{y_1, y_2, ...y_m\}$ and connections provided by the set of edges E so that $e_{i,j} \in E$ with $i \in X$ $j \in Y$. We will call it $G_{bip}(\mathfrak{N} = X, Y, E)$ graph. It can be represented by the adjacency matrix $A = (a_{ij})^{nXm}$.

$$a_{ij} = \begin{cases} 1 & if \quad e_{ij} \in E \\ 0 & in \quad other \quad case \end{cases} \quad (2.7)$$

Recall that it $G_{bip}$ is bipartite so that no element in $E$ connects two nodes of the set $X$ or two nodes of the set $Y$.

The simplest projection of a single-part network into a single-part network will be given by

$$w_{i,j} = \sum_{l=1}^{m} a_{il} a_{jl} \quad (2.8)$$

which, written in a matrix, is

$$W = AA^t \quad (2.9)$$

This projection results in a one-part, weighted graph and undirected $G_x(X = \{x_1, x_2, ... x_n\}, E_x, W_x)$ of nodes $X = x_1, x_2, ... x_n$, edges $E_x = e_{i,j}$ with $i, j \in 1, 2 ... n$ which they have associated weights $W_x = w_{i,j}$. Each edge $e_{i,j} \in E_x$ The weight between nodes $i$ and $j$ is simply the number of typical neighbors $Y$ that they share in the original network.

Another possible bipartite projection of the graph $G_{bip}$ on nodes X, was defined by Zhou [24], and is known as ProbS. This projection results in a one-part, weighted directed graph $G_x(X = \{x_1, x_2, ... x_n\}, E_x, W_x)$ of nodes $X = x_1, x_2, ... x_n$, edges $E_x = e_{i,j}$ with $i, j \in 1, 2 ... n$ which they have associated weights $W_x = w_{i,j}$. Each edge $e_{i,j} \in E_x$ takes values

$$\begin{cases} 1 & si \quad w_{i,j} \neq 0 \\ 0 & si \quad w_{i,j} = 0 \end{cases} \quad (2.10)$$

and the weights $w_{i,j} \in W_x$ are defined according to

$$w_{i,j} = \frac{1}{k_{x_j}} \sum_{l=1}^{m} \frac{a_{il} a_{jl}}{k_{y_l}} \quad (2.11)$$

where the sum runs over all nodes $y_l$, l = {1, 2, ... m}. Note that if the nodes $x_i, x_j$ do not have any common neighbors $y_l$, then the sum 2.28 will be zero and there will be no connection between these nodes in $G_x$. 02/28 gives expression elements of weight matrix W. Note that in general in this graph $w_{ij} = w_{ji}$ projected.

This weight matrix W can be obtained from the adjacency matrix. Given $G_{bip}$ with adjacency matrix A, we define the column normalization operation as the division of each element $a_{ij}$ by the sum of the elements of the j-th column and we will notice it $\hat{a}$. That is to say that for each element $\hat{a}_{ij} \in \hat{A}$ we have

$$\hat{a}_{ij} = \frac{a_{ij}}{\sum_j a_{ij}} \quad (2.12)$$

Notice that the denominator of Eq. 2.12 is the degree of the j-th node $k_{y_j}$. Then we can rewrite Equation 2.11 as

$$w_{ij} = \sum_{l=1}^{m} \frac{a_{il} r_l a_{jl}}{k_{y_l} k_{x_j}} = \sum_{l=1}^{m} \frac{a_{il} r_l}{k_{y_l}} \frac{a_{jl}}{k_{y_j}} \quad (2.13)$$

$$= \sum_{l=1}^{m} \hat{a}_{il} \frac{a_{jl}}{k_{y_j}} = \sum_{l=1}^{m} \hat{a}_{il} \frac{(a_{lj})^t}{k_{y_j}} \quad (2.14)$$

$$w_{ij} = \sum_{l=1}^{m} \hat{a}_{il} \hat{a}_{ij}^t \quad (2.15)$$

$$W = \hat{A} \hat{A}^t \quad (2.16)$$

where the exponent $(a_{ij})^t$ in Eq. 2.15 indicates the transpose operation. The matrix expression of Eq. 2.16 gives a direct relationship between the weight matrix of the one-part graph projected as a function of the adjacency matrix A of the bipartite network $G_{bip}$. On the other hand, expression 2.37 can be trivially extended to expression

$$W = \hat{A} I \hat{A}^t \quad (2.17)$$

where I represents the identity matrix I of size mxm.

## 2.4.2 Projection by statistical validation

The projection method by statistical validation (StatVal)[25] consists of assigning links between those vertices of type A that have a statistically significant amount of links to vertices B of a given degree. Namely, let the subgraph be bipartite $S$ with $N_A$ vertices of type A and $N_B$ vertices of type B, for each $k$ in the degree distribution of the vertices of type B the induced subgraph $S_k$ of the $N_B^k$, elements B of degree $k$, and of all elements A is constructed attached to these. Therefore, the only heterogeneity that exists in $S_k$ is that due to nodes of type A. If i and j are nodes of type A in $S_k$, $N_{i,j}^k$ is the number of first neighbors in common, and $N_i^k$, $N_j^k$ their respective degrees in $S_k$. Under these conditions, if the links between nodes A and B were produced randomly, the probability that $i$ and $j$ share $N_{i,j}^k$ neighbors is given by a geometric distribution. This identifies a p-value for each $i$ $j$ pair. The pairs $i$ $j$ whose p-values (adjusted by the FDR method) exceed a threshold will be considered as statistically validated pairs and a link between the two will be assigned. This method is repeated for $k$ all, if a pair $i$ $j$ is validated for more than a subgraph $S_k$ will be assigned a weight corresponding to the number of times that was validated.

Let the induced bipartite graph be $S_2$, where the nodes on which we project have degree 2 like the blue ones in **figure 2.2**. The black nodes of the set in this subgraph have a variety of degrees, in particular, nodes 4 and 5 have quantities $N_4^2 = 6$, $N_5^2 = 5$ and $N_{4,5}^2 = 5$. With which the hypergeometric distribution $p(N_{4,5}^2; N_4^2, N_5^2, N_B^2) = 0.0004$, and if the threshold value is chosen, $0.0005$ the edge $e_{4,5}$ is validated and will be part of the projected graph.

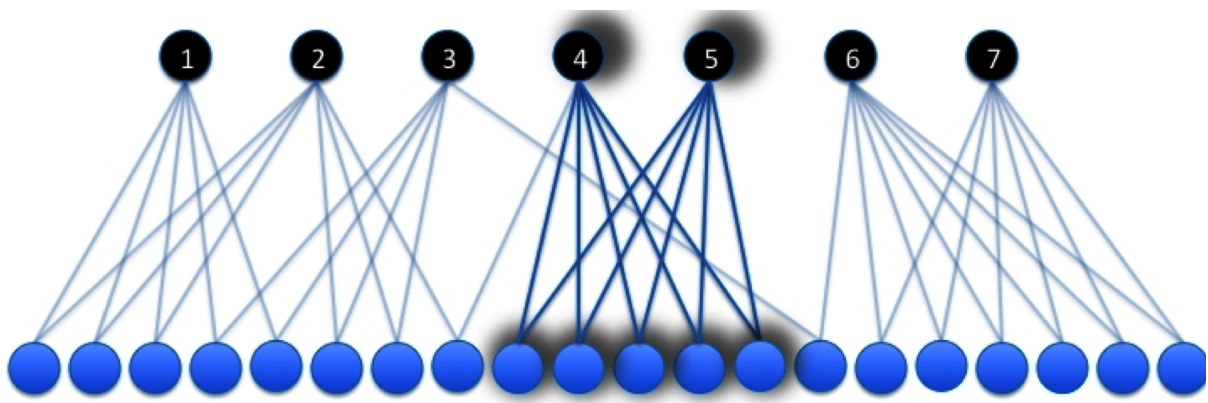

**Figure 2.2**. Subgraph $S_2$ for the projection by statistical validation. The blue nodes are on which it is projected, and they all have degree 2, while the black nodes are the ones that will integrate the projected network. Nodes 4 and 5 present quantities $N_4^2 = 6$ $N_5^2 = 5$ (proper degree of node 4 and 5 in $S_2$) and $N_{4,5}^2 = 5$ (neighbors shared by nodes 4 and 5). With which the hypergeometric distribution is given by $p(N_{4,5}^2) = 0.0004$, and if the threshold value is chosen, $0.0005$ the edge $e_{4,5}$ is validated and will be part of the projected graph.

Alternatively, there is another way to calculate StatVal that we call Restricted StatVal. The greatest restriction is a result of the statistic used increasing the p-values obtained. This is accomplished by nothing that each subgraph, $S_k$ was itself a separation in $p$ disconnected components, which leads to subgraphs $S_k^p$. When validating the links within these components instead of in the subgraph, the number to incorporate in the hypergeometric function is $N_B^{kp}$ (number of links in the entire subgraph component), instead of $N_B^k$ (number of links in the entire subgraph), which is a larger number, effectively running the distribution to the right, therefore increasing the p-values and reducing the number of links that are validated at the end.

### *2.5 Multilayer Networks*

In many real systems, the use of networks such as the ones we have defined so far can result in an oversimplification of the problem under study. In particular, the fact of thinking that the interactions between objects always occur at the same level of importance is inappropriate for many cases and can even lead to incorrect conclusions in the study of the dynamics of the system under study. [26]. A generalization of the classical theory of networks called multilayer networks consists of thinking about systems composed of a set of networks in different planes or layers of abstraction interconnected with each other, with edges of different nature and levels of relevance.

Let us consider as an example, the classic and historical paradigm of complex networks: social systems. Let's think of a social network like Facebook, where the nodes represent users and the edges represent connections between them. A user usually has connections of a very diverse nature, he may be connected to other users through work relationships, family relationships, friendships, partners in certain sports or cultural activities, etc. In this sense, it may be appropriate to think that links of different nature or nature are located in different levels of abstraction, instead of all being treated at the same level of hierarchy.

If one wants to study, for example, the spread of a rumor on this social network, it is logical that each user does not spread the rumor uniformly throughout all their links, but rather, in principle, with greater probability towards users potentially interested in the particular rumor. Furthermore, it may not spread it to contacts in a certain field. This example would be particularly suitable for dealing with multilayer networks, where each layer of the network can contain a specific type of relationship and users can be simultaneously in different layers, so that the probability that a user will spread the rumor to their neighbors it depends on the layer or nature of the connection you have with these neighbors.

Another example that concerns more the thematic axis of this thesis is that of gene co-expression networks. A classic approach to these is to think of the set of genes of a given organism and draw connections between two genes if there is some kind of correlation in the level of expression of the same in a given experiment. However, these experiments can be of a very varied nature, they can even be experiments carried out in different tissues, or under different experimental conditions. A fact currently accepted in the literature is that the treatment of this type of systems considering all the interactions simultaneously can result in noisy models, since the interactions can occur in very disparate contexts.

It is usual to carry out the construction of these networks by limiting the interactions to a tissue of interest, or to a given set of experimental conditions. From the point of view of multilayer networks, this type of system is particularly appropriate for thinking of each tissue or experimental condition in a different layer, so that each gene can belong to more than one layer and in fact have different environments and levels of connectivity in each of them.

Another case of particular interest for this thesis that we will expand on in chapter 5, is that of networks of proteins and chemical compounds used for the search, prioritization, and repositioning of drugs. These networks can be thought of as layers composed of nodes of a different nature, such as chemical compounds, proteins, metabolic processes, protein-specific functional domains, etc. The connections between nodes are also very diverse in nature, being able to represent the structural similarity between compounds, evidence of activity of a given compound on a protein target, belonging of two proteins to the same functional domain or the same metabolic pathway, etc. This example will be expanded in greater detail in Chapter 5, where the construction of a network with these characteristics will be carried out.

### 2.5.1 Notation

Multilayer networks are essentially a generalization of traditional network theory. A multilayer network can be thought of as a set of networks at different levels or layers related to each other. Formally, a multilayer network can be represented by a pair $M = (G, C)$, where it $G = {G_\alpha, \alpha \in \{1, 2, ...M\}}$ is a family of graphs $G_\alpha = G(X_\alpha, E_\alpha)$ (directed, undirected, heavy or not heavy) that we will call layers of $M$, and $C = \{E_{\alpha,\beta} \subseteq X_\alpha \times X_\beta; \alpha, \beta \in 1, 2, ...M, \alpha, \beta\}$ is the set of connections between different layers $G_\alpha, G_\beta, \alpha \neq \beta$. The elements of C are called cross or transverse layers [47].

Each layer $G_\alpha$ contains the set of nodes $X_\alpha = x_1^\alpha, x_2^\alpha, ..., x_{N_\alpha}$ and their connections can be represented by an adjacency matrix $A^{[\alpha]} = (a_{ij}^\alpha) \in \Re^{N_\alpha X N_\alpha}$ whose elements are defined

$$a_{ij}^\alpha = \begin{cases} 1 & si \quad (x_i^\alpha, x_j^\alpha) \in E_\alpha \\ 0 & in \quad other \quad case \end{cases} \quad (2.18)$$

On the other hand, the transversal layers can also be represented by their adjacency matrix $A^{[\alpha,\beta]} = (a_{ij}^{\alpha,\beta} \in \Re^{N_\alpha X N_\beta})$, whose elements are defined by

$$a_{ij}^{\alpha\beta} = \begin{cases} 1 & si \quad (x_i^\alpha, x_j^\beta) \in E_{\alpha\beta} \\ 0 & in \quad other \quad case \end{cases} \quad (2.19)$$

In addition, each node $x_i^\alpha \in X_\alpha$ lives in a layer and can be connected through two kinds of edges, some called intra-layer connections $E_\alpha$ that join them to nodes of the same layer, and others called inter-layer connections $E$, which join nodes located in other layers.

It should be noted that many other kinds of networks can be represented as multilayer networks. For example, temporal networks (whose nodes and edges are time-dependent) can be mapped to multilayer networks by interpreting each temporal step as a new layer. An example that is of particular interest for this thesis is the multiparty networks defined in the previous section.

# Chapter 3

Network structure

## *3.1. Introduction*

In this chapter, we will describe and characterize the multilayer network that we will use in this study. It is made up of a drug layer, a protein layer (which we will denote equivalently as targets), and a layer that includes nodes that represent three types of concepts that we think are relevant to our problem. These involve the ideas of PFAM domains, orthologous groups, and metabolic pathways, related to the presence of subunits protein structures, sequence similarity due to the existence of a common evolutionary origin, and the participation in similar chemical reactions, respectively. Basic characteristics and a diagram of the organization of our network are included in **table 3.1** and in **figure 3.1** respectively.

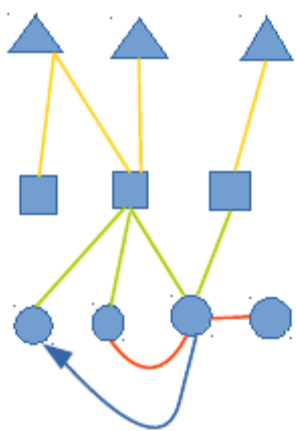

**FIGURE 3.1:** Diagram of the multilayer network. The triangles represent nodes in the annotation layer that can be either pfam domains, orthology groups, or metabolic pathways. The squares represent the or proteins, which are attached to nodes of the previous layer. The proteins that are connected to the drug layer (circles) are said to be dopable and the link represents bioactivity. The drugs are related through two types of similarity substructure (blue arrow) and Tanimoto (red arrow).

|  | **Nodes** | **Links** |
|---|---|---|
| **Drug** | 14,888,034 | 15,950,829(sub) <br> 23,028,356(Tanimoto) |
| **Bioactive Drugs** | 177,506 | 576,147(substructure) <br> 2,067,256(Tanimoto) |
| **Protein** | 168,622 | 325,843 (a drug) |
| **liableaddicted Proteins** | 6,051 | 325,843(drug) |
| **PFAM** | 2252 | 219 313 |
| **Metabolic pathways** | 145 | 77,389 |
| **Orthologous groups** | 2,789 | 51,702 |

**TABLE 3.1:** Number of nodes by type and number of links in our network.

### 3.2.1 Similarity between drugs

The drug layer has 14,888,034 nodes associated with drugs of known effect and other chemical compounds, characterized by their molecular weight and chemical structure. At the same time, we have information on compound similarity from two complementary notions of structural similarity: Tanimoto-type and substructure similarity. The Tanimoto relationships are determined by the similarity of molecular fingerprints that the drugs share. The molecular fingerprint is a set of bits that describe in a biomolecule the presence of characteristic chemical groups or structures so that if a characteristic is present, the corresponding bit will be 1, otherwise, it will be 0. Substructure relationships, on the other hand, are defined when one molecule contains another as a structural subset.

**Figure 3.2** shows a set of four drugs with Tanimoto equal to 1, they share all molecular characteristics marked with circles, however as seen none of the molecules is a structural subset of the other. This illustrates the complementary aspect between both relationships of similarity that we will discuss below.

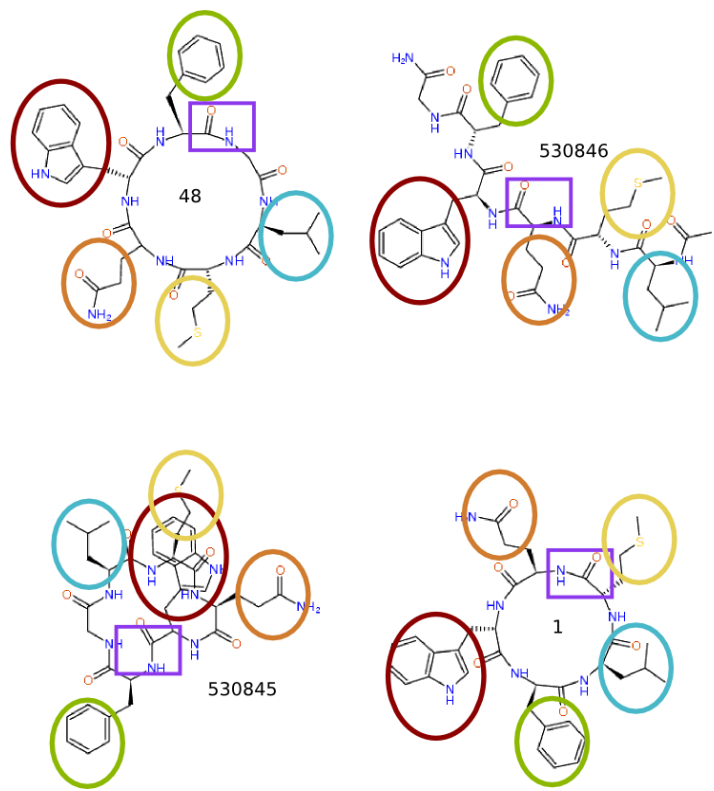

**FIGURE 3.2:** Molecules with Tanimoto 1 similarity are observed in the figure. Chemical groups used in the molecular fingerprints are marked in different colors. It is seen that in all the compounds there are the same chemical groups, which results in the Tanimoto similarity equal to 1. As can be seen, they are not substructures of each other.

### 3.2.2. Tanimoto and substructure

As introduced before, the drug layer is composed of bonds of two types: Tanimoto and substructure. In order to calculate the Tanimoto relationship, first how many and which characteristics to include in the molecular fingerprint are established. In the case of the TDR network, the footprints only contain information about two-dimensional structures. The size of the footprint associated with each compound is 512 bits.

Once the type of fingerprint to represent the molecule has been defined, it is necessary to define a notion of similarity. These depend on the amount of 1s in the footprint of each compound, being the pair of compounds $AB$, we denote $a$ and $b$ the number of ones in $A$ and $B$, respectively. In turn, we will call *c* the number of ones in common (ie shared characteristics). For the calculation of Tanimoto similarity, the Jaccard index established between a pair of footprints is considered, which is defined $sim(A,B) = \frac{c}{a+b-c}$. For the case of our network, only Tanimoto relationships with a value greater than 0.8 were taken into account, which is a reasonable value according to what was reported in [27]. This is important since it guarantees that the size of the network is suitable for work, and allows studies of its structure, avoiding computational bottlenecks. With this criterion, are established 23,028,356 links between drugs (**table 3.1**).

Regarding the substructure relationship between two drugs, A and B, a value of 1 drug A is the substructure of the B complex. It is interesting to note that, contrary to what happens with the Tanimoto similarity, this type of association is directed, and in principle, it does not have a weight associated with it. In the network, we were able to establish 12,076,297 substructure relationships between chemical compounds.

Given the similarity metrics adopted, we set out to identify two types of relevant structures: strongly *connected* sets defined from substructure links (ie sets where the directed relationship of is-substructure-of allows us to establish a path between any pair of drugs in the set) and drug *cliques* for links with Tanimoto = 1.

We call these structures S-clusters and T-clusters respectively. These make up clusters *identities*, in the sense that their components are indistinguishable from the point of view of one or another similarity. It is important to note, however, that since both relationships are based on simplified representations of the real chemical structure, they may not contemplate the total molecular degrees of freedom. In particular, the fact that there is an identity between two molecules according to the Tanimoto metric does not necessarily imply that one is a substructure of the other (see **Figure 3.2**).

It is important to note that the clusters thus defined, it is corroborated that the values of the similarities between nodes of different clusters are the same for any pair. For example, be it $x_A x'_A$ nodes in cluster A and $x_B x'_B$ nodes in cluster B, it is always checked $T(x_A, x_B) = T(x'_A, x'_B)$. In this way, the Tanimoto relationship between two T-clusters is well defined from the Tanimoto association of any of the inter-cluster pairs. Something similar happens for S-clusters in the sense that the substructure relationships of elements of the same S-cluster towards external elements remain identical. Finally, we assume that drugs not included in S-type or T-type identity clusters made up their own cluster of size 1.

The description in terms of S-clusters and T-clusters allows us to implement a significant reduction in the size of the network without loss of information. A more detailed study of the characteristics of these structures is included in the next section. Here we simply mention that by replacing all the elements of a cluster by a single node, a substructure graph results in a reduction in the number of nodes of almost 4% while the reduction in the number of edges is 17%. to 103,550,339 reduction With respect to Tanimoto there are 1,330,235 clusters and 23,028,356 edges. This implies a reduction in the number of nodes of 10% and a reduction in the number of edges of 47%.

Finally, to move forward we decided to assign a weight to each bond in the substructure space, which takes into account the asymmetry in the molecular weight of the associated vertices. We define then:

$$Sub(X,Y) = \frac{\acute{min}(w_x,w_y)}{\acute{max}(w_x,w_y)} \tag{3.1}$$

where *x* and *y* are S-clusters of molecular weights $w_x$ and $w_y$ respectively.

Maximum and minimum values are taken since due to the problems of incomplete representation of chemical compounds it could happen that $w_X > w_Y$, even when $x \subset y$. In this way, we are giving more weights to a relationship of similarity by substructure, the more similar the sizes of the associated molecules are.

### 3.2.2 Annotations

In this subsection, we will introduce in some detail the components of the annotations layer considered to structurally characterize the proteins of our network.

Pfam domains:

Proteins are amino acid sequences that present a structure that in general can be characterized according to four levels of description. The first level involves the chain of amino acids, that is, the sequence. When it is folded, it gives rise to the appearance of a secondary structure that includes arrangements in helixes and sheets (see figure 3.3, in particular, the red structure is markedly formed by sheets). The tertiary structure corresponds to the chain already completely folded. The PFAM domains can be identified in the tertiary structure (colored portions of the protein in figure 3.3) and correspond to structural units that fold more or less independently of each other and that in certain cases can be associated with protein functions. specific that are preserved throughout different species [28]. In our case, we consider the 2252 PFAM domains described in http://PFAM.xfam.org/.

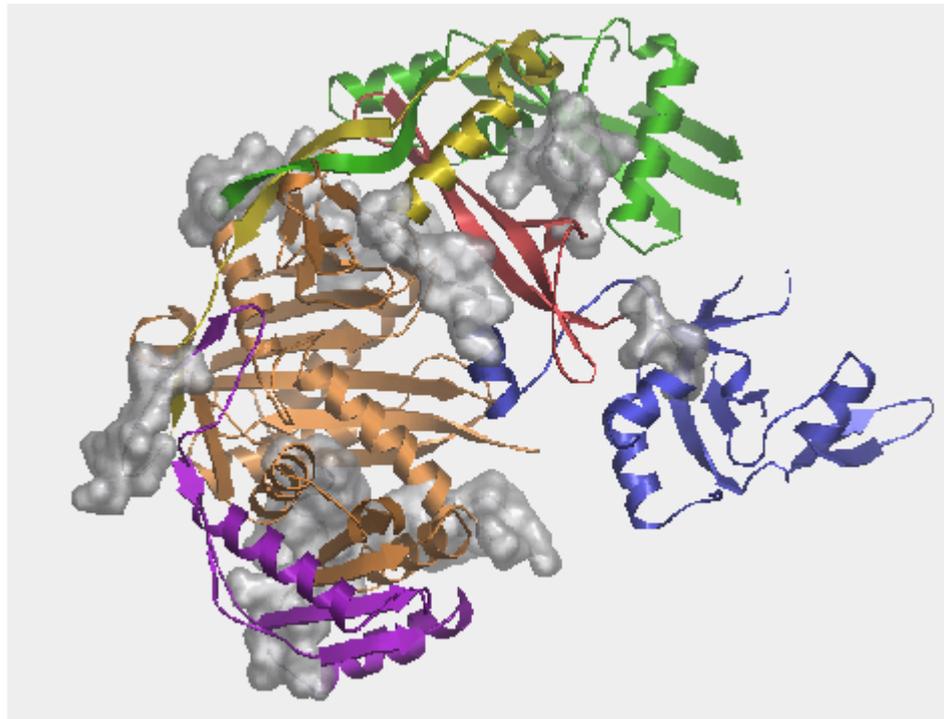

**FIGURE 3.3:** protein *ago1_human*, its PFAM domains are illustrated in colors while areas of the protein that have not been assigned to domains are shown in gray. PFAM domains are structures that are repeated in various proteins. The secondary structure of the proteins can also be observed, formed by helices and sheets.

orthology:

Two proteins are homologous if they descend from the same ancestral protein. It can usually be concluded that two sequences are homologous on the basis of their high sequence similarity. When Proteins thar are *homologous* belong to the same species, they are said to be *paralogues*. *Homologous* Proteins present in different species are called *orthologous*. We consider 2789 group of orthologies defined in version 4 of orthoMCL (http://orthomcl.org/orthomcl/home.do) that uses Markovian clustering methods on the sequences of each protein to recognize groups of proteins of similar sequences.

Finally, we consider 145 metabolic pathways, which we will also denote *pathways*. They correspond to chained enzymatic reactions that occur within a cell. In them, an initial substrate is transformed and gives rise to final products that in turn can be used as a substrate for subsequent reactions. The information of belonging to a given metabolic pathway (pathway) corresponds to that extracted from http://www.genome.jp/kegg/pathway.html.

# Chapter 4

Characterization of the drug layer

In this chapter we will deal with describing the drug layer and its relationship to the protein layer. The links between the different types of similarities are established in the drug layer, Tanimoto and substructure. The clusters that the two similarity relationships generate are compared and their homogeneity is analyzed in order to establish whether these clusters were well chosen and whether they represent different types of information on drugs. In addition, the relationships between these bonds of similarity and the bonds of bioactivity on proteins are studied.

## *4.1 Substructure vs. Tanimoto pairs in clusters*

As mentioned before, given that the substructure and Tanimoto relationships are based on simplifications of the molecular structure, it is appropriate to study the overlap between the two types of clusters formed in order to analyze the coincidences and differences found.

For this purpose, we will consider persistence $P_{er}$, a measure that allows us to quantify the relationship between the composition of S-clusters and T-clusters. Given a cluster of substructure *clusA* of partition S, we define the persistence coefficient per cluster $P_{er}(clusA)$,, as the fraction of drug pairs found in clusA that also appear together in the same cluster-T:

$$P_{er}(ClusA) = \sum_{i,j \in ClusA} \sum_{ClusB} \frac{\delta_{i,j}^{ClusB}}{N_{ClusA}(1 - N_{ClusA})/2} \quad (4.1)$$

with $\delta_{i,j}^{ClusB}$ the delta which is one if the pair i, j appears in cluster B and $N_{ClusA}$ the number of drugs in Cluster A. Likewise, we performed this calculation to estimate the persistence of T-clusters in S-clusters.

The panels in **figure 4.1** show the persistence results of S-clusters in T-clusters and vice versa.

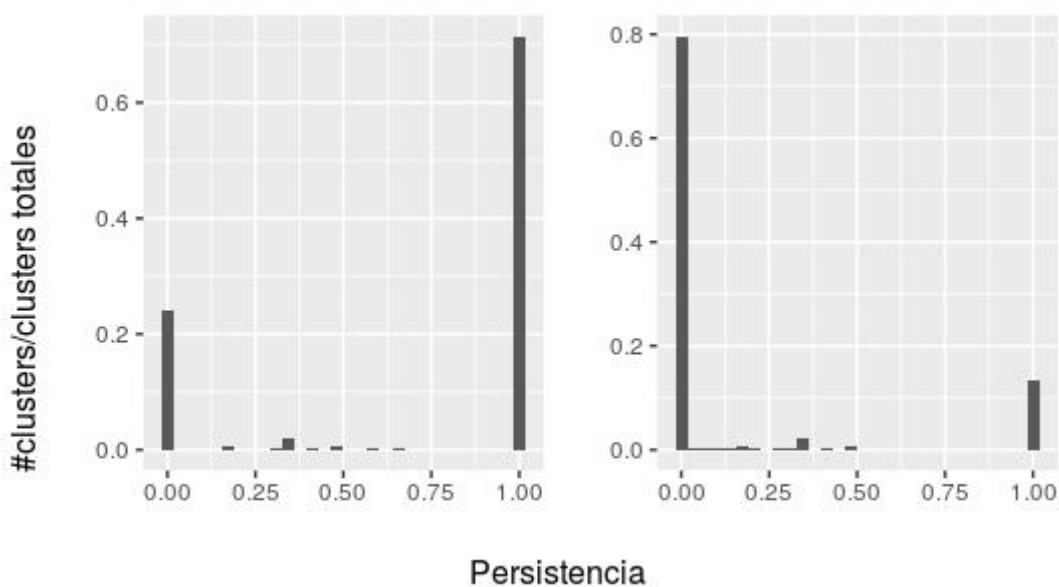

**FIGURE 4.1:** Right panel bar graph with the fraction of clusters S (*#clusters/clusters totales*) that present a given persistence (*Persistencia*) of T in S, as defined in equation 3.1 Left side proportion of clusters S for each persistence of S in cluster T. Large values of persistence indicate that several of the drug pairs in a given S cluster are also together in a T cluster.

Knowing the persistence values of all S-clusters it was possible to estimate an average persistence value (or partition persistence) on All of them, to quantify to what extent the partition structure is preserved in Cluster S when making the cluster T. Similarly, the average persistence of clusters T in S. The values obtained were P (S-> T) = 0 , 73 and P (T-> S) = 0.15 which would imply that S-type clusters tend to stay together in the partition induced by Tanimoto clusters.

This makes sense since the identity requirement per substructure is stronger than that induced by maximum levels of the Tanimoto index, in which it only matters which functional groups are repeated, regardless of order or spatial position, for example. The "substructure" condition, on the other hand, is a strong condition that implies that several constituent elements appear contiguously. In fact, it is known that isomeric compounds have the same chain characteristics (that is, they will have Tanimoto 1) but different spatial configurations, so they are not generally substructures of each other. Since it is this same difference that is often responsible for the fact that compounds possess significantly different biological activities, from a chemical and biological point of view it is desirable, and in fact it occurs in the network, that the substructure clusters separate isomeric drugs into different clusters.

### *4.2. Homogeneity of S and T clusters*

So far we have grouped the types of Tanimoto relationships and substructures in identity clusters by biochemical similarity. As already indicated, neither of the two criteria captures all the degrees of freedom that exist in a molecule. That is why it is interesting to study some properties of the S and T clusters and to investigate how homogeneous these structures are with respect to other conditions such as molecular weight and shared bioactivities.

**Molecular weight 4.2.1. Análisis**

**Figure 4.2** shows the molecular weight distribution of structures T and S.is noted that the distributions are similar, with a heavy tail towards large weights. For the cluster-T, an average molecular weight of 427 g / mol was obtained with a standard deviation of 251g / mol, while for the weight distribution of clusters S, an average of 442 g / mol and a standard deviation of 262 g / mol were obtained. The standard deviations found imply that there is a great variety of clusters with very dissimilar average masses. We further note that the sizes of the drugs we are considering (reported by average weights) are typically much smaller than that of proteins, which is in the order of 10,000 g / mol.

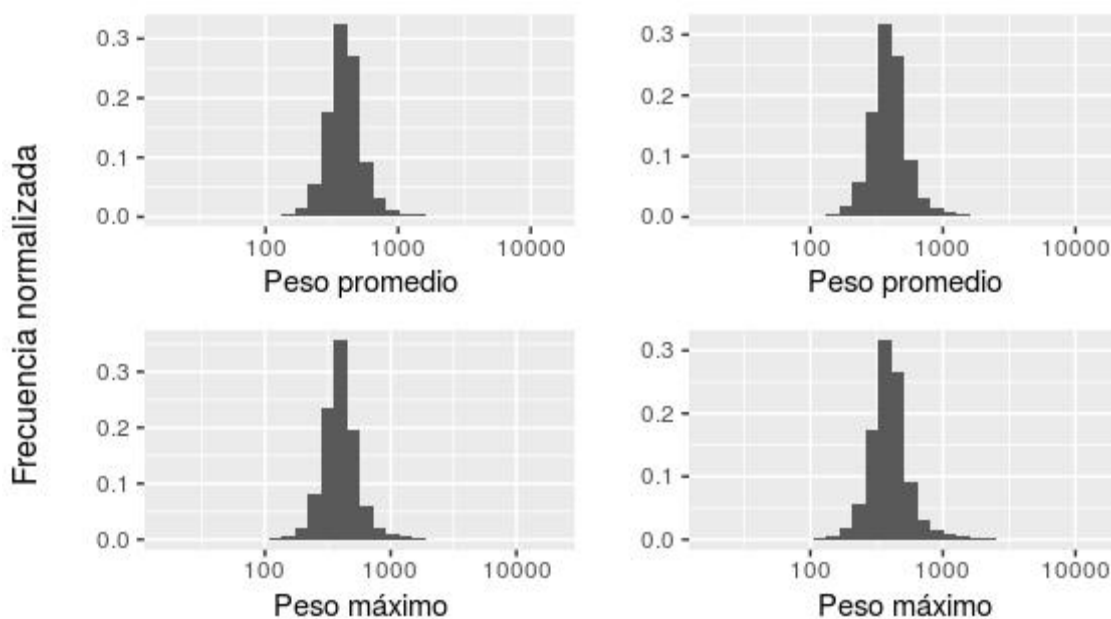

**FIGURE 4.2:** The left panel corresponds to the distributions of weights of the Tanimoto clusters, the right corresponds to the distributions of weights of the S clusters.

We also found a high degree of homogeneity of the clusters in terms of the molecular weight of the elements that compose them. 98% of the S clusters have maximum differences in weight less than 1, and the situation of clusters with greater differences is explained as a result of the fact that the substructure relationship is inaccurate since it starts from a simplified representation of the molecule. The situation for Tanimoto clusters is different. The difference in weights between the largest and smallest element is greater than 1 g / ml, for more than 3/4 of the T clusters, and greater than 50 g / ml for 1/4 of them. In this case, we find that typically the difference between the maximum and minimum weight found for molecules of the same cluster T is of the order of 10% of its average weight (see **figure 4.3**). In [29,30,31] it is stated that the Tanimoto similarity tends to be greater for relationships that involve large molecules, because in general they have a greater variety of characteristics and have a greater chance of sharing bits with another protein. This bias implies that it $Tani(A,B)$ will be greater if $A$ or they $B$ are a pair of molecules of dissimilar sizes, in contrast to the case in which $A$ and $B$ are molecules both of small and similar sizes. It is observed that the maximum weight correlates with the standard deviation of weights, as shown in **figure 4.3**. The interquartile distance corresponds to 10% of the median, which means that in all cases the clusters include drugs with similar molecular weights.

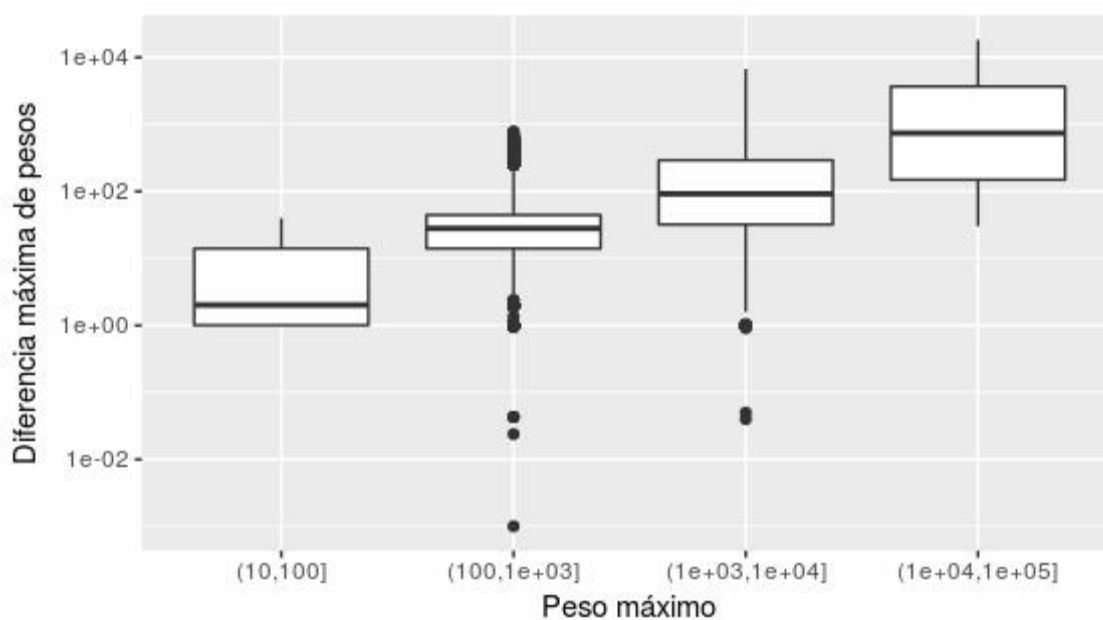

**FIGURE 4.3:** Graph that characterizes the standard deviation of the weights in the T clusters as a function of their maximum weight. It is observed with logarithmic axes that the weight variation measures increase with the maximum weight of the cluster.

### 4.2.2. Homogeneity of targets

As mentioned above there are 177506 drugs in the network with activities on proteins. Given that the clusters reflect some type of chemical identity between drugs, it is desirable to investigate whether this is reflected in the existence of similar bioactivities for drugs belonging to the same cluster. This will allow us, from the associations between chemical compounds, to infer new bioactivities and will be particularly useful when recommending new drugs in which some activity is already known.

With this in mind, the shared target rate of all drugs (with at least one reported link to ais analyzed *target* protein) from the same cluster. This is quantified by taking the Jaccard index of drug pairs in the clusters, and averaging this magnitude over all pairs. That is, for each pair of drugs, AB, with blanks $T_A T_B$ is taken $\frac{T_A \cap T_B}{T_A \cup T_B}$ and then averaged over the total pairs of drugs in the cluster.

We found that although more than a third of the Tanimoto clusters present null Jaccard, 50% of the Tanimoto clusters present an average Jaccard value of 1. For S clusters this occurs for 80% of the clusters. This first result suggests that the co-appearance in a cluster S can be strongly associated with the fact of sharing protein targets.

To analyze the statistical significance of the associations found between co-appearance in identity clusters and the number of coincidences in associated targets, the Jaccard frequency analysis was performed among all possible pairs in the same cluster and contrasted with pairs chosen at random. For this, drug pairs selected at random from all the available ones were considered, respecting that they have the same number of bioactivities as those of the intra-cluster pair data. In particular, for each pair of bioactive drugs within a cluster, 10 other pairs formed at random with drugs of the same grades as the original pairs, but in different clusters, were taken.

For Tanimoto clusters there are 62,241 drug pairs in the same cluster. Among all of them (**figure 4.4** upper panel), 86% have a Jaccard index greater than 0 (at least one shared target), 65% greater than 0.5, and 60% equal to 1 (exactly the same targets). These values were significantly higher than those obtained from pairs of drugs taken at random, respecting the original number of bioactivities. For this control, it was obtained that 42% of control pairs have on average a Jaccard index greater than 0, 9.99% greater than 0.5, and only 4% equal to 1.

Similarly, for the 9,271 pairs of drugs belonging to clusters of substructure (**figure 4.4** lower panel), we found that 85% had a Jaccard index greater than 0, 54% greater than 0.5, and 42% equal to, while the random control reported that 78% had an average Jaccard index greater than 0 but none greater than 0.5.

These observations indicate that, although belonging to the same cluster does not guarantee that two drugs have exactly the same bioactivities, it does guarantee that these, on average, will have more bioactivities in common than any other two taken at random from different clusters. In this way we see that the structures generated from these complementary notions of structural similarity, prove to be relevant with respect to the reported bio-activities. In particular, the fact that random pairs in different S clusters do not exceed the Jaccard of 0.5, reinforces the importance of identity substructures based on this notion of similarity to define bioactivities.

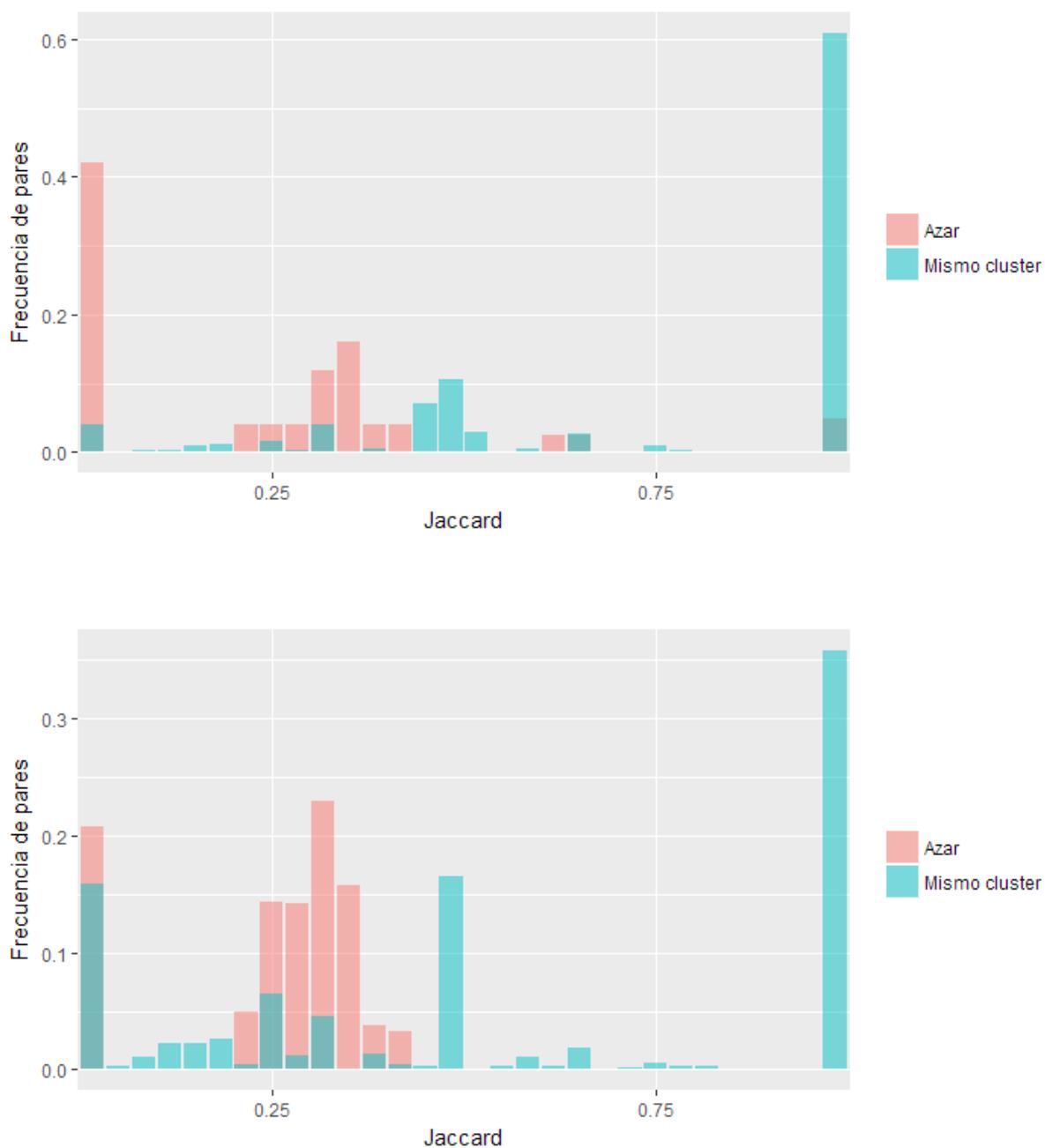

**FIGURE 4.4:** Frequency of pairs normalized for the Jaccard index between bioactive drugs from the same cluster and taken at random. In the upper panel Tanimoto clusters, below clusters substructure

### *4.3.Degree of Chemical Similarity and shared bioactivities*

So far the analysis has shown us that there is a positive relationship between shared targets and belonging to the same type S or type T cluster, or in other words, between drugs with similarities of value 1. We will now address the question of whether, in general, the degree of chemical similarity provided by the Tanimoto indices and / or substructure, it correlates with the tendency to present common targets.

To analyze this, we estimate the fraction of shared targets (using Jaccard indices) between drugs linked through Substructure or Tanimoto links of a given value. Clearly, we observe in Figure 3.9 an increasing monotonic relationship between the indices of chemical similarity between pairs of chemical compounds and the number of shared targets. (left and center panel). We observe that in both cases, Tanimoto and substructure, a similarity value between compounds close to 0.7 - 0.8, corresponds to a fraction of shared targets of the order of 40%. The right panel finally shows the linear relationship that acts as a calibration curve between the two similarity indices.

These results show in particular that both notions of similarity, although complementary alternatives, are consistent with each other. Furthermore, this reinforces what has been seen before that the substructure relationship is stronger than that of Tanimoto, since a lower weight of the substructure relationship is needed to achieve the same Jaccard value. It should be noted that since substructure similarities are addressed, it is necessary to consider edges in both directions, only exit edges, or only entrance edges. Since no significant differences are observed between these cases, we show the result for the case in which the addresses are not considered.

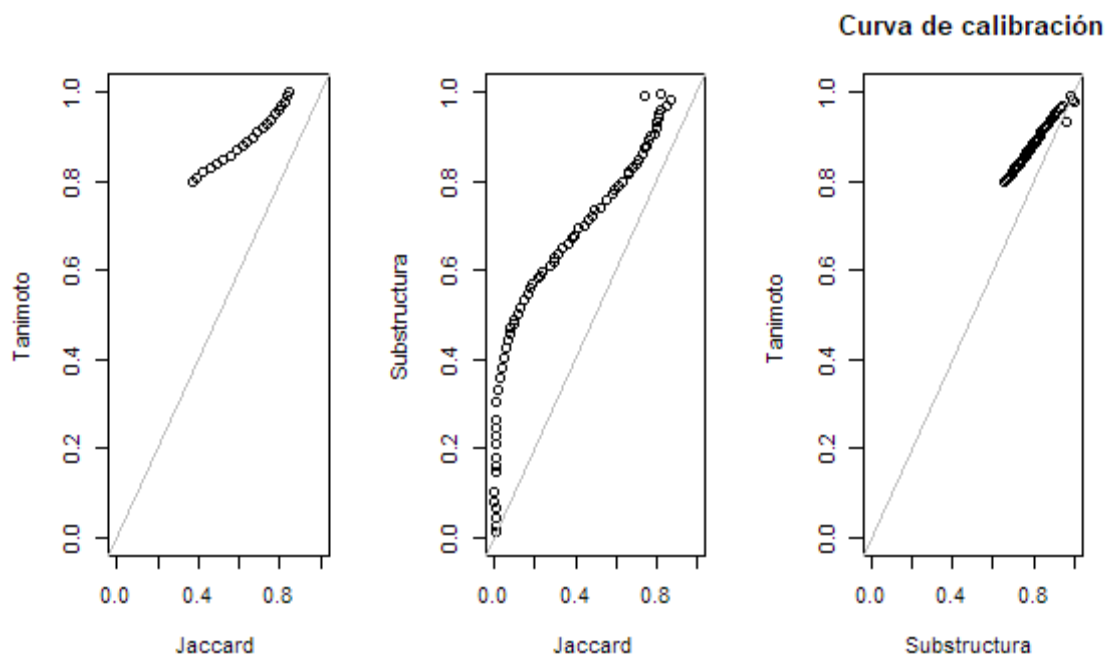

**FIGURE 4.5:** On the left is the Tanimoto ratio averaged for drug pairs with a certain Jaccard of whites. The center graph is the same as the previous one but for the substructure similarity. The graph on the right shows the result of calibrating the two graphs based on the Jaccard of blanks.

## *4.4. Conclusion*

As seen previously, the analysis of identity structures from links with $Tani(A,B)=1$, allowed us to recognize clusters (T-clusters, S-clusters) that allow a description in terms of 'coarse grained' structures and thus reduce the effective size of the network in both types of similarity metrics. In this chapter we established the relationships between these two types of clusters. We observe that on average the substructure relationship is stronger than that of tanimoto, in the sense that persistence is much greater when going from a partition of the layer into clusters S to T than vice versa, which indicates that the clusters- S are contained to a greater degree within T-clusters than vice versa. Finally, we established a comparison and calibration criterion between the numerical value of the similarity by substructure and Tanimoto. The calibration scale between both measurements (right panel of **figure 4.5**) allows us to appreciate that for the same similarity value, the substructure relationship corresponds to a greater coincidence of shared targets.

# Chapter 5

Targets and Annotations Layer

In this chapter, we will study the protein layer and its relationship to the annotation layer. In the same sense, the capacity of the annotations of PFAM, metabolic pathways and orthology, to mediate drugability information between proteins in different species, will be analyzed. Finally, the projected layer of annotations will be studied in order to find specific, densely connected groups of annotations that reflect the properties of dopable proteins.

## *5.1. Characterization of the distribution of degrees of affiliation sets to affiliates*

We begin the analysis of the bipartite network: targets-annotations by studying basic aspects of connectivity from the observed degree distributions. The distributions of the number of domains per target and the number of targets per domain were obtained in **Figure 5.1**. As can be seen in the left panel, each target has at most 10 associated PFAM domains. This is natural since the domains represent molecular substructures and the existence of a limit to the number of domains is a natural consequence of the finiteness of the protein chains.

The right panel of the figure shows that most PFAMS have few associated proteins. However, there are some particularly promiscuous PFAM domains, present in more than 500 proteins. Said PFAMs correspond to general functional domains often associated with structures that carry out biological functions such as: signal transduction, metabolism and recognition of RNA motifs, among others.

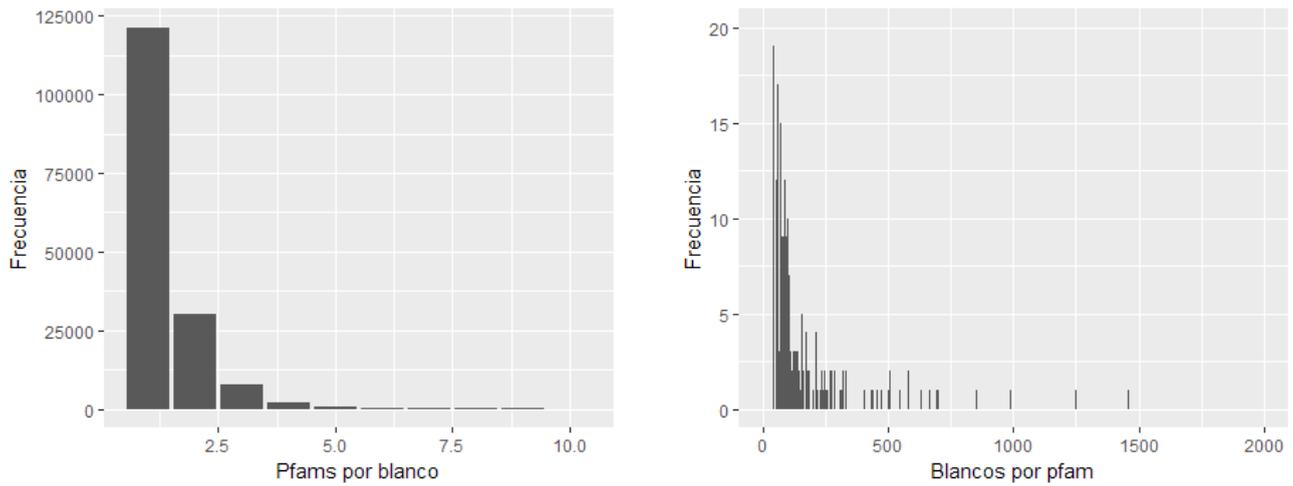

**FIGURE 5.1:** Left: histogram of the number of PFAMs per protein. Right: histogram amount of proteins affiliated by PFAM.

In the case of orthologous groups, by definition each target is associated with only 1 group. Likewise, **Figure 5.2** shows that most orthologous groups have no more than a few dozen proteins. On the other hand, there are groups with more than a hundred proteins, which must correspond to the oldest and most essential cellular functions for life. The orthologous group with the most proteins is OG4_126558, which is related to ATP metabolism, which is of fundamental importance for cellular respiration in all living beings.

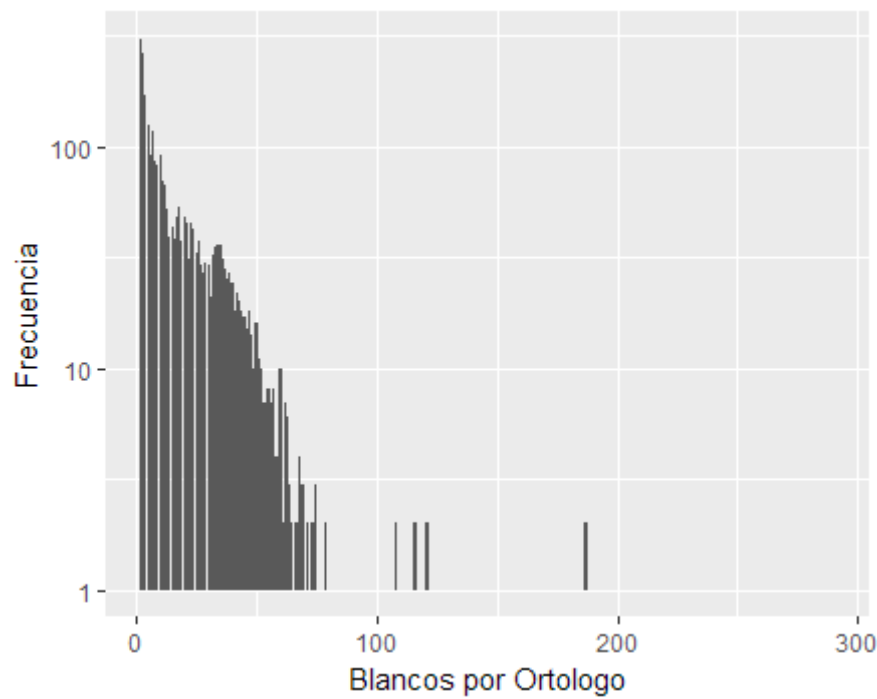

**FIGURE 5.2:** Distribution of protein quantity per orthologous group.

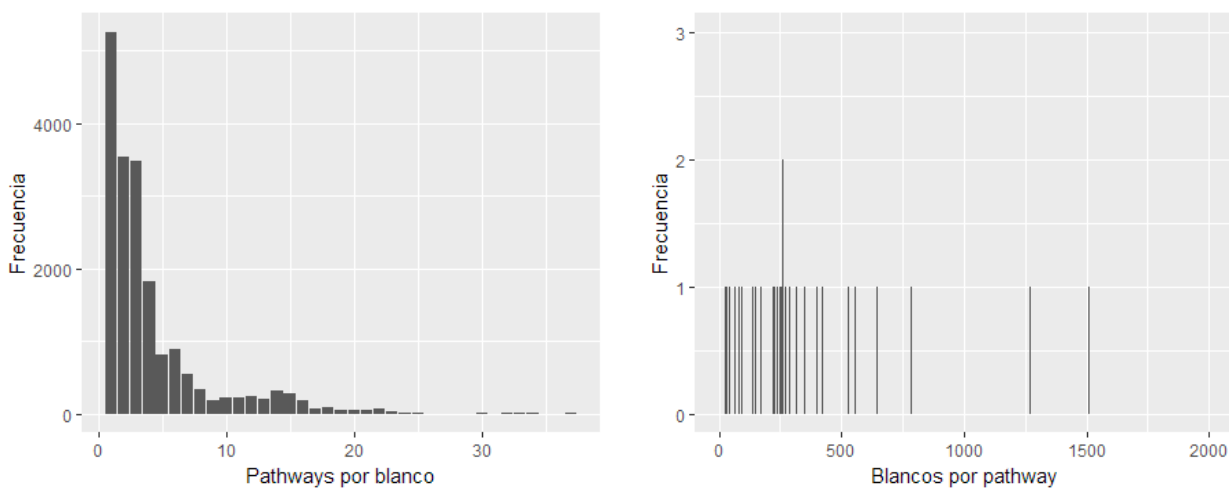

**FIGURE 5.3:** Left: histogram of the number of pathways to which the proteins are affiliated. Right: histogram of the amount of proteins affiliated per pathway.

The number of pathways per targets (figure 5.3) is to be expected since, in general, proteins participate in a few connected metabolic pathways. The pathway distribution of proteins is quite broad and relatively uniform, reflecting the variety and varying degrees of complexity and hierarchies of metabolic pathways.

## *5.2 Relevance of annotations*

### 5.2.1 Druggability and annotations

As mentioned above, one objective of our work is to prioritize proteins through a first neighbor algorithm. For this, we will project the explicit relationships in the bipartite target-annotations network to introduce similarity relationships between targets. In this way, the targets will be linked to the extent that they share characteristics and annotations that we consider relevant, possibly related to their *drug-ability. As* we already mentioned, the characteristics that we take into account are related to concepts such as having a common evolutionary origin (sequence similarity) or share structural building blocks (ie pfam domains).

It is important to note that we can actually anticipate that not all categories will be equally relevant in terms of acting as a drug-ability proxy. For example, **Figure 5.4** shows that different pfam domains have different fractions of annotated dopable proteins.

In **figure 5.4** , those annotations with the highest quantity of druggable proteins are labeled with names. Among the domains with the most associated proteins reported as *druggable,* we find that there are, for example kinases (*Pkinase, Pkinase_tyrion*), channels (*Ion_trans*), g-coupled receptors (*7tpm_1*) and nuclear receptors (*Anf_receptor*).

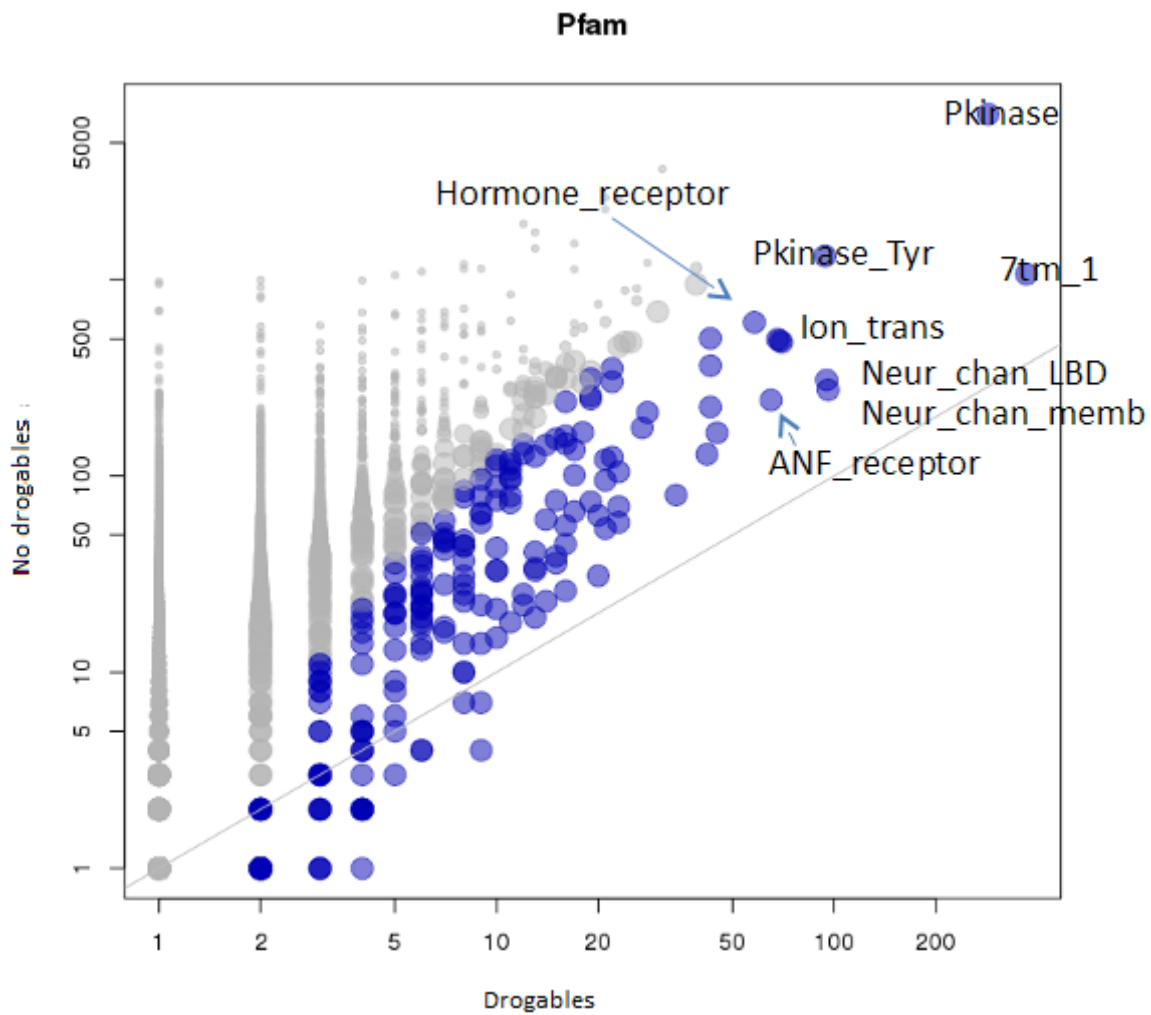

**FIGURE 5.4:** Quantity of druggable (*Drogables*) and non-druggable (*No-drogables*) proteins per PFAM. The size of each circle corresponds to the fraction of druggable proteins with respect to the number of non-druggable ones. The PFAM domains with the highest quantity of druggable proteins are indicated.

Kinases are enzymes that participate in phosphorylation reactions that are typically associated with the activation of molecules in intracellular signaling pathways. Nuclear and G-protein coupled receptors are involved in the reception of transmembrane messages. Ion channels regulate the passage of ions and small molecules through cell membranes.

These 4 types of proteins participate either in intracellular signaling chains or they can inhibit the passage or alter the sensitivity to factors external to the cell and for these reasons they are the type of proteins most sought after when starting the development of a new drug. since they increase their success rates [32]. Our analysis shows that this trend is reflected in the data.

### 5.2.2. Druggability and p-values

As seen above, not all annotations are associated with the same amount of druggable proteins. It is then possible to think of assigning to each annotation an index that talks about the relevance of knowing that a given protein is affiliated with a given category, in relation to the druggability condition.

For this, we use a statistical measure related to the Fisher's Test, which allows us to quantify the degree of association between being druggable, or not being, and being listed, or not, in a given category. The importance of each category in particular is then given by the 2 x 2 contingency table that aggregates the information of how many proteins $VP$ are (or are not) listed in it, and how many of them are (or are not) drug targets.

The network contains a total of N proteins of which $B$ are targets of some drug. On the other hand, the affiliation node $V_B$ contains in total $V_B$ annotated proteins of which $V_p'$ are targets of some drug.

The probability that by chance the node VB has exactly that number of annotated targets follows a hypergeometric distribution that is calculated exactly by the equation

$$P_{V_B}(V_p') = \frac{\binom{\#V_B}{V_p'}\binom{N-\#V_B}{\#B-V_p'}}{\binom{N}{\#B}}$$

(5.1)

The importance of an annotation is then associated with the p-value that the contingency table has for said distribution. Smaller p-values indicate that it is more unlikely that the distribution of druggable proteins of an annotation was randomly generated. Figure 5.4 shows in blue those annotations that are significant, that is, with a p-value lower than the 20th-percentile.

In previous works [33] it was decided to define for the purposes of the correct operation of the recommendation system and to incorporate the p-values. The way in which the importance of each annotation $j$ in this work is established through the "Relevance score".

$$R_{score}(j) = \begin{cases} 1 & si \quad p_j \leq q_{0.2} \\ [\frac{-log_{10}(p_j)}{máx_j(-log_{10}(p_j))}]^\alpha & si \quad p_j > q_{0.2} \end{cases} \tag{5.2}$$

where $p_j$ is the p-value of the Fisher test and $q_{0.2}$ denotes the lower 20-percentile of the distribution of p-values.) As you can see, the structure of the function is a $log(p)$, normalized so that the function is bounded between 0 and 1. Where the relevance value 1 corresponds to 20% of the annotations with the lowest p-value.

For each type of annotation, there is a different distribution of p-values, observing **table 5.1** it can be seen that the category with the most relevant annotations is that of PFAM domains. It is also the one with the smallest average p-value and the lowest p-value annotation with which this will be the category of greatest importance when prioritizing.

|  | Number of significant | Fraction of significant |
|---|---|---|
| Pfam | 335 | 0.148 |
| Orthology | 347 | 0.124 |
| Pathway | 18 | 0.124 |

**TABLE 5.1**: Characteristics of annotations by layers.

It is important to distinguish that defining the relevance through the p-value is very different from the simplest characterization of the druggable protein fraction, in particular, this difference is notable in annotations with high degrees but a druggable protein fraction less than 0.5 . If we take an annotation with 1000 proteins and 50 druggable, the p-value is 0.0130, while for a more specific annotation but with a higher fraction of druggable proteins, as is the case of an annotation with 50 proteins and 5 druggables, it is reported a p-value of 0.0330. This effect is evidenced in **Figure 5.4** where it is noted that at higher values of the p-value the normalized frequency of pairs of high fraction of druggable proteins per annotated protein decreases.

### 5.2.3 Entropy of annotations by species

In the previous section we introduced a way to quantify how relevant a category can be considered a priori for the process of suggesting new druggable proteins. This was done on the basis of existing evidence, which mostly corresponds to data obtained from model organisms. Our focus is on identifying new molecular targets for species associated with neglected diseases. These generally involve little-studied proteomes, and it is necessary to know to what extent shared annotations serve us to "transfer" drug target information between species. For this we proposed to analyze the inter-species promiscuity of each annotation, quantifying it by means of the entropy defined according to

$$S = -\sum_{i=1}^{k} p_i log(p_i),$$  **(5.3)**

where $p_i$ is the proportion of proteins of the species $i$ associated in the annotation of interest.

In general, the entropy is zero if the annotation presents proteins from a single species and is maximum if the proteins affiliated with it are homogeneously distributed between species. In the case of an annotation that presents associations with proteins of $k$ species in the same proportion (frequently $1/k$) it is $S_{max} = -\sum_{i=1}^{k}(1/k)log(1/k)$ and then $S = -log(1/k) = log(k)$ . Therefore the maximum entropy varies according to the number of species to which a PFAM is affiliated. This allows us to define a quantity that only indicates how uniformly the proteins of annotation are distributed (regardless of the number of species, which we denote normalized entropy $S_{normalizada} = \frac{S}{S_{max}}$.

The following figures show bar graphs for the normalized entropy distribution by annotation and the number of species recorded. Also included in the graph of entropy as a function of species reached. Logarithmic horizontal axes are used to show the fitted line and how it compares with the case of maximum entropy.

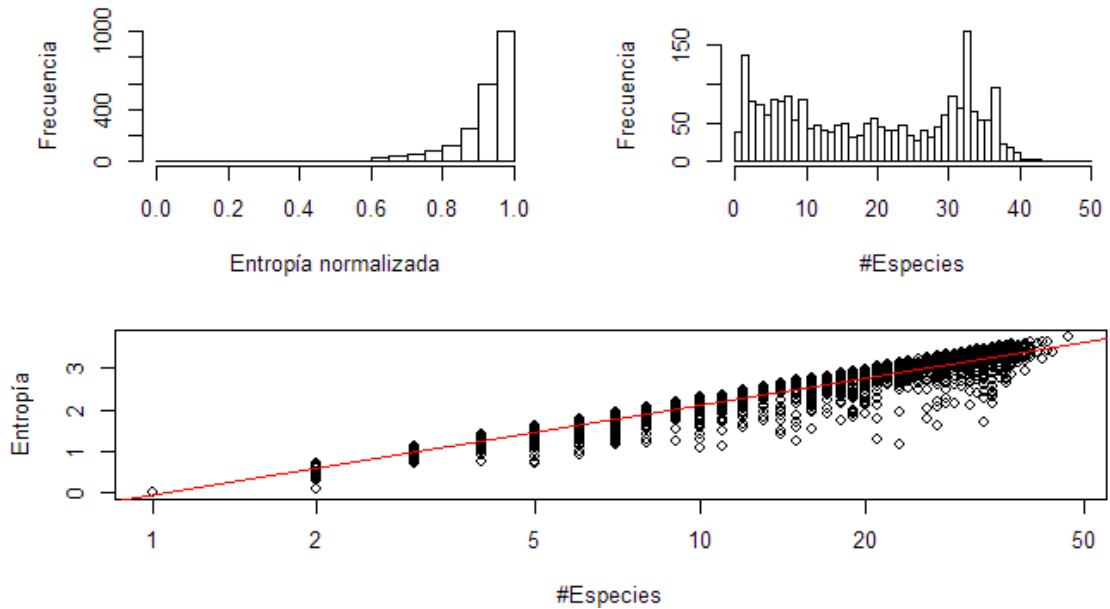

**FIGURE 5.5:** Upper panel: Bar graph of PFAM annotations with an entropy and a number of affiliations to given species. Lower panel: entropy vs number of species with affiliations. The line obtained has a gradient $0.94\pm0.01$.

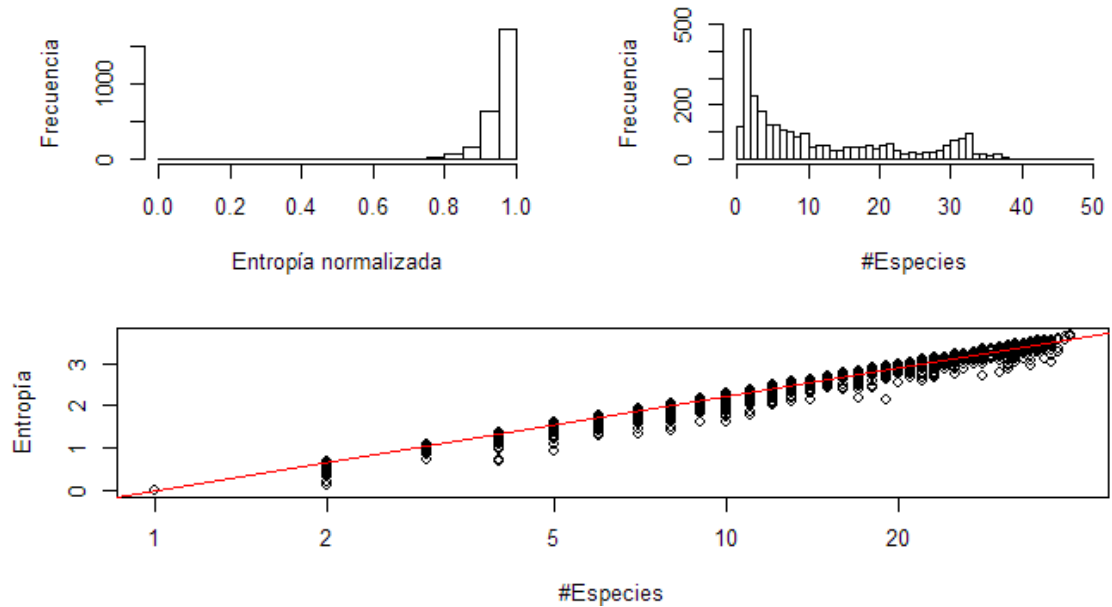

**FIGURE 5.6:** Top panel: Bar chart of orthologous groups with a given entropy and number of species affiliations. Bottom panel entropy vs number of species with affiliations. The slope of the adjustment made is $0.970 \pm 0.002$.

In the case of PFAM class annotations (**figure 5.5**) and orthology (**figure 5.6**) it can be seen that the normalized entropy distributions are similar, more bulky for high values in the second (the same characterization can be done for the number of species) . When adjusting the entropy as a function of the logarithm of the number of species per annotation, both categories have slopes close to 1, and therefore it is noted that the entropy distribution is close to the maximum value regardless of the number of species, so they play a role important for connectivity and dissemination processes in the information network between species. In the case of pathways (**figure 5.7** ) or metabolic paths, the correlation between species numbers and entropy is similar to that of the other types of annotations but less, as it can be observed by the dispersion of the points in the lower panel of the figure. 5.7.

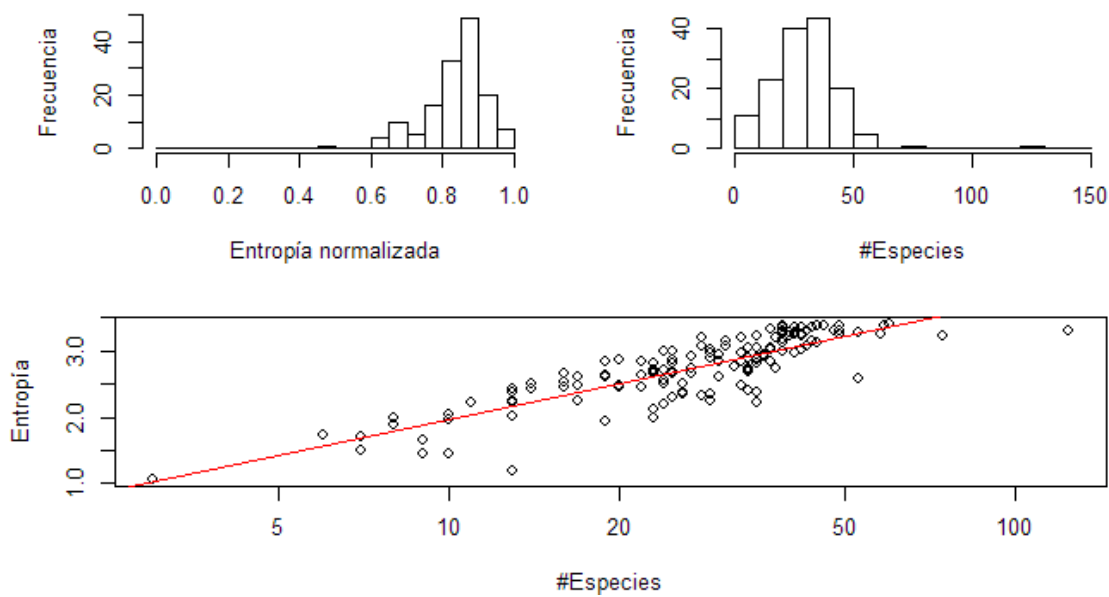

**FIGURE 5.7:** Upper panel: Bar graph of Paths with an entropy and a number of affiliations to given species. Lower panel: entropy vs number of species with affiliations

### 5.2.4. P-values and entropy

Once we have defined our two quantities of interest, a first one that characterizes the over-representation of druggable proteins in each annotation (p-value of the fisher test or Rscore) and a second associated with the annotation's ability to transfer information between species (entropy S), questions arise about the relationships between them since they will affect our prioritization methods. Figure 5.8 reports the obtained Rscore distributions for given inter-species entropy cuts for the case of PFAM annotations.

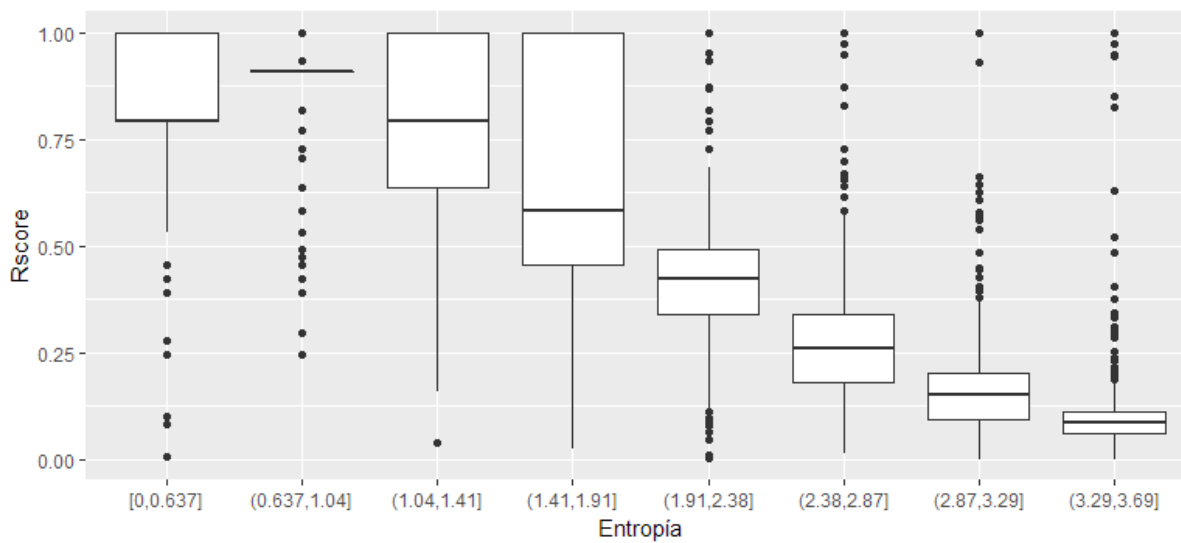

**FIGURE 5.8:** Entropy versus Rscore plot for orthology annotations (a similar plot is obtained for PFAM). The Rscore measures the importance of annotations to define the druggability of a protein, while entropy measures how easily it transmits information between species. On average this shows that there is a trade-off between the ability to disseminate information and biological relevance. In particular, the outliers by entropy excess will be of interest.

It is observed that in general, for annotations with higher entropy (ie inter-species connectors), relevance scores of lower value are obtained. However, it is interesting to analyze which are the particular annotations that present high relevance and at the same time high inter-species entropy. These annotations (table 5.5) will be those that will make it possible to disseminate information on the 'dopable' nature of one species to another through the network.

The domains of interest that appear in Table 5.5 are related to enzymes that either synthesize or serve to repair DNA. For example *DHFR_1* is found in all species and has a central role in the synthesis of nucleic acid precursors, while *Thymidylate synthase* participates in the first stages of DNA biosynthesis, and is selected as a target to control the growth and development of various types of cancer, by chemotherapy [34]. Ribonucleotide reductase (RNR), is an enzyme that catalyzes the formation of deoxyribonucleotides whose function is conserved in all living beings, regulates the rate of DNA synthesis, and has been used as a target to prevent the replication of herpes simplex virus (HSV)[35] . As we have just shown, these are all essential processes in many species. Furthermore, because they are essential, they are generally chosen as drug targets.

| PFAM | Protein# | Fraction of druggable targets | #Especies (#NTD) | Rscore |
|---|---|---|---|---|
| PFAM186 *DHFR_1* | 61 | 0262 | 47 (21) | 1 |
| PFAM303 *Thymidylat_synt* | 53 | 0188 | 40 (22) | 1 |
| PFAM317 *Ribonuc_red_lgN* | 43 | 0139 | 36 (21) | 1 |
| PFAM2852 *Pyr_redox_dim* | 170 | 0088 | 40 (24) | 1 |
| PFAM2867 *Ribonuc_red_lgN* | 45 | 0.133 | 37 (22) | 1 |

**TABLE 5.2**. Characteristics of PFAM annotations with maximum Rscore and entropy.NTD refers to the species that generate neglected tropical diseases.

The first of the groups is associated with domains *Ribonuc_red_lgN*, *Ribonuc_red_lgC*, *ATP-cone* that generally appear in ribonucleotide proteins. The second is associated with *tRNA-synt_2d*, *Phe_tRNA-synt_N* they are linked to catalyze the attachment of an amino acid to its tRNA (transmitter RNA), for the formation of new proteins in ribosomes. The third group in the table corresponds to the one with the highest Rscore, and is associated with the PFAM *DHFR_1*, *Thymidylat_synt domains* that are of interest in their own right (**Table 5.2**). These domains are conserved throughout the majority of living beings, and are related to the synthesis of nucleic acids. The fourth group has as its main domain *Prenyltrans*, which is associated with the creation of precursors for cholesterol, steroid hormones and vitamin D in vertebrates. For the first and third orthologous groups, in **table 5.3**, we find agreement with the aforementioned for the PFAM domains of **table 5.2**.

| | # Proteins | druggable Fraction #species | (#NTD) | Rscore |
|---|---|---|---|---|
| ORT10318 | 45 | 0.133 | 38 (22) | 0.82 |
| ORT10500 | 40 | 0.100 | 40 (24) | 0.51 |
| ORT10927 | 49 | 0.204 | 49 (22) | 1 |
| ORT11497 | 40 | 0.150 | 36 (22) | 0.94 |

**TABLE 5.3.** Characteristics of the orthology annotations with maximum Rscore and entropy. NTD refers to the species that generate neglected tropical diseases.

### 5.3. Analysis of similarities of layers

Many spaces coexist in the network who inform us of chemical similarities of drugs, of structure and sequence of proteins, and of bioactivities. In the previous sections, we have referred on the one hand to the drug layer and on the other hand to the protein layer in conjunction with the annotations layer. Here we will elucidate how consistent is the information that the drug layer and annotations give on the protein layer. We take advantage of the fact that the scheme of interrelationships between drugs, targets, and annotations allows us to infer different notions of similarity between the entities that make up our network. For example, we can establish a criterion of similarity between proteins based on the similarity of the drugs that target them, or alternatively, based on the PFAM annotations they share or the orthology groups to which they belong.

To analyze the coherence of similarities between proteins induced by pfam and orthology annotations, we will consider the partition of proteins into groups based on their membership in orthology groups and compare them with clusters detected through different algorithms, in the trivial protein network. projected by PFAM annotations and pathways alternately. The values obtained for the comparisons are detailed in the following table. The measurements are used to compare partitions, NMI and Rand as defined in section A.3.2.

| | NMI | Rand |
|---|---|---|
| fast_greedy | 0.80 | 0.97 |
| infomap | 0.83 | 0.98 |
| louvain | 0.81 | 0.97 |

**TABLE 5.4**. Values obtained for PFAM projection with different clustering methods

|  | NMI | Rand |
|---|---|---|
| fast_greedy | 0.36 | 0.71 |
| infomap | 0.27 | 0.44 |
| louvain | 0.37 | 0.74 |

**TABLE 5.5.** Values obtained for Pathways projection with different clustering methods

In all cases We found a close relationship between the orthology classification and the PFAM domains that is much greater than between the first and the metabolic pathways. This occurs regardless of the clustering method or the comparison index, so we can say that it reflects an intrinsic property of the network of relationships between annotations and proteins. Likewise, this allows us to establish that the PFAM domains and the orthology groups in general provide complementary but coherent information since, despite representing different characteristics of the proteins, they do not present contradictory information. If the division into communities differed greatly between the different knowledge domains considered, there would be a situation in which typically the similarity / difference information between proteins induced by one and the other knowledge domain would differ, which when integrating the information into a prioritization it would be unprofitable.

Once the similarity induced by sets of proteins between the different types of annotations has been described, it is interesting to evaluate to what extent the similarities between proteins, inferred by annotations, coincide with the similarities that can be inferred from their bioactivities.

In **Figure 5.9** we show the probability that a pair of dopeable proteins share a drug as a function of the number of shared PFAM domains. It can be observed that the greater the number of shared PFAM annotations, the probability of having a drug in common (blue line) or having drugs that belong to a common cluster (red line) increases up to 3 annotations and then oscillates around 0.7. As expected, the red line is related to higher probabilities that a protein pair will share a drug, and presents less variation for more than 4 annotations, since for 0 common annotations the probability that a drug is shared is $1,05 \times 10^{-6}$ In view of the above, it is established that the number of shared PFAM domains generates greater similarity between drugs that are active in proteins.

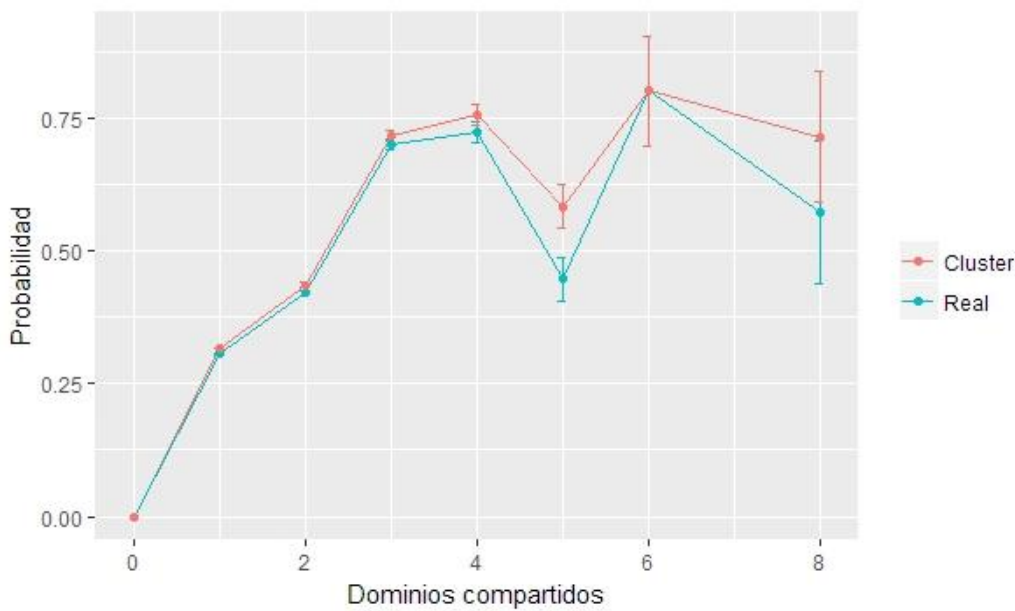

**FIGURE 5.9:** Number of shared PFAM domains *(Dominios compartidos)* per druggable protein versus the probability that the pair also shares a drug. In blue we take the event of sharing a drug in a "strong" way, that is, there is a bioactivity from the drug to both proteins in the pair. In red the event of the relaxed condition of having some drug in the same S or T cluster.

## *5.4.Conclusions*

In this chapter, we detected categories that present a preference for scoring proteins with reported bioactivities. We saw that this is a reflection of biases that actually exist in the area of development and search for new drugs. Many categories have a very important role in the interconnectivity between proteins of different species, and we quantify this through the entropy of the distribution of proteins by species. This measure is very relevant to understanding how knowledge can be transferred between species.

We also analyze the coherence between the different types of annotations through similarities to first neighbors in the network. We established that the information that the orthologous annotations layer imposes on the protein layer is similar to that of PFAM. Furthermore, we established the existence of a link between common drugs, proteins, and the number of shared PFAM annotations. A clear growing relationship between both magnitudes can be observed and that approximately 60% of proteins that share 3 or more domains in their architecture are reached by at least one drug in common.

# Chapter 6

Network projected in annotations

In the previous sections, the drug, target, and annotation layer were studied and characterized. This section will advance the analysis of the interconnectivity between them. In particular, we will be interested in elucidating whether linkages induced by dopable targets reflect non-trivial structure in the biological annotations space.

For this, we will make use of the concept of projection of a bipartite network into a one-part network introduced in section 2.4. Through this procedure we will establish associations between annotations (PFAM or orthology) induced by the existence of proteins in common that have pairs of annotations. Thus, the fact that two annotations are strongly connected in the projected network will imply that the annotations have a high level of overlap in terms of the proteins that each one has associated with. Furthermore, if for this estimation we only consider druggable proteins, strong links between PFAM or orthologous categories, they could suggest that they are associated with biological functions or structures that appear in molecules that have been repeatedly used as targets. In general, the networks that result from both types of projections (ie considering all proteins or only dopeable ones) will have different structural properties, as illustrated by the panels in **figure 6.1**, where the giant components of the PFAM network projected by proteins and by proteins reached by drugs respectively. We will make a brief characterization of these differences in the following sections of this chapter.

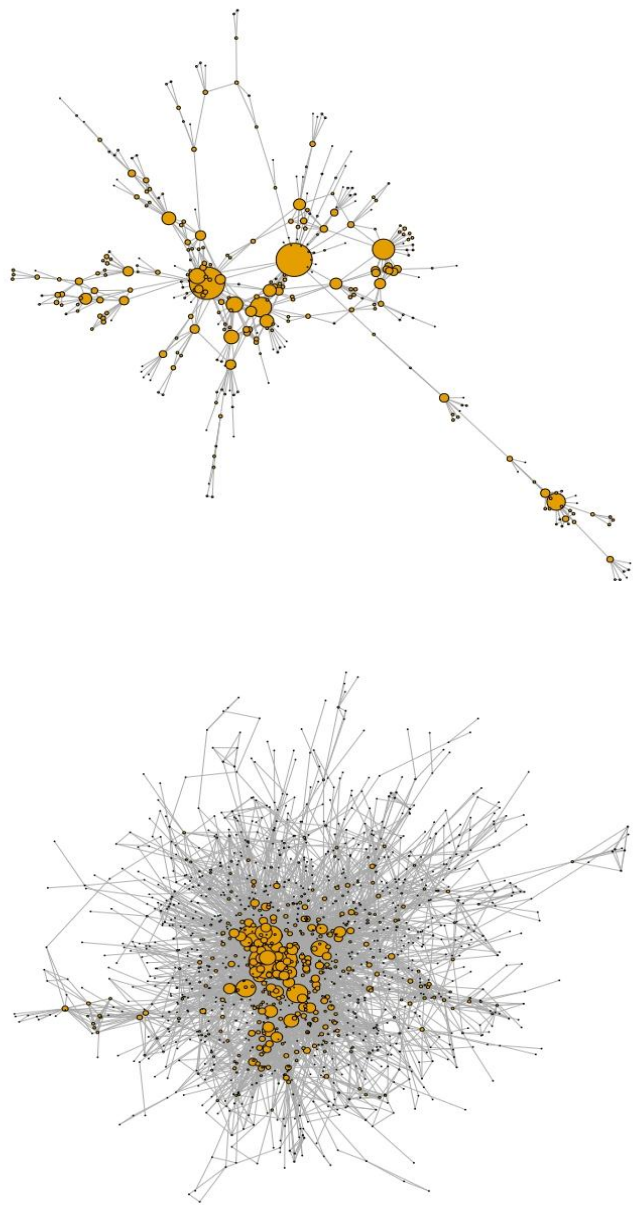

**FIGURE 6.1:** In the upper panel is the giant component (CG) projection of the drug-drug targets affiliation graph to PFAMs, the size of the nodes is associated with the amount of drug-drug proteins noted. The fan structure related to the negative

assortativity coefficient is visible. In the lower panel the same projection but taking into account all the proteins, the size of the nodes is associated with the amount of proteins annotated.

## 6.1. Projection methods

It is possible to consider different methodologies to examine the relevant structures in the annotation network, related to different projection alternatives with which to obtain a one-part network of annotations (see section 2.4). We will consider the three projection methodologies introduced in Chapter 2: trivial projection, diffusion probability projection, Probs [24] and projection by statistical validation, StatVal [25].

Under the trivial projection, the weight of the bond that will join two annotations is equal to the number of proteins in common that they have. As we saw before, eq (2.8), is calculated according to

$$w_{i,j} = \sum_{l=1}^{m} a_{il} a_{jl} \tag{6.1}$$

According to the ProbS methodology, the adjacency matrix of the annotation network will be given by the expression of eq (2.11).

$$w_{i,j} = \frac{1}{k_{x_j}} \sum_{l=1}^{m} \frac{a_{il} a_{jl}}{k_{y_l}} \tag{6.2}$$

Now, the contribution of a protein common to a given pair is down-weighted (ie penalized) by its degree of connectivity $k_{y_l}$. At the same time the adjacency matrix is no longer symmetric, since the weight of the link will depend on the connectivity of the 'arrival' annotation $k_{x_j}$. It is important to note that these two projections present the same characteristics of underlying connections in the sense that what varies is the weight of the links and the directionality. For each link in the naive network there are two links with different directions and different weight in Probs.

Finally, in the statistical validation projection (StatVal), the links are determined by the existence of common proteins, of a given degree, in a proportion significantly higher than that expected by chance. This means, for example when taking PFAM annotations, that the links between two annotations of this type are defined by systematically analyzing, for each group of proteins that present a given number of PFAM domains, if there is an unexpectedly high number (in the statistical sense ) of common proteins. To gain insight into this procedure, **figure 6.2** shows the number of validated links between annotations considering projections armed with proteins of a given degree. It is possible to observe that most of the associations come from the validation of linkages induced by proteins of grade less than 4. The difference observed between the absolute values of the number of validated linkages between those obtained considering all proteins or only druggable ones arises from the fact that druggable proteins are only 6% of the total proteins, which results in a greater penalty when making the selection of edges with the hypergeometric distribution.

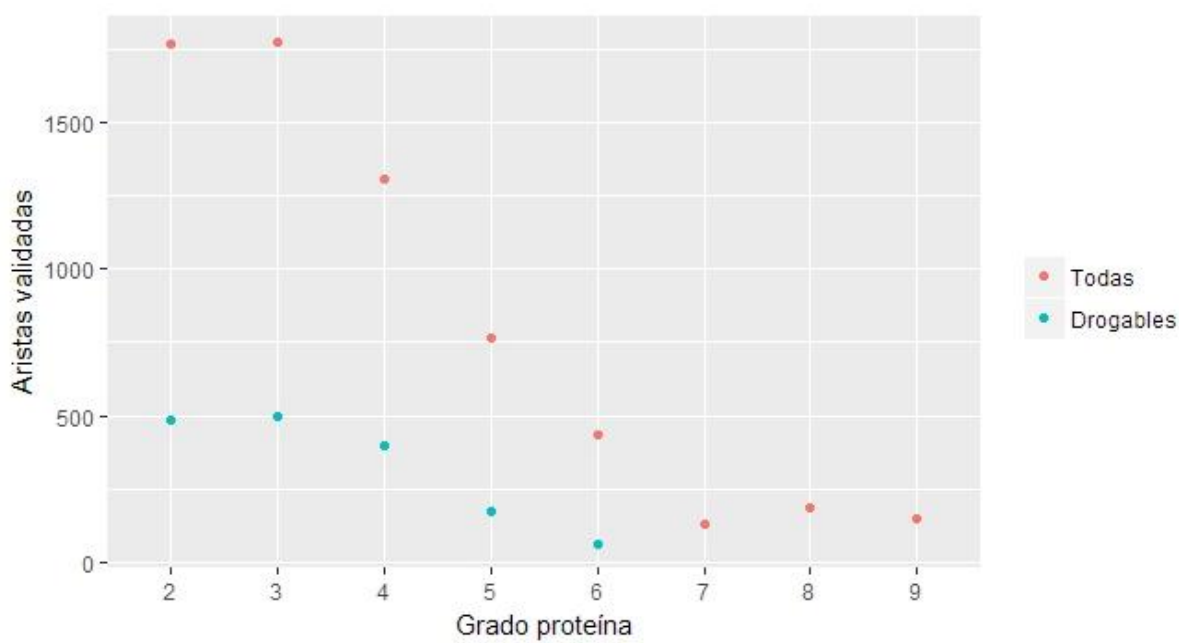

**FIGURE 6.2:** Graph for the number of validated links (Aristas validadas) by degree of annotation (Grado de proteina) using a threshold of p-value 0.01. As the statistical validation projection depends on the subgraph induced by all proteins of a certain grade, it is seen that for low grades both projections differ by more number of bonds than for high grades.

## 6.2 Networks obtained

In **TABLE 6.1.** interesting data on the structure obtained for each of the projections made are shown. It is interesting to note that due to the characteristics with which the validated links are obtained in the statistical validation method, the number of nodes in the projected network is significantly reduced, since the annotations for which no relevant link was found by this method are extracted from the net. In all cases it happens that when all proteins are considered to be projected, the main component obtained by projecting only on druggable proteins is enlarged. In other words, the effect of taking into account non-druggable proteins only creates intra-component bonds (i.e., within already connected structures) or between components and the giant CG component, see A.1.2.

| Projection | projected with proteins | #components | giant component | #nodes | #links / stregth total |
|---|---|---|---|---|---|
| Naive / Probs | all | 729 | 1340 | 2252 | 4653 / 606.28 |
|  | dopeable | 1272 | 335 | 2252 | 1732 / 704.87 |
| StatVal | all | 208 | 868 | 1422 | 2296 |
|  | dopeable | 141 | 12 | 287 | 366 |

**TABLE 6.1**. Characteristics of the graphs obtained through projections.

**Table 6.1** reflects the size and scale of the structures obtained with the different methods of projection. We see that in statistical validation it is presented as a more restrictive methodology, in the sense that the number of links is substantially less. In particular, for the projection of only druggables, this involves 1/10 of the nodes and 1/7 of the links of those obtained in the other projections.

What's more. When taking only druggable proteins, the amount of links decreases in Naive and ProbS by 63%, while in StatVal it decreases by 85%. Due to the loss of links, the decrease in the size of the giant component can be observed for all the projection methods (in particular, for the StatVal projection, it leaves nodes without links, which we remove from the network). In general, the projected network with dope drugs results in a much more fragmented distribution, with a giant component that comprises only 15% of the network's mass (5% in the case of StatVal), compared to the 60% obtained in projections made. considering all targets.

## *6.3 Topological measures of interest*

### 6.3.1 Transitivity

Transitivity or clustering coefficient, as can be read in section A.2.3, refers to the probability that two nodes that have a common neighbor are in turn neighbors between them. In our case, for example, it answers the question: if two PFAMs each have some common protein with a third PFAM domain, how likely is it that they will score the same protein together? Obviously, this number may not be the same depending on whether we are considering exclusively druggable proteins or if we analyze the complete set of available targets. In any case, it is possible to test the statistical significance of the transitivity values found by establishing a comparison with what was obtained considering a random network, in which the degree of each node of the projected network is maintained, but which nodes are selected at random. they are linked to each other by an edge.

We present in **table 6.2** the global transitivity values of the projected protein network considering the trivial projection (similar results are obtained for the other projection methodologies). The first and second columns of the table show the value obtained without considering, or considering, the weight of the links respectively.

|  | Global (random) | Barrat (random) |
|---|---|---|
| dopeable | $0,45(0,06 \pm 0.02)$ | $0,819(0,200 \pm 0,032)$ |
| proteins | $0,20(0,0 \pm 0,01)$ | $0,619(0,0910 \pm 0,033)$ |
| liableaddictedCG | $0,28(0,14 \pm 0,01)$ | $0,753(0,068 \pm 0,034)$ |
| proteinsCG | $0,20(0,18 \pm 0,01)$ | $0,586(0,037 \pm 0,021)$ |

Table 6.2.Coefficients of clustering graphs obtained through projection and trivial compared with the network chance.

As a first comment, it is possible to notice from **table 6.2** that in general the giant components exhibit lower clustering coefficients than those of the complete network. This is due to the fact that the complete network includes, in addition to the giant component, numerous very small clique-like structures (less than 4 elements). These small components, densely connected internally, increase the mean value of transitivity reported in the case of the entire network.

Likewise, we see that considering drug-resistant proteins, projections are obtained with coefficients greater than those obtained in control networks (ie, rewired at random with identical degree distribution) and greater transitivity than when all proteins are used (with all proteins, the global transitivity is 0.206 while in the one originated through drug targets, 0.459 is reported). This suggests that drug target-related PFAMs tend to form more locally interconnected clusters. That is, we found that in drug-only protein networks, linkages induced by drug-drug co-annotation generate more interconnected local environment structures for pfam annotations. This may reflect a cooperative effect of the domains in the sense that proteins that include such domains together have a greater chance of being druggable.

### 6.3.2. Centrality

In network studies, the concept of centrality is associated with a measure of the importance of a node with respect to its neighbors and is not uniquely defined but depends on the algorithm applied. The simplest measures of centrality of a node to formulate are the degree and the strength (see appendix). In our case, the PFAM domains with the greatest strength in all networks were found to be PFAM69 and PFAM1. PFAM1 represents GPCR receptors like *Rhipdosin*. PFAM 69 represents the kinase domain. These domains are related to structural characteristics usually used to guide the search and design of new drugs [32]. Furthermore, for drug-only projections, we find that strength positively correlates with Rscore (see **figure 6.3**). For example, considering the strength vs Rscore relationship, we obtained correlation values of 0.29 and 0.28 for projections obtained with druggable proteins using the trivial methodology and that of statistical validation, respectively. Both are significant when compared with those obtained from randomly rewired networks: $(0,07 \pm 0.03)$ and $(0.0 \pm 0.1)$ respectively.

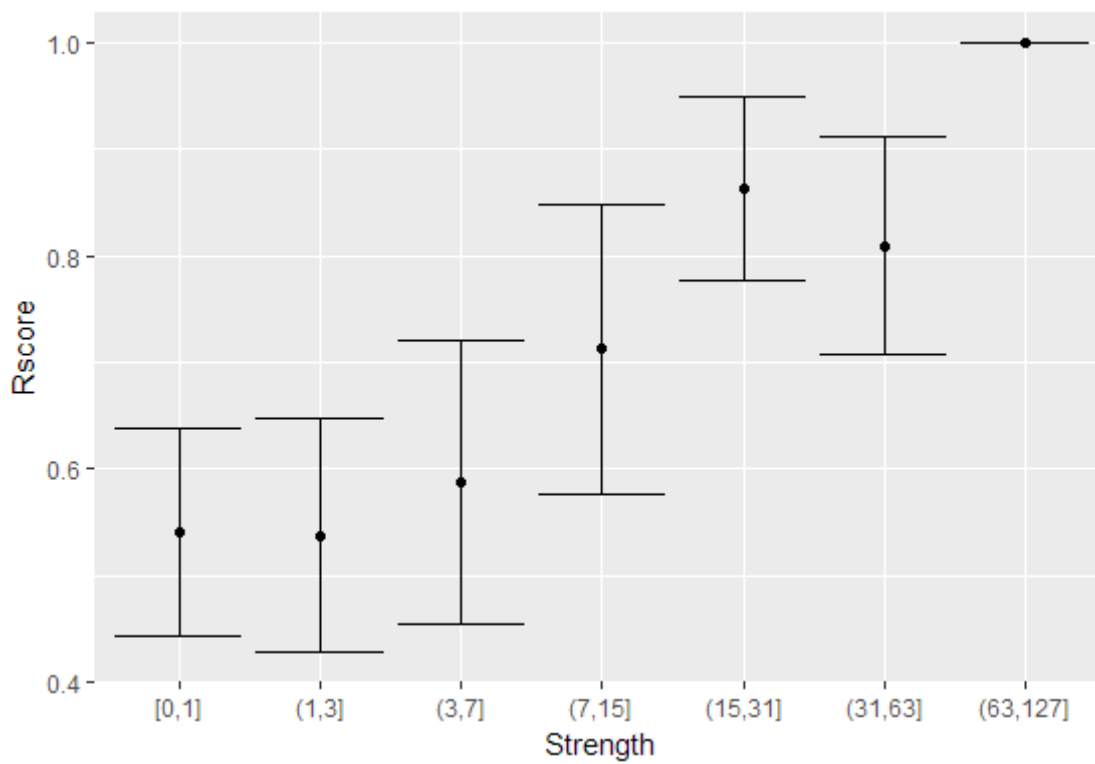

**FIGURE 6.3:** Relationship between Strength and average Rscore for sets of vertices taken by logarithmic intervals in Strength. These correspond to the trivial projection of dopeable.

Additionally, various notions of centrality can provide us with different and complementary information about the network. The measures that were analyzed alternately are the Kleinberg centralities (hub and authority), and eigen centrality (defined in A.2.4 of the appendix) in the context of PFAM annotation networks.

In **tables 6.5** and **6.6** we report PFAM domains of interest, identified as those that presented high values of centrality and Rscore~ 1. The first table includes results obtained for ProbS projections, while the second for trivial and statistical projections. For the latter, it was found that the maximum values obtained corresponded to the same nodes, regardless of the centrality measure chosen.

|  | Hub | Authority | Eigenvector |
|---|---|---|---|
| All | PFAM69 * PFAM433 * PFAM169 * | PFAM659 * PFAM12474 * PFAM11629 * | PFAM69 * PFAM433 * PFAM659 * |
| Bioactive | PFAM69 * | PFAM659 * | PFAM69 * |

|  |  | PFAM433 * PFAM130 * | PFAM12474 * PFAM11629 * | PFAM433 * PFAM130 * |
|---|---|---|---|---|

**TABLE 6.5** Centrality values obtained with the Probs projection. The nodes that are of greatest interest are marked with * since they present centrality and Rscore values close to 1. PFAM 69 kinase PFAM659 *Pole kinase*, PFAM12474 *TGF beta receptor 2* , PFAM11629 *C-terminal SARAH*. AND PFAM 69 PFAM 433 kinases PFAM 2931 ion channel neurotransmitter PFAM 2932 ligand binding domain.

| Projection type | All centralities |
|---|---|
| Trivial Bioactive CG | PFAM69 * PFAM433 * PFAM130 * |
| Trivial Bioactive | PFAM2931 * PFAM2932 * PFAM413 * |
| Trivial all / CG | PFAM271 PFAM270 PFAM176 |
| StatVal Bioactive / CG | PFAM520 * |
| all | PFAM65V Stat1PFAM1 |
| PFAMFAM1PFAM1 | Stat10 PFAM10 * PFAM1012 |

**TABLE 6.6**: Table for the centrality values obtained with the StatVal and trivial projection. The nodes that are of greatest interest are marked with * since they present centrality and Rscore values close to 1. PFAM 69 PFAM 433 kinases, PFAM 2931 ion channel neurotransmitter PFAM 2932 ligand binding domain, PFAM413 C1 domain, PFAM 413 *Peptidase_M10,* PFAM 1 *7 transmembrane receptor* , PFAM 10 *Helix-loop-helix DNA-binding* domain and PFAM520 Ion_trans.

In general, it can be seen that for the trivial projection with all its proteins there are no clear cases of annotations that simultaneously present high centrality and Rscore. However, in other cases, the annotations of greater centrality have a high Rscore and it is observed that some of these domains belong to the four main types, seen in section 5.2.1, which are usually searched as targets. Namely, these are kinases: Kinase PFAM 69 PFAM659 *Polo*kinase, *PFAM433* Protein Kinase. Protein-coupled receptors *PFAM12474* g *TGF beta receptor 2,* and the ion *channel PFAM2931 Neur_chan_LBD, PFAM* **520** *Ion_trans.* In this way we see how the structure of interconnections in our data reflects the bias of the industry in the search for drugs towards signaling proteins (kinases), channels, receptor proteins, etc.

Finally, it is interesting to mention that the annotations of maximum centrality reported in **tables 6.5** and **6.6** are different from those reported in chapter 5 and that they presented maximum entropy and high Rscore. This is because centrality gives an idea of how annotations are related to others through shared proteins, while entropy measures the ability of annotations to transfer information between species, to the extent that high values of the introduced entropy In the previous chapter, it allows identifying annotations associated with a structural characteristic common to proteins from different species.

### 6.3.3. Associativeness

To understand the structure of the network on a global scale and detect non-trivial relationships between annotations, it is sometimes useful to be able to measure whether similar nodes tend to link to similar nodes. This is done by calculating the assortativity of the network (see appendix). In our case, we are interested in knowing whether annotations with many druggable proteins tend to link in the projected network with other annotations that have a similar amount of druggable proteins (or similar Rscore values). For this, the assortativity of Rscore and the self-grade in the different projections were measured. The proper degree refers, in the case of projections on bioactive proteins, to the amount of bioactive proteins that the PFAM domain records, and in the case of projections on all proteins, to all the proteins that this domain records.

|  | Own grade (Azar) | Rscore (Azar) |
|---|---|---|
| PFAM bound by CG | $-0,111(-0,094 \pm 0,0211)$ | $0,350(0,259 \pm 0,025)$ |
| dopables PFAM bound by CG proteins | $-0,064(-0,059 \pm 0,030)$ | $0,302(0,288 \pm 0,045)$ |

**TABLE 6.7**: Associations obtained for probs / trivial projections. The assortativities are indistinct due to what was observed before, between one projection and another, only the weight of the edges and the direction changes.

In general, it is noted from **Table 6.7** (and it is found for the rest of the projections) that the projections that take into account only the set of druggable proteins are the ones that present the highest degree of disassortativity. It is seen that this is reflected in **figure 6.1** , where the characteristic fan-type structure of networks with this property is discerned. This implies that domains with many annotations (ie large ones in the figure) tend to co-appear in druggable proteins with other less common domains. However, we cannot say that the assortativity is significantly different from that obtained in a randomly rewired network, which maintains the original degree distribution. For this reason, we cannot rule out that the values obtained arise as a consequence of said distribution and do not come from non-trivial correlations of two bodies.

On the other hand, for the projection taking druggable proteins into account as observed in **table 6.7** (and it is verified for the rest of the projections) the assortativity of the Rscore is appreciably large, compared to the random rewiring, maintaining the distribution of degree and with the obtained in the network of projected with all the proteins. This implies, as occurs with the Rscore-transitivity relationship, that druggable proteins tend to nucleate around PFAMs with high relevance score, and also that domains with low Rscore tend to share druggable proteins with others with low Rscore.

## *6.4 Prioritization of drugs from strongly related domains*

As we have seen in the previous section, the connectivity between annotations induced by drug targets allows us to reveal interesting structural characteristics that present proteins that are drug targets. In this section, we will be interested in investigating whether there is any relationship between any of the structures detected and certain specific drugs. To carry this out, we consider the projected network, considering only druggable proteins, by restricted statistical validation (see section 2.4.2).

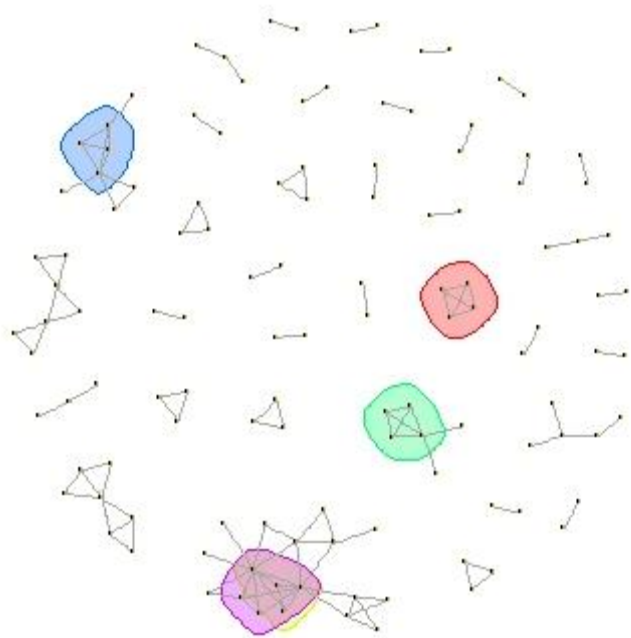

**FIGURE 6.4:** Projected graph with statistical validation taking into account only druggable proteins and with the calculation correction only on components.

Figure 6.4 shows a diagram of the network of PFAM annotations obtained, where 4-cliques (that is, completely connected components of 4 elements) were identified. These structures represent strongly connected PFAM domains thanks to the existence of druggable proteins in common, and what we wondered was whether it was possible to identify the existence of some specific group of drugs responsible for the bioactivities involved.

To investigate this, each of these 4-cliques of PFAM annotations was considered, at a time, as seeds with unit scoring for a prioritization process using a voting methodology towards the protein layer. The score for each protein was then propagated by making VS back to the drug layer. In this way we assign a score to each drug according to

$$W = PAV \qquad (6.3)$$

where $A$ is, as before, the protein-annotation adjacency matrix, $P$ is the drug-protein adjacency matrix, and $V$ is the seed vector, that is, it has 1s in the PFAM of the 4-clique that is analyzed and 0's in the rest.

**Table 6.9** consign a summary of the results of this procedure. In the second column, the PFAMs belonging to the clique from which it is prioritized are recorded. The third column shows the set of drugs that obtained the 5 highest prioritization values. The fourth includes a brief description of their common spectrum of action. In the fifth column, it is mentioned which diseases and which actions had the 5 highest rated drugs (which also targeted a protein with the 4 domains). Columns 5 and 6 show the results of the validation exercise of the obtained drugs, which will be explained below.

To validate the significance of the ranking that allows identifying relevant drugs, two alternatives were considered. In the first instance, it was considered to generate rankings using a null model that included prioritizing drugs according to equation 6.3 from 4 PFAM annotations chosen at random. By repeating this procedure 1000 times, drug rankings were made, recording the percentage of times that the same 5 top-ranked drugs appeared. The other method we use is based on quantifying the "ranking power" of the scores produced, that is, the inhomogeneity with which scores are distributed among the drugs from a given set of seeds. For the ranking and therefore the selection of the most scored elements to be good, there should be a clear difference between the highest score levels and the rest of the elements, that is, it differs from the null model. The null model proposed to analyze statistical validity assumes that the normalized score $x$ generated by the PFAM clique is assigned for each element (in this case drug) ranked with a uniform distribution. This gives a probability value $\alpha = (1-x)^{k-1}$ that there is a higher score than $x$ when the score is uniformly distributed among *k* ranked elements [36]. In our case, *k* is the cardinality of the set of drugs that receive a non-zero score from a given set of seeds. Low values of $\alpha$ will indicate unusually high scores with respect to this control. As a measure of confidence in the ranking that produced them, Table 6.9 shows the maximum value $\alpha$ of the best-ranked drugs selected.

The results are shown in **table 6.9**. The 4-clique 3 and the 4-clique 4 were essentially indistinguishable when taking into account the 5 most scored drugs. For each case the selected drugs show common action spectra, although for none of the cases the most scored belonged to the same Tanimoto cluster. or substructure. The information thus derived is not trivially contained in the chemical similarity information and, in general, it is noted that the drugs thus found have common mechanisms of action, as reported in the PubChem and Chembl databases.

| Annotations | Drugs | Joint action of drugs | Probability with random seeds | Alpha validation (maximum) |
|---|---|---|---|---|
| PFAM69 (*Pkinase*) PFAM433 (*Pkinase_C*) PFAM168 (*C2*) PFAM130 (*C1_1*) | D 34576 * D 453394 D 453399 * D 32183 * D 40110 | Kinase inhibitors, related to diabetes and cancer treatments candidates for | 2.9% | $2,707042 \times 10^{-09}$ |
| PFAM520(*Ion_trans*)PFAM11933(Na_trans_cytopl) PFAM612(IQ) PFAM6512(Na_trans_assoc) | 34576 D * D 453391 D 453394 D 370532 D 363932 D 511441 | Drugrelated vasodilation and epilepsy. Mainly calcium blockers. | 0.6% | $5,561837 \times 10^{-13}$ |
| PFAM41 (*fn3*) PFAM7714 (*Pkinase_Tyr*) PFAM1030 *Recep_L_domain* PFAM757 (*Furin-like*) | D 3571 D 30211 D 116 479477 D 154D 36752 D 62971 D 35548 D 38936 | Drugs associated with different types of cancer | 2.4% | $9,796857 \times 10^{-03}$ |
| PFAM41 (*fn3*) PFAM7714 (*Pkinase_Tyr*) PFAM536 (*SAM_1*) PFAM1404 (*Ephrin_lbd*) | D 34576 * D 453391 D 453394 D 453399 D 511441 | Drugs associated with different types of cancer | 1.3% | $1,562485 \times 10^{-06}$ |
| PFAM13855 (*LRR_8*) P1462 PFAM1 (*LRR_8*)) P1462_1*LRRNT*) PFAM12369 (*GnHR_trans*) | D 2174 * D 757 * D 7724 * D 4112 D 3571 D 33147 | Psychotropic drugs | 2.6% | $1,153909 \times 10^{-07}$ |

**TABLE 6.9**. Cliques used as seeds, with the type of drugs obtained by doing VS in the drug layer. The common uses of the drugs obtained are noted and the quality of the validation is characterized using a comparison with prioritization from 4 random annotations and the method seen in [36]. Smaller numbers of validation characteristics imply that the drugs obtained cannot have been obtained from random

seeds. D43576 staurosporine, D 453399 Vandetanib, D 32183 canertinib, D 2174 Closapine, D757 Mianserin, D 7724 Risperidone.74### 6.5. Drug prediction from domain architecture

Of the 40 links of the annotation network inferred from drug targets (**figure 6.2**), only two do not appear when all proteins are considered in the projection by statistical validation. So we wanted to investigate if they encoded some type of information related to the drug-ability of the proteins that gave rise to it.

To do this, for each of the two links, we identify, through the propagation procedure described above, drugs that could be associated with the respective PFAM domains. Then we identify targets that present an architecture of domains given by the pair of PFAMs involved with the idea of analyzing the possibility that they are druggable and that there could be a link between them and the drugs identified above.

| Domains | Drug | $\alpha$ validation | Random network |
|---|---|---|---|
| "PFAM780" "PFAM621" *CNH RhoGEF* | D 549548 (potassium metabisulfite) D 17322 (adenosine a 2 a3 receptor modulator antagonist) D 581648 (latrunculi a) | $2.163809 \times 10^{-10}$ | 2% |
| "PFAM51" "PFAM24" *Kringle PAN 1* | D 173221 D 172257 D 176 011 | $1.153909 \times 10^{-08}$ | 3% |

**TABLE 6.10** . Pairs used as seeds, which appear in the projection of proteins but not druggable proteins, used to make a VS in the drug layer. The common uses of the drugs obtained are noted and the quality of the validation is characterized using a comparison with prioritization from 4 random annotations and the method seen in [36]. Smaller numbers of validation characteristics imply that the drugs obtained cannot have been obtained from random seeds.

When performing the described prioritization procedure, the results of **table 6.10 are obtained.** It is observed that the drugs that have been validated show a very low probability of having appeared in prioritizations with random seeds and also a $\alpha$ much lower than 0.05. For the first bond that involves thedomains *CNH RhoGEF*, the drugs obtained are related to heart diseases such as *adenosine a2 a3* receptor modulator antagonist and *latrinculin a* , but to a lesser extent the effect of potassium metabisulfite has been studied and is of interest. which is a widely used preservative.

In the case of the domains, *Kringle PAN 1* the validated drugs are related to coagulation factors and plasminogenesis. Interestingly, it was found in the literature that one of the 7 proteins that present both domains in their architecture and did not report bioactivies in our network, is actually the target of the TDRdrug D 173221 , prioritized by the PFAMs of the corresponding link [37].

## *6.6. Conclusion*

We have projected to the PFAM annotations layer giving rise to 6 different graphs. We have found that the projections using only bioactive proteins were more transitive than those of all proteins, which allows us to say that the domains associate to allow activity on the proteins. We reached similar conclusions when studying the assortativity of Rscore. We established that the most central domains do not coincide with those with the highest entropy, which can have a negative effect when prioritizing.

We also recognized sets of 4 totally related domains, 4-*cliques,* for the StatVal projection of bioactive drugs. We did a prioritization exercise from the cliques to the drug layer, and we found that by doing this, drugs with a similar spectrum of action are obtained. On the other hand, we find two links with domains that appear in the projection of specifically druggable proteins but not when considering the complete set of proteins. In both cases it was found that they acted on common mechanisms, although for the results of the first there are no studies that support the action of one of the three drugs selected in the mechanism. Finally, it was established that one of the proteins that had two of the selected domains in its structure had a bioactivity from one of the selected drugs, which was not in the network.

# Chapter 7

Network growth and prioritization

There is an aspect not yet explored in our data related to the temporal dimension. The analysis of the chronological evolution of this type of chemo-genomic networks is interesting because it can serve to understand characteristics and modes of operation of both the pharmaceutical industry and the dynamics of biomedical research projects [38].

In our case, we will use the time dimension to analyze particularities of the growth of our network, such as: how and with what frequency new drugs are incorporated into the network as time passes or what relationship the new reported bioactivities have with which the drugs precede. Understanding these mechanisms will help us to contextualize the in-silico prioritization procedures that we will implement to *grow the network* through suggestions of new possible bioactivities between known drugs and possible new targets. Consistent with the general vision of this work, we will focus primarily on targeting organisms associated with neglected tropical diseases.

## *7.1. Analysis of drug-protein connections*

### 7.1.1 Timeless analysis of drug-protein connections

As a first step, we relegate the chronological description and begin our analysis by considering all the bioactivities reported in our network that involve proteins reached by a single drug. We will call the same 'unique targets' and the set of drugs that reach them 'unique drugs' (see figure 7.1). Likewise, within the group of 'unique targets' we recognize a subset of them reached by drugs of grade $k = 1$, which we will call 'hyper-unique targets' (these are proteins reached by a single drug that does not have other targets). At the same time, we will call 'hyper-unique drugs' those mono-directed 'unique drugs', that is, grade $k = 1$.

As already mentioned in Chapter 3, there are a total of 177,506 bioactive drugs and 6051 druggable proteins in the network. Of these, there are a total of 386 'unique targets' reached by a set of 262 'unique drugs'. Only 1 in 3 (86 of 262) drugs directed at unique targets have a single target (see figure 7.2). This suggests that only this fraction of drugs is possibly qualitatively novel and mono-targeted.

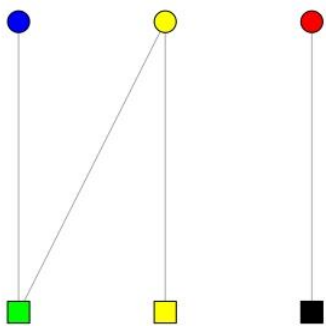

**FIGURE 7.1** Schematic of a set of drug and protein nodes. The circles represent proteins and the squares represent drugs. A single target is shown in blue, it receives only one link. The drug (green) that receives the link is therefore a "single drug" is promiscuous, since it has two links. A "hyper-unique" target is marked in red, this has a single drug associated with it, which in turn has only one link. The hyper-unique drug is marked in black.

In the same sense, in **Figure 7.2** we show the fraction of unique targets reached by unique-drugs with more than one associated target. We see that in 80% of the cases the 'unique targets' are reached by drugs that also present bioactivities towards targets that are not unique. This could be reflecting the importance of drug repositioning as an expansion strategy for drug targets.
.

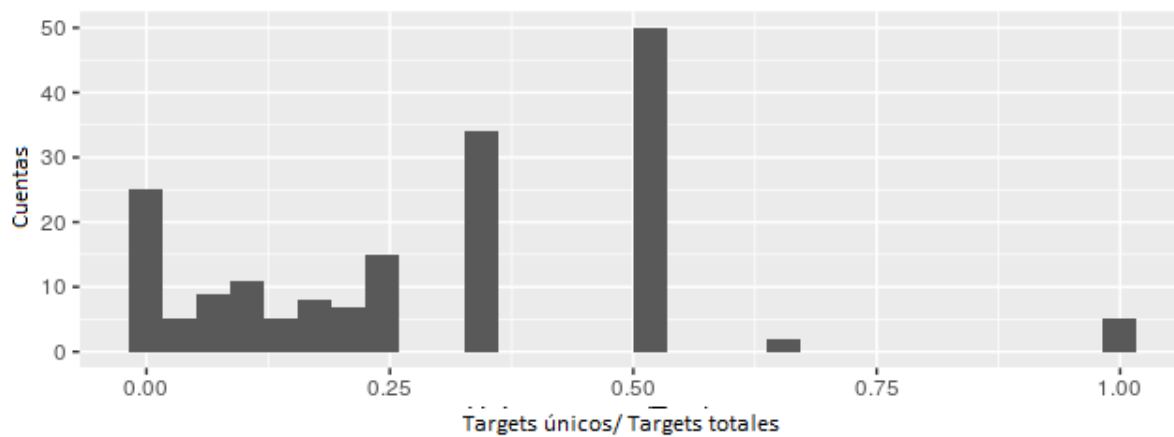

**FIGURE 7.2**: Number of drugs (cuentas) with a certain proportion of unique targets over total targets (Targets únicos/targets totales), while taking into account only drugs that target at least one "unique" target. In other words, a target that only has one drug associated with it.

As we show in **Figure 7.3**, it is interesting to note that more than 80% of 'unique drugs' target at least one protein from a model organism. This supports the idea of the importance of drug repositioning as a drug development strategy. For non-model organisms, the bioactivities associated with these drugs would generally arise from previous tests in other organisms.

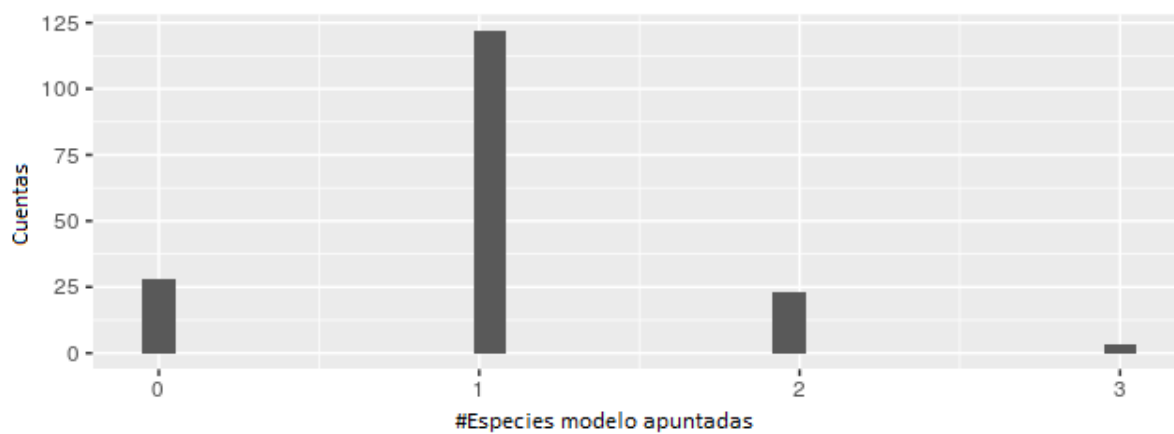

**Figure 7.3.** Number(Cuentas) of distinct model organisms targeted (#Especies modelo apuntadas) by unique drugs.

### 7.1.2 Temporal evolution

The results of the previous section seem to suggest that a large part of the new bioactivities that involve targets not yet explored arise from already known drugs. However, in order to advance in the characterization of this exploration trend, it is necessary to evaluate how bioactivity links are established to new proteins over time.

The drugs in the TDR network were extracted from 3 databases DrugBank, Chembl, and Pubchem. In order to establish the dates on which each bioactivity was published, it was necessary to split these databases, mine the information and unify it with the network. This was done for 97% of the drugs with known bio-activity from the network. That is, only for 14,926 drugs out of a total of 494,294, a publication date could not be entered.

In **figure 7.4**, using the dates of the first publication in which each drug is reported, the growth rate of the universe of drugs explored in biomedical research can be observed. The data acquisition of the TDR version used to build our network occurred mainly until 2009, which explains the decrease observed as a technical artifact.

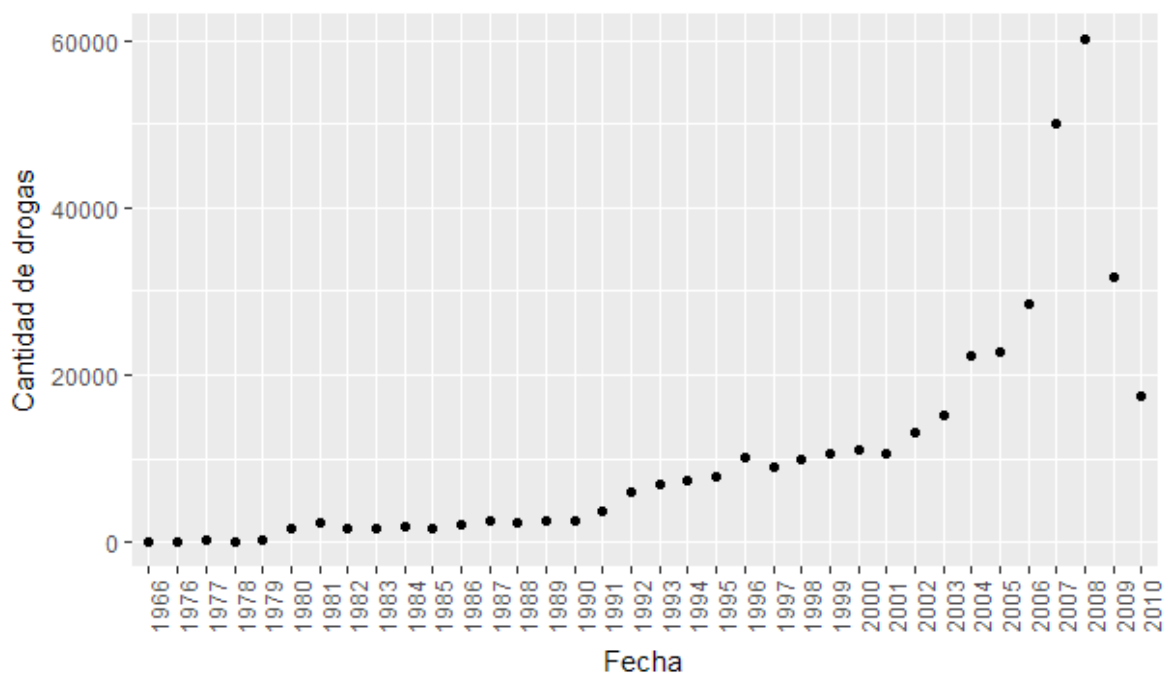

**FIGURE 7.4**. Graph with the number of new drugs (Cantidad de drogas) per year (Fecha). In other words, those for which its first published bioactivity was found in that year.

With the temporal data collected, it is possible to study patterns that describe the growth mode of the network, that is, how new bioactivities arise in the context of those already reported. To do this, collect information when each bioactivity was reported, following concepts introduced by Yildrim et al [37], classify drugs our network into two categories: *saltadoras (hoppers)* or *crawlers (crawlers)* (see **Figure 7.5**). The former include drugs for which their first reported bioactivity does not involve targets already reached by a drug previously introduced into the network. The second, on the contrary, do involve knowledge in some way already embedded in the network because they refer to new bioactivities that link: (a) novel targets (not yet reached by any drug) with drugs already incorporated into the network, and (b) new drugs with targets that already report associated bioactivities (upper-right and lower-right panels of **figure 7.5**, respectively)

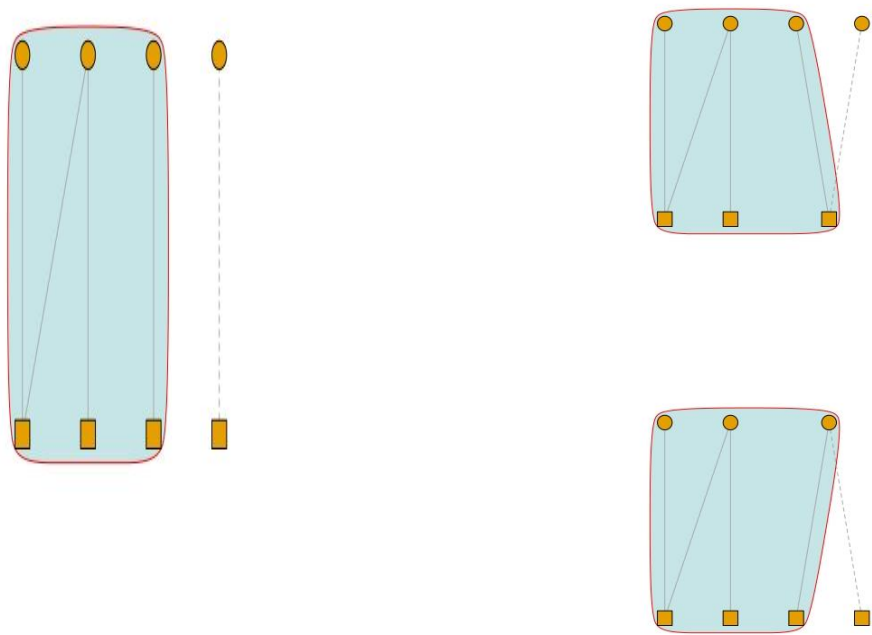

**FIGURE 7.5**. Growth modes of the drug network (squares) proteins (circles) with the original network marked on a celestial globe. The left panel shows growth by drugs *jumping* (hopping), a drug is added to the network in such a way that it is bioactive to a novel drug (new drug not previously linked to the rest of the network), and they form a disconnected entity from the network. rest of the graph.mechanisms are illustrated on the right panels *Crawling*. In the upper part of the right panel, growth is observed through the addition of a novel protein, while in the lower panel, growth is via the addition of a drug that has at least one bioactivity to old proteins (ie already bound to the rest of the network).

The upper panel of **figure 7.6** shows the evolution of the number of drugs in each category as a function of time. It is possible to appreciate that most of the incorporations to the network involve drugs of the type *crawling*. The *saltadoras* moreover, not only represent a small fraction of the total additions, but their incorporation rate remained relatively constant during the period. In the lower panel of the same figure, we include information on the growth of bioactivities reported year by year, differentiating contributions that point to targets not yet reached by any drug.

From the data, it can be seen that 6,339drugs were reported *jumping(left panel,* **figure 7.5***)* for the period 1970–2010, which implies that only 10% of all drugs appeared for the first time associated with novel targets. In turn, the targets of such drugs constitute 56% of all new targets (see left panel), while the other 44% correspond to crawling (see **figure 7.5,** lower left panel) introduced during the entire period under study. 74% of the bioactivities added correspond to old targets, therefore, in general, most drugs are added by creeping on the network, mostly increasing links to old targets.

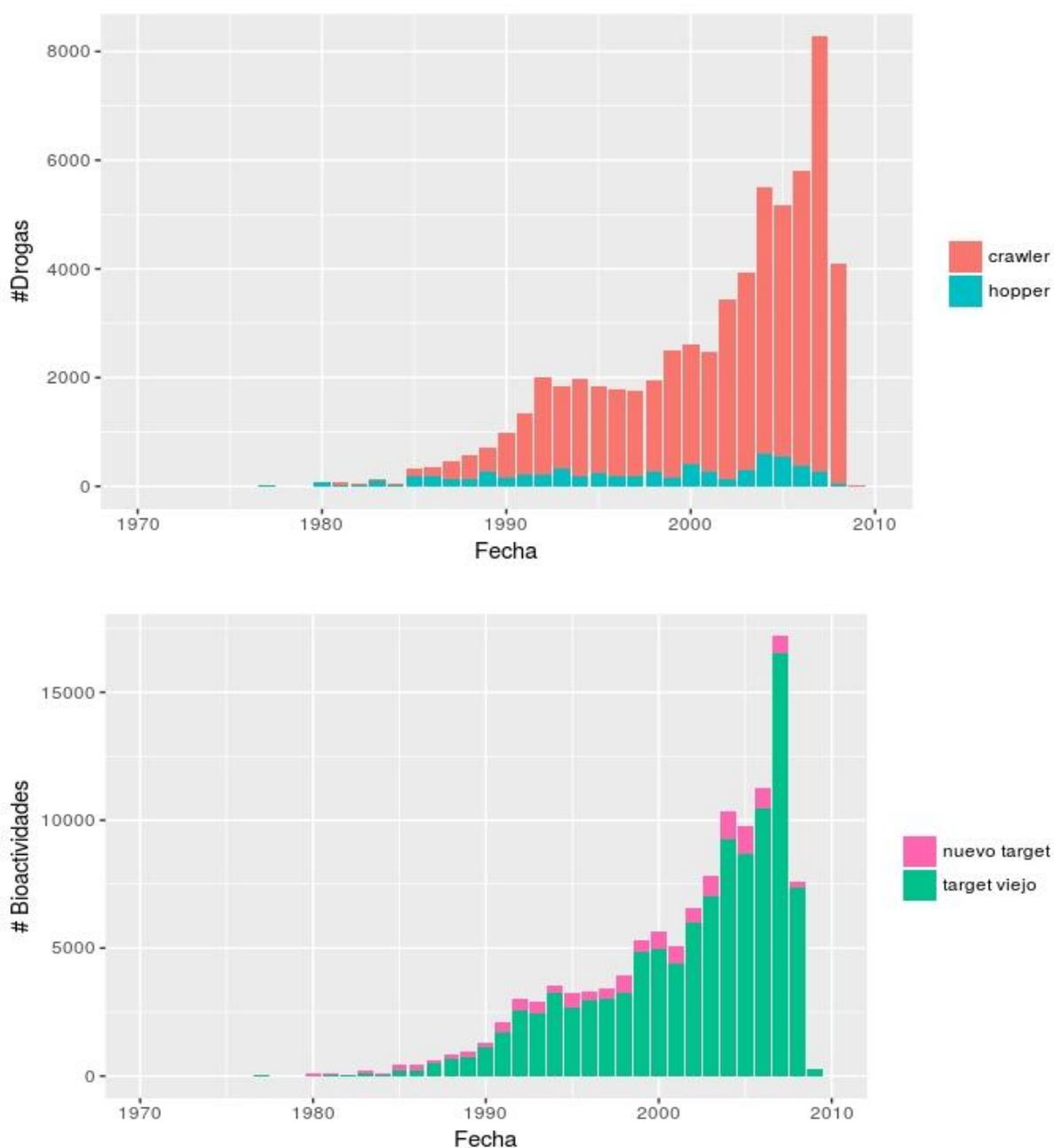

**FIGURE 7.6**. Graphs obtained taking into account all the drugs and proteins of the dated TDR network. In the upper panel, the number of new drugs (Drogas) per year is shown and a distinction is made if they are hoppers (they share a bioactivity on the date of inclusion with a previous drug) and crawlers (that if they share targets with a previous drug). The lower panel shows the amount of new bioactivities per year, and a distinction is made between target bioactivities, with previous bioactivities or not.

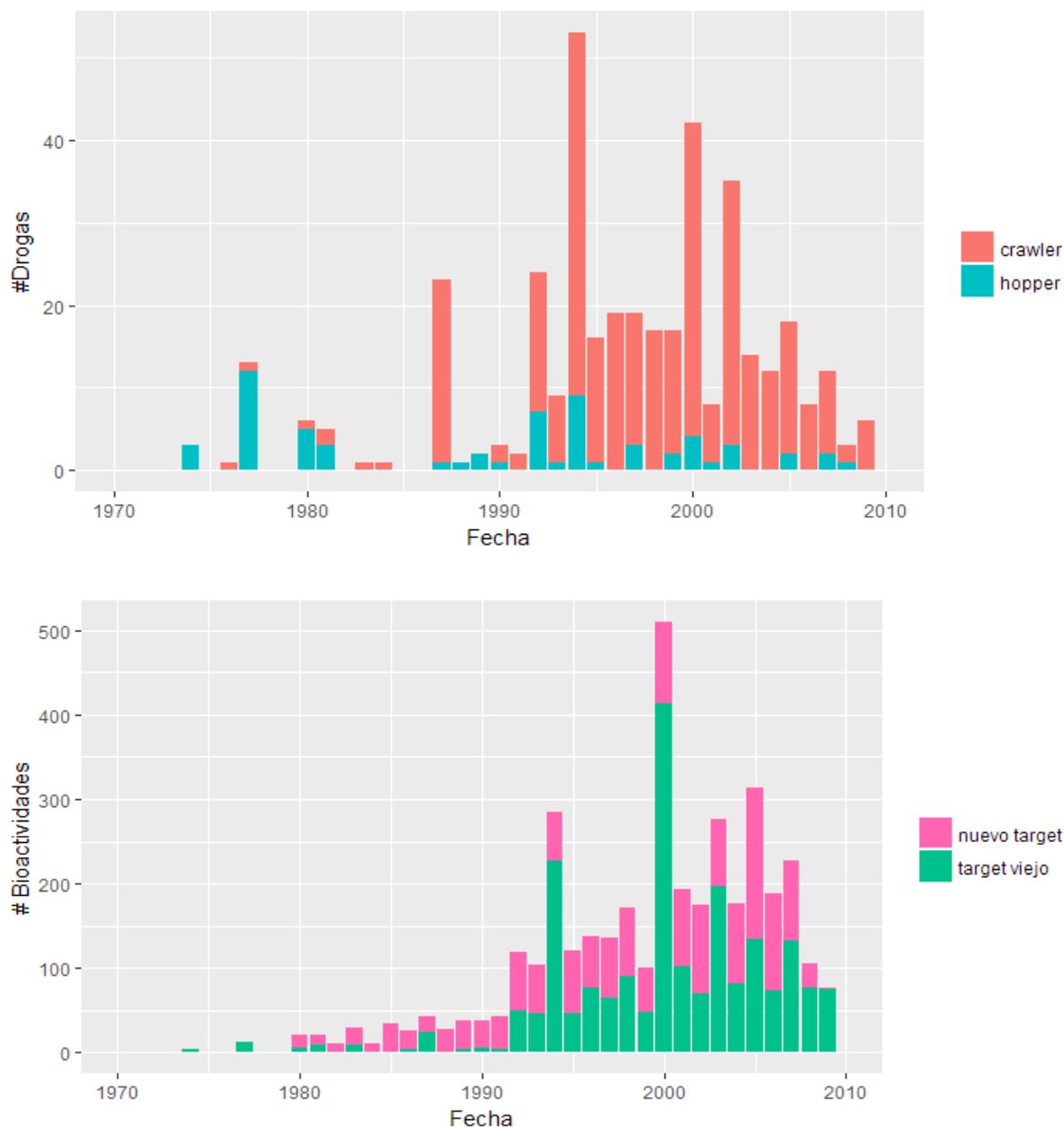

**FIGURE 7.7**. Graphs obtained taking into account all drugs approved for use in humans and dated TDR network proteins.

Finally, **Figure 7.7** shows the temporal evolution of the number of drugs and bioactivities approved by the FDA for use in humans. What has been observed allows us to suggest a scenario for the development of drugs associated with TDR. On the one hand, the abundance of 'crawler' type drugs suggests that when bioactivity was introduced into the species associated with TDR, the drug already presented bioactivity to a target from another species. On the other hand, the fraction of new targets compared to known ones is higher taking into account the graph of only these organisms, in particular 0.269 for the case of only approved bioactivities in humans, versus 0.097 for the original network. All of this is compatible with a drug repositioning strategy to reach novel targets in TDR pathogens.

## *7.2. Prioritization*

From what has been seen in the previous section, the typical growth mode of the chemogenomic network involves a dynamic where the new associations between drugs and targets are not entirely new, but are established, for example, involving targets that already present reported bioactivities with other molecules. This is compatible with the idea that, in practice, the industry advances using a conservative strategy, type *crawling*, from the knowledge already acquired.

In the same sense, it is then possible to think of developing a strategy *in-silico* that, based on the knowledge of known bioactivities, predicts new associations between targets and drugs. This is exactly what we plan to do in this chapter by introducing prioritization methodologies over our network. For this, we will consider the prioritization scheme introduced in section 2.3.1, to carry out prioritization tasks for complete species. In other words, we are going to remove all the evidence of bioactivity related to a species of interest and, using computational techniques, we will score each target of that species with a druggability score obtained using the rest of the information embedded in the network. We will use the performance metrics presented in 2.3.2 as the evaluation strategies for our predictive procedure. Finally, we will try to contextualize the results obtained in light of the structural properties and information flow of our network.

To carry out our objective, we consider the protein network obtained by means of the ProbS projection from the bipartite target-annotation network and we identify drug targets of different species as seeds that we are interested in analyzing. To make the projection, we took into account the information on the apriori relevance of each annotation captured in their respective Rscore values (5.2). We define the protein-protein matrix of the projected network connectivity, W, by:

$$W = \hat{A} R \hat{A}^t \tag{7.1}$$

where $A$ is the adjacency matrix and R is the diagonal matrix of elements $r_{ll}$, that is, the Rscore of the annotation $l$.

As a prioritization strategy we used a voting scheme or, as it is known in the jargon, KNN (K Nearest Neighbors) with K = 1, on the protein network projected with ProbS- The validation method we considered was to evaluate the AUC0.1 on the ROC curve, which compares the FPR and TPR rates.

### 7.2.1. Validation of the prioritization and target-species linkage methods

The analysis protocol presented was applied to carry out the prioritization of the entire genome of 5 different species: 3 species of model organisms and another 2 species of pathogens. The pathogens considered were: *trypanosoma Cruzi* (cause of Chagas disease) and *Plasmodium falciparum* (cause of malaria). *Homo Sapiens*, Mus musculus (mouse), and *Saccharomyces cerevisiae* (yeast) were the models considered. *Mus Musculus* is used primarily for laboratory testing prior to enabling the use of a drug in humans. While *Saccharomyces Cerevisiae* is an organism that has been studied exhaustively because it is an organism of easy handling for which there are numerous experimental techniques developed.

In **Table 7.3** the result of validating the predictions obtained in each caseis

|         | tcr  | pfa  | mmu  | hsa  | sce  |
|---------|------|------|------|------|------|
| AUC 0.1 | 0851 | 0705 | 0745 | 0641 | 0495 |

**Table 7.3** AUC 0.1 .Results for validating whole organism *Homo sapiens* (hsa)presented,*Mus musculus*(), mmu*Plasmodium falsiparium* (pfa), *Saccharomyces cerevisiae*(sce), and *Trypanosoma crucis* (tcr).

As we can see, the prioritization in pathogenic organisms with the proposed methodology was very satisfactory. Good mouse validation was also obtained. However we see that for human and yeast the results were not of the same type. In particular, values of AUC.01 close to 0.5 were obtained for this last species, which, as we saw in section 2.3.2, corresponds to making a random prediction on the drug-ability condition. In the next sub-section we try to understand why this difference occurs.

### 7.2.2 Structure, flow in the network and prioritization

In order to understand the score achieved in our recommendation exercise, let us remember that we remove all the information of a given species $sp$, so we use the knowledge embedded in the network regarding bioactivities the rest. For this reason, it is worth studying the interconnection pattern and information flow between different species induced by the prioritization process. For that we introduce the notions of: participation, *P*, and of extra-specificity, *O*:

$$P(i) = 1 - \sum_j \left(\frac{S_{i,j}^{sp}}{\sum_j S_{i,j}^{sp}}\right)^2 \tag{7.2}$$

$$O(i) = \frac{\sum_{j|j \neq sp} S_{i,j}^{sp}}{\sum_j S_{i,j}^{sp}} \tag{7.3}$$

with, $S^{sp}$ is the scoring matrix for validation in species *sp*, whose component *ij* corresponds to the score for a protein $i$ en $sp$ generated by the proteins of species *j*.

We then see that while participation is associated with the diversity of species from which a given protein receives a score, extra-specificity quantifies the fraction of the total score that comes from contributions from other species.

In our case, the total score $\Sigma(i)$ receiving a protein of a given species $\Sigma(i) = \sum_{j \neq sp} S_{i,j}^{sp}$ is.

To understand the result of the validation exercise reported in **Table 7.3**, **Figures 7.7**, **7.8** and **7.9** show the relationship between participation and extra-specificity for each of the druggable proteins of the different species considered. Each symbol represents a druggable protein, the size is proportional to the score received and its position in the plane reflects the flow of score received from the interspecies connection. It can be shown that the inner parabolic shape corresponds to the case in which there are two species that contribute scores in part equal to the protein. In this case it is obtained that the participation as a function of the extra-specificity results $P = -2(O^2 - O)$.

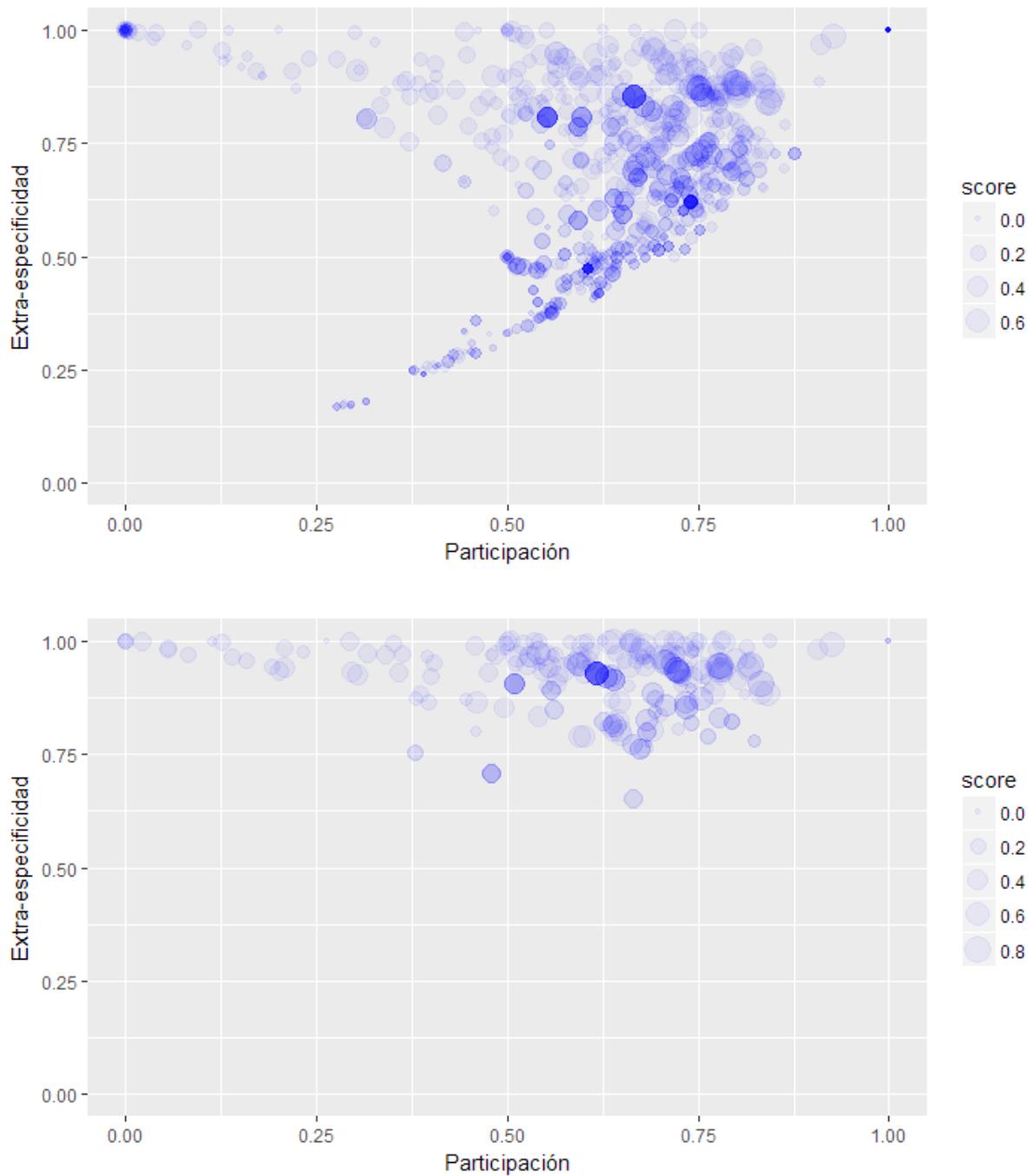

**FIGURE 7.7**. Graph with the scoring results for dopeable proteins and border parameters for *Homo Sapiens* (upper panel) *Mus musculus (lower panel).* The extra-specificity(Extra-especificidad) refers to how much score comes from another

species and the participation (participación) to how it is distributed.

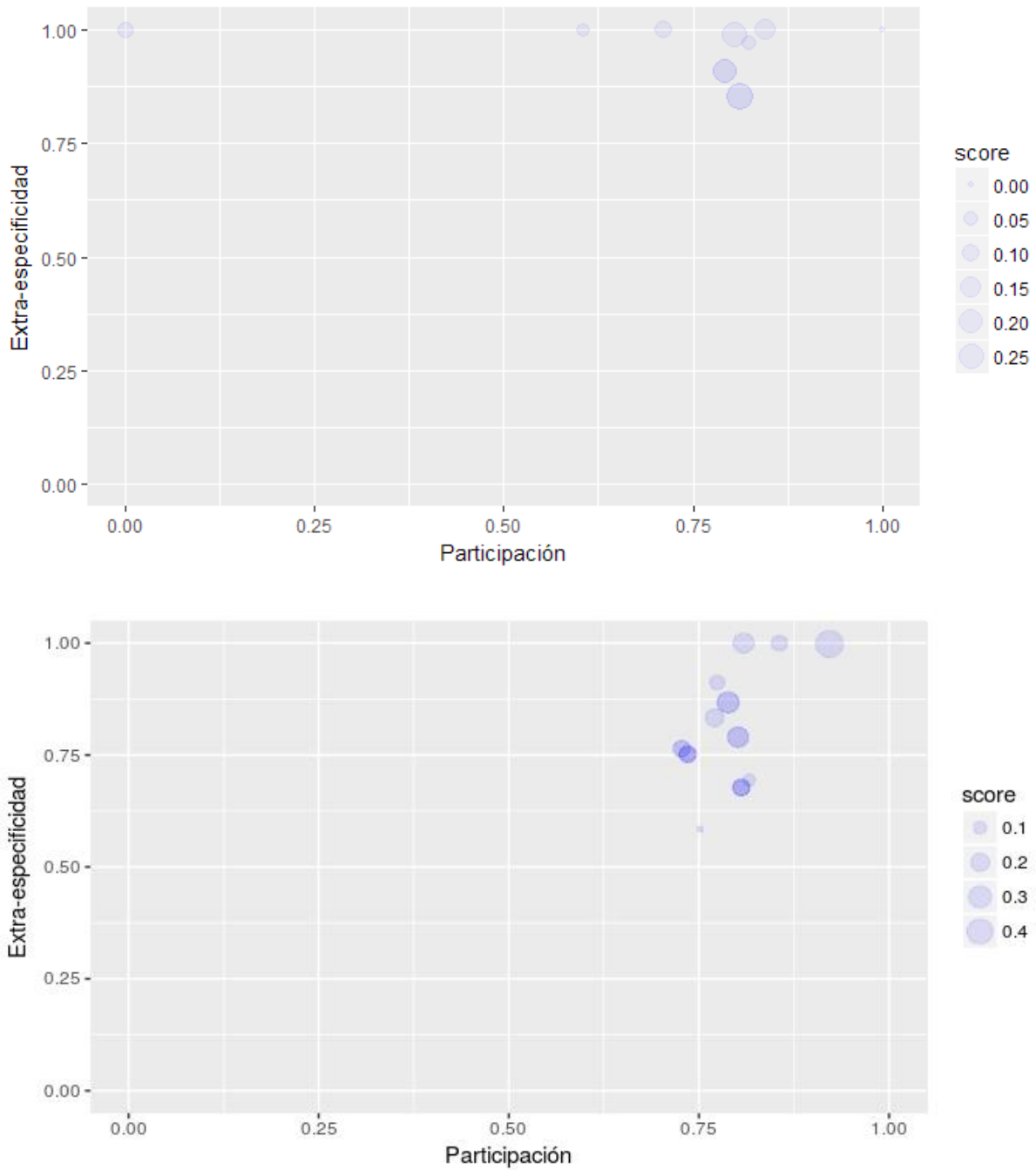

**FIGURE 7.8**. Graph with the scoring results for dopeable proteins and border parameters for *Plasmodium falsiparium (upper panel)* and *Trypanosoma cruzi* (upper panel). The extra-specificity (Extra-especificidad) refers to how much score comes from another species and the participation (participación) to how it is distributed .

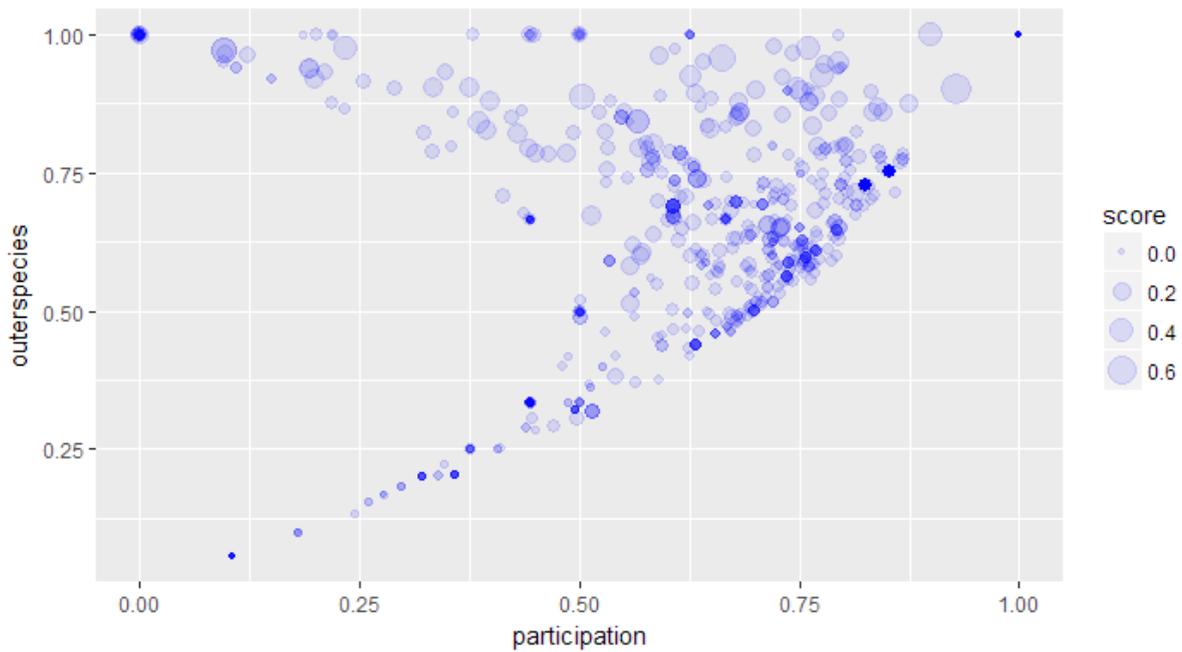

**FIGURE 7.9**. Graph with the scoring results for druggable proteins and border parameters for *Saccharomyces cerevisiae*.

In **figures 7.7**, **7.8,** it is observed that for species that had high values of AUC01 (MUS, pfa, TCRs), the values of extra-specificity of their dopable proteins remain above 0.6, which favors obtaining high score values in the whole genome prioritization process. Furthermore, almost all of the proteins of the analyzed pathogens also present high participation values that show that the score received comes from a flow of information from various species.

The situation for human and yeast is very different. In both cases, low levels of AUC0.1 are reported in the validation process and this is due, as we see in the upper panel of **figure 7.7** and in **figure 7.9**, to the fact that there is a considerable fraction of their druggable targets that present extra-specificity values less than 0.5. This indicates that they are "difficult to reach" proteins from other species in the prioritization, so they do not receive a good score. Furthermore, it is possible to note that in this case several druggable proteins are found in the borderline case of only two scoring species.

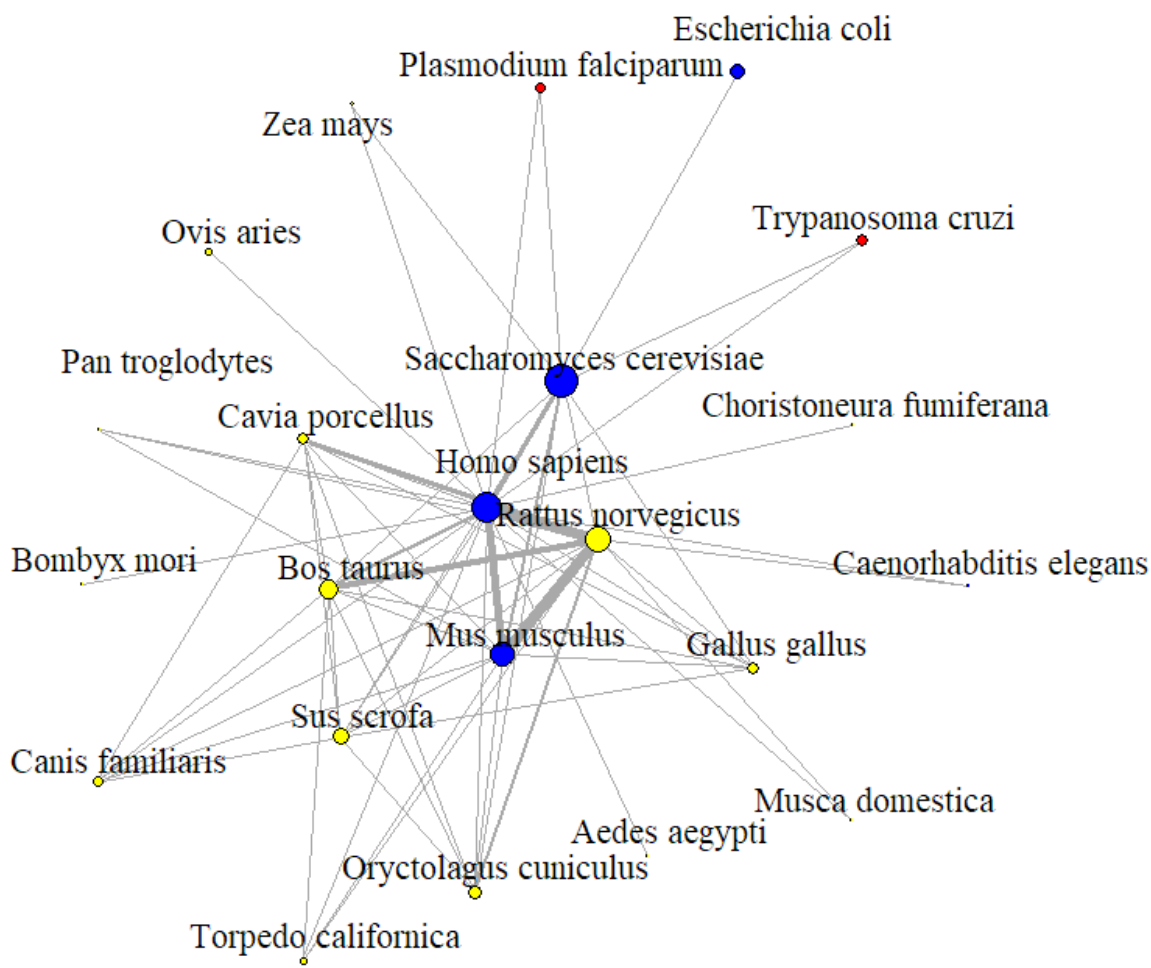

**FIGURE 7.10**. Graph of connection mediated by annotations between species with druggable proteins and with a weight greater than 50. The width of each edge is given by the logarithm of the amount of proteins and annotations shared by the species at their ends normalized by the amount of proteins drugs of said species. Model organisms are shown in blue, and NTD-related pathogens in red. The size of the nodes is given by the logarithm of the number of druggable proteins in each species.

It is interesting to note that the range of prioritization score values received by targets of different species is different (see the scales in **figure 7.7-7.9**). For example, species pfa and tcr have values within the range 0-0.4 while human and mmus within 0-0.8. This is also a reflection of the interconnectivity structure of the network. In order to understand it, we show in **figure 7.10** the species network, with the width of each edge given by the logarithm of the number of pathways mediated by druggable proteins and annotations between species, normalized by the amount of druggable proteins of the species. This scheme therefore serves to illustrate the inter-species interconnectivity pattern put into play in prioritization processes. It is possible to recognize that there is a core of strongly interconnected species that includes model organisms such as *Homo sapiens, Mus musculus, Rattus norvegicus, Bos taurus, Saccharomyces cerevisiae* and *Cavias porcellus*. And then it is understood that the information dissemination process reaches targets of different species with different intensity, and in particular, to a lesser extent, tcr and pfa, which are on the periphery of this network.

### 7.2.3 Drug and flux

In the previous section, we saw how the structure of connections between drug targets of different species affects the whole genome prioritization process. It is now worth asking, are the connection characteristics that we observe for the set of druggable proteins that a species currently has, is it particular in any sense? Or does it represent what one would expect for any other set of proteins of that species?

In other words, are dopable proteins in a privileged situation that makes them easier to reach by external scoring or, on the contrary, are they more difficult to reach than the rest of the proteome of that species? To answer this question for each species analyzed we made 1000 random selections of targets, with the restriction that they have the "same" number of annotations of each type as the original set. We consider the targets selected in this way to be drugged for that iteration (seeds to the uses of our prioritization). Then we recorded the scoring received, as well as the average externality and the average participation received by the set of seed proteins considered in each iteration. Using the means and dispersions of these observables, we finally estimate standardized parameters Z for the externality and mean participation of the set of druggable proteins in our original dataset.

$$Z_x = \frac{x_{real} - <x>}{DS(x)}.$$  (7.4)

where $x$ is a random variable (in our case the results of each seed selection, averaged over all proteins) $<x>$ and $DS(x)$ are the average value and standard deviation over all the realizations, and $x_{real}$ is the value obtained when calculating $x$ with the seeds of the real network.

**Table 7.4** shows that -z values of externality for model species are much greater than for pathogenic species. This would in principle give an explanation of why all scoring methods work better for pathogens: drug targets that are intended to be recovered in the whole species prioritization exercise tend to appear at the interface with other species, possibly because the identification of the respective bioactivities took place via some form of *crawling*, based on knowledge available from other species.

|  | Z participation | Z externality |
|:---:|:---:|:---:|
| *Homo Sapiens* | 37.75 | -48.8 |
| *Mus musculus* | 11.24 | -26.36 |
| *Plasmodium Falsiparium* | -0.208 | 1.58 |
| Sacharomyces-*cerevisae* | 146.98 | -142.93 |
| *Trypanosoma cruzi* | -2.58 | 0.54 |

**TABLE 7.4**. Data obtained using equation 7.4 by selecting proteins at random and treating them as druggable in the validation process seen above.

When we analyze **Table 7.4** it is seen that the following pattern exists: negative Z_participation values with positive Z_externality values, indicate that the real distribution has more connections to the outside than if seeds were taken at random, in **Figures 7.7-11** this would coincide with points in the upper left region.

## *7.3 Conclusion*

Analyzing the structure of bioactivity bonds, we have established that in the vast majority of cases that drugs that act on unique proteins have activities on proteins in other organisms. Furthermore, when taking into account the date of first publication of the bioactivities, we have established that the "new" links in the drug-protein network appear mainly by "crawling", that is, they use information already existing in the network.

We also use a first-neighbor prioritization method, as we proposed at the beginning of this work, and we have validated it in 5 different species when trying to make a complete proteome recovery. The validations were notably better for pathogenic species than for the rest of those used. Finally, we proposed two quantities that evaluate how the score reaches the proteins and we use it to explain the development of the prioritizations.

# Conclusions

In Chapter 3 of this work, the TDR network was introduced and it was explained how its size was reduced when taking clusters for two types of similarity: Tanimoto and substructure, based on the analysis of identity structures. These allow a description in terms of 'coarse grained' structures and thus reduces the effective size of the network in both types of similarity metrics. In Chapter 3, the coherence between these two types of clusters was analyzed by leveraging the persistence values and comparing weight distributions. The persistence indicated that the partition generated by substructure clusters is much stronger than that of Tanimoto, in the sense that the persistence is much greater when going from a partition of the layer in clusters S to T than vice versa. Finally, we established a comparison and calibration criterion between the numerical value of the similarity by substructure and Tanimoto. The calibration scale between both measurements allows us to appreciate that for the same similarity value, the substructure relationship corresponds to a greater coincidence of shared targets.

Then we studied the protein layers and annotations. We detected categories that show preference for scoring proteins with reported bioactivities, through the use of the fisher statistic and the definition of the Rscore. We saw that this is a reflection of biases that actually exist in the area of development and search for new drugs. Many categories play a very important role in the interconnectivity between proteins of different species, which we associated with the entropy of the distribution of proteins by species. A measure that is very relevant to transmit information through the network about knowledge acquired with a bias towards certain species. We also analyze the coherence between the different types of annotations through similarities to first neighbors in the network. We established that the information in the orthologous annotations layer is similar to that of PFAM. Furthermore, we established the existence of a link between common drugs between proteins and the number of shared PFAM annotations. A clear growing relationship between both magnitudes can be observed and that approximately 70% of proteins that share 3 or more domains in their architecture are reached by at least one drug in common.

In the Chapter 6 we projected to the PFAM annotations layer, giving rise to 6 different graphs, by 3 different projection methods and by selecting all or only druggable proteins. We have found that the projections using only bioactive proteins were more transitive than those of all proteins, which allows us to say that the domains associate to allow activity on the proteins. We reached similar conclusions when studying the assortativity of Rscore. We established that the most central domains do not coincide with those with the highest entropy, which can have a negative effect when prioritizing. We also recognized sets of 4 totally related domains, *cliques,* for the StatVal projection of bioactive drugs. We did a prioritization exercise from the cliques to the drug layer, and we found that by doing this, drugs with a similar spectrum of action are obtained. On the other hand, we find two links with domains that appear in the projection with the total of proteins, but not in the specific bioactive. In the two cases it was found that they acted on common mechanisms, although for the results of the first, there are no studies that support the action of one of the three drugs selected in the mechanism. One of the prioritized drugs was also found to have non-lattice bioactivity on a protein with the structure of the two domains mentioned above.

In Chapter 7 we established that it is important in the vast majority of cases that drugs that act on single proteins have activities on proteins in other organisms. Furthermore, when taking into account the date of first publication of the bioactivities, we have established that the "new" links in the drug-protein network appear mainly by "crawling", that is, they use information already existing in the network.

We also used a first-neighbor prioritization method, as we proposed at the beginning of this work, and we have validated it in 5 different species when trying to make a complete proteome recovery. The validations were notably better for pathogenic species than for the rest of those used. Finally, we proposed two quantities that evaluate how the score reaches the proteins and we use it to explain the development of the prioritizations.

# Appendix A

### *A.1. Subgraphs, Connexity and Giant Component.*

Given a graph, it $G(\mathfrak{N}, \mathfrak{E})$ is possible to consider a subset of its nodes and edges to define another new graph $G'$. We will say that it $G'(\mathfrak{N}', \mathfrak{E}')$ is a subgraph of $G(\mathfrak{N}, \mathfrak{E})$, if it is verified that $\mathfrak{N}' = n_1, n_2, n'_N \subseteq \mathfrak{N}$ y $\mathfrak{E}' = e_1', e_2', ..., e_m' \subseteq \mathfrak{E}$. On the other hand, if it $\mathfrak{E}'$ contains all the possible edges $G$ that join nodes of the set, it $\mathfrak{N}'$ is said to $G'$ be an induced subgraph or complete subgraph and is simply denoted by $G' = G(\mathfrak{N}')$.

An important concept in graph theory is the *connection* between pairs of nodes and the complete graph. A *path* from node $i$ to node in $j$ the graph is a sequence of adjacent nodes and edges leading from node $i$ to node $j$. The *length* of a path is given by the number of edges of the path. If each node of the path is visited only once, we are in the presence of a *single path*. If the path begins and ends at the same vertex, it is said to be a *cycle*, and if the length of the cycle is also unitary, we say that we are in the presence of a *loop*. We say that two nodes $i$ $j$ are *related* if therea path that connects, otherwise we will say that the nodes $i$ and $j$ are *disconexos* or *disjointed.* We say that a graph is connected if all its pairs of nodes are connected. Finally, a component of a graph G is defined as an *induced subgraph* that fulfills two conditions: be connected and contain the maximum possible number of edges common to $G$. A *giant component* of a graph $G(\mathfrak{N}, \mathfrak{E})$ is a component whose size (number of nodes) is of the same order as $N$ [39].

### A.1.1. Heavy Graphs

There are numerous systems where it is possible to associate a given magnitude or intensity to each interaction. Such systems can be described beyond a set of binary interactions. For example, consider a network of co-authors, where each node represents a researcher and an edge between two researchers denotes if they share a publication in common. In this case, it is natural to think of a measure of the intensity of the interactions, which could be defined as proportional to the number of publications that both authors share. The networks that these systems represent are of special interest for this thesis and are called *heavy graphs*. A heavy graph can be directed or undirected. In general, a weighted graph $G^W = G(\mathfrak{N}, \mathfrak{E}, \mathfrak{W})$ consists of a set of $N$ nodes $\mathfrak{N} = \{n_1, n_2, ....n_N\}$, a set of $m$ edges $\mathfrak{E} = \{e_1, e_2, ...., e_m\}$ (whose elements will be ordered pairs if the graph is directed), and a set of $m$ weights $W : E \to R, W = w_1, w_2, ....w_m$, each associated with the corresponding edge in the set E. Usually the set W takes positive values, but it is important to note that such a condition is not strictly necessary (see examples [40]).

### A.1.2. Adjacency matrix and weight matrix

A particularly useful representation for graphs is by matrix notation. Given an unweighted graph $G(\mathfrak{N}, \mathfrak{E})$ $N$ nodes, its *adjacency matrix* $A \in NxN$ It is a square matrix of binary elements $A = \{a_{ij} \mid i, j \in 1...N\}$, so that $a_{ij} = 1$ if the corresponding edge exists $e_{i,j}$ and 0 otherwise. The adjacency matrix A will be symmetric or asymmetric, depending on whether they are undirected or directed graphs, respectively. Diagonal elements must be null in order to satisfy the absence of loops required by the given graph definition. On the other hand, in the case of heavy graphs $G_W = G(N, E, W)$, the corresponding matrix representation is usually referred to as a weight matrix $W \in N \times N$. In this case, its element $w_{ij}$ is the weight *w* of the arc that connects node *i* with node *j* if it exists, and otherwise $w_{ij} = 0$. Again, the weight matrix W will be symmetric only if the graph is undirected.

## *A.2. Principal Topological Observables*

### A.2.1 Distribution of degree, assortativity and disassortativity

Let us consider an unweighted graph $G(N, E)$ with an adjacency matrix $A$. We define the degree $k_i$ of a node as the number of first neighbors or adjacent nodes it has,

$$k_i = a_{ij} \quad j \in Ne(i) \tag{A.1}$$

where $Nei(i)$ denotes the set of nodes of the graph that are first neighbors of the node $i$, and $a_{ij}$ is the element $(i,j)$ to the adjacency matrix A of the graph. In the case of heavy graphs $G(W) = G(N, E, W)$, with a weight matrix given by $W \in NXN$, the weights of the corresponding edges can be included, extending definition 2.1 to the observable usually denoted as strength $s_i$ of a node:

$$s_i = w_{ij} j \in Nei(i) \tag{A.2}$$

with $w_{ij}/inW$.

These measurements allow to establish the simplest possible characterization for a graph, which is its degree distribution $P(K)$. It corresponds to the probability that a node *i* taken at random from the graph has degree $k_i = K$. The degree distribution $P_k$ fully characterizes the statistical properties in uncorrelated networks [40], that is, those where the connections between nodes are defined randomly. In contrast, in many cases of networks that represent real systems there are higher-order correlations, in the sense that the probability that a node of degree $k$ has an edge pointing to another of degree $k'$ (which we will denote $P(k|k')$) depends on $k'$. In real cases the effects of finite size introduce noise in the direct study of the conditional probability $P(k|k')$. Therefore, an observable related to this probability of greater practical utility is defined, the mean degree of first neighbors. For a node *i* this magnitude $k$

$$k_{nn,i} = \frac{1}{k_i} \sum_{j \in Nei(i)} k_j \tag{A.3}$$

then, averaging over all nodes of degree k of the network, the mean degree of nearest neighbors for nodes of degree k, $k_{nn}(k)$, is calculated which can also be expressed in terms of the conditional probability $P(k|k)$ according to

$$k_{nn}(k) = \frac{1}{n} \sum_{k'} k' P(k'|k) \tag{A.4}$$

expression that under the absence of degree correlations, does not depend on k.

### A.2.2 Assortativity

Similar to how weights were defined for the edges, different characteristics can be defined for the vertices depending on what they represent. For example, for a social network these could be: ethnicity, gender, socio-economic level. Given this characterization, there is the possibility that the vertices tend to join to a greater extent with vertices with values of similar characteristics, in which case it is said that the graph is *assortative*. If, on the other hand, the vertices tend to have edges with vertices of very dissimilar characteristics, the graph presents *disassortativity*. This can have great effects on the structure of the graph, since having strong assortativity, highly connected groups with similar values can be generated.

A quantity can be defined $e_{xy}$, which is the fraction of edges that join nodes with vertex values $x, y$ for some scalar variable of interest the matrix $e_{xy}$. Using, the assortativity measure is defined. First you notice that it $e_{xy}$ satisfies

$$\sum_{xy} e_{xy} = 1 \quad \sum_{y} e_{xy} = a_x , \quad \sum_{x} e_{xy} = b_y \tag{A.5}$$

where $a_x, b_y$ are, respectively, the fraction of edges that start and end at vertices with values $x$ e $y$. (If the graph is undirected and monopartite $a_x = b_x$ ,.) Then if there is no assortative mixing. $e_{xy} = a_x b_y$ If there is assortative mixing, it can be obtained by calculating Pearson's correlation, therefore:

$$r = \frac{\sum_{xy} xy(e_{xy} - a_x b_y)}{\sigma_a \sigma_b} \tag{A.6}$$

where $\sigma_a$ and $\sigma_b$ are the standard deviations $a_x$ and $b_y$. This assortativity coefficient is known as $r$ Newman's, and it will be used in the rest of the work. The value of $r$ assumes values in $-1 \leq r \leq 1$, with $r = 1$ indicating perfect assortativity and $r = -1$ indicating perfect *disassortativity*.

Similar to how the assortativity coefficient was defined for a general characteristic of the vertices, the degree assortativity is defined for a graph as

$$r = \frac{\sum_{jk} jk(e_{jk} - a_j b_k)}{\sigma_a^2 \sigma_b^2} \tag{A.7}$$

where $j$ and $k$ are the values of the *degree in excess* of the vertices and $a_k$ $b_j$ are, respectively, the fraction of edges that start and end at vertices with values $k$ y $j$. The *excess degree* is the number of all the outgoing or incoming edges in a vertex, discounting the edge by which said vertex was reached or left. The networks that present degree assortativity in the real world are the co-authorship networks, people who are active in the publication of academic works tend to work with people of the same tendencies, while an example of dissortativiity is the metabolic networks [41].

### A.2.3. Clustering coefficient

A fundamental topological measure arises from quantifying how connected the environment of a node is connected. The most basic measure for this purpose is the local clustering coefficient ci of a given node. This measure compares the number of connections present between the first neighbors of node i with the maximum number of connections that could exist between them. Thus, the local grouping coefficient is defined according to

$$c_i = \frac{2}{k_i(k_i-1)))} \sum_{j,m \in Neigh(i)} a_{ij} a_{jm} a_{mi} \qquad (A.8)$$

Equivalently, this topological observable can be interpreted as a relationship between the number of triangles that make up node i, and the total number of possible triangles that could occur between it and its first neighbors $(k_i(k_i-1)/2)$.

This measurement gives an intuitive idea of how connected the environment of a node is. For example, if the induced subgraph $G_i$ has a star-like structure (the neighbors of the node are $i$ not connected to each other but through the latter), its local clustering coefficient will be null ($c_i = 0$). In opposition, if all the neighbors of the node are completely connected to each other, we will be in the presence of an induced subgraph called clique, in which case we will have $c_i = 1$. The total clustering coefficient of the graph is the mean of $c_i$ for all nodes in the graph. Alternatively, the global clustering coefficient is

$$C = \frac{\# \text{ closed triads}}{\# \text{ total triads}} \qquad . (A.9)$$

En the case of heavy networks, a possible generalization for the coe fi cientede grouping a node $i$ was proposed in [42]

$$c_i^w = \frac{1}{s_i(k_i-1)} \sum_{j,m \in Neigh(i)} \frac{w_{ij}+w_{im}}{2} a_{ij} a_{jm} a_{mi} \quad \text{(A.10)}$$

is, that each triangle is counted under a weight factor given by the average of the arcs of the same that include the node $i$. Note that this definition is reduced to the case of non-heavy nets when the weights are all uniform. In addition, the normalization factor $s_i(k_i-1)$ ensures that $c_i^w \in [0,1]$ since only the weights involving the node are considered $i$. In this way it is possible to define the total clustering coefficient in a weighted graph $C_w$ as the average of the $c_i^w$ and compare it with the total clustering coefficient that does not consider the weights w. Si $C_w > C$ means that the triangles in the graph are typically made up of heavy arcs. On the contrary, if it $C_w < C$ means that triplets are typically made up of low weight arches.

### A.2.4. Centrality

A lot of works have been presented in the study of the concept of centrality that arises when making the question "What vertices are important in the network?". There are therefore different definitions of centrality in graphs. Each definition uses a different heuristic, and comes from different notions of what is an important vertex in the graph. In particular, here we will consign two methods, namely: the *hub and authority score* by Kleinbergand and *eigenvector centrality.*

The eigenvector centrality gives a centrality score to a vertex proportional to that of its first neighbors. The vertices with the maximum score will then be those that have high score neighbors or that simply have many neighbors. Bonacich[43] shows that by assuming initial scores $\mathbf{x}(0)$ on the network and iteratively propagating them $t$ times according to

$$\mathbf{x}(t) = \mathbf{A}^t \mathbf{x}(0) \quad \text{(A.11)}$$

for $t$ sufficiently large the scores tend to the equilibrium value

$$x_i = \sum_{ij} \kappa_1^{-1} A_{ij} x_j \qquad (A.12)$$

with $\kappa_1$ the highest eigenvalue of the matrix $\mathbf{A}$.

The heuristics behind the Kleinberg centralities[44] is as follows, a vertex will be important in the authority sense, if it has edges leading to many important vertices in the hub sense. And a hub will be important if it has many protruding edges towards vertices that are important as authorities. Thus, the algorithm for its calculation is given by two scoring diffusions with parameters $\alpha$ and $\beta$

$$x = \alpha \mathbf{A} y, \quad y = \beta \mathbf{A}^\mathbf{T} x \qquad (A.13)$$

The final result is independent of the choice of coefficient.

The important thing about this second method to define centrality is that it lacks the problem of the first, in which the score of vertices outside strongly connected components or only with salient edges is null. In particular, a vertex with numerous protruding edges will have a high hub score, and may induce a high authority score on the vertices it points to.

### A.3 Modular structure and partitions

#### A.3.1 Modular structure and quality of a partition

Many networks present a high degree of inhomogeneity in their connectivity patterns, reflecting the presence of a non-trivial level of order and organization in the network [45]. In general, the edge distribution of the graph is neither global nor locally uniform, so it is common to find areas of the network with a high density of edges connecting different groups of nodes and a low density of edges between these groups. This type of structure, usually present in networks that represent real systems, is known as a community structure or modular structure.

Numerous examples of communities in different types of networks could be mentioned. In the case of social graphs, it is intuitive to think of modular structures that represent family groups, groups of friends, work groups, etc. In metabolic or protein interaction networks such as the one we will present in the next chapter, these modular groups can represent and / or correlate with functional groups or some other set of interest.

In general, given a graph $G(N,E)$, a community or module can be thought of as a subgraph whose nodes are strongly connected to each other and weakly connected to other nodes in the graph.It is important to note that however, there is no formal definition of community in graphs universally accepted. Furthermore, the modulus definition usually depends on the specific system and application in mind[45] .

A *partition* is a division of a graph into modular structures so that each node of the graph belongs to a single module. In many problems it is of special interest to define communities in such a way that a given node can belong to more than one of them. Such a division into overlapping communities is usually called *coverage*. In this work, only network divisions into disjoint module partitions will be the object of study. There are numerous algorithms to detect possible partitions of a graph. Each algorithm is usually based on its own definition of community, so it is expected that qualitatively different partitions will be obtained depending on the algorithm used. To compare the performance of different algorithms and their resulting partitions, it is necessary to define some quality function that allows us to quantify how good a given partition is. The most widely used quality function is *modularity* $Q$ by Newman and Girvan [46]. This measure is based on the idea that a random network does not have a modular structure. Therefore it is feasible to measure the quality of a given module by comparing the density of internal arcs that it has with which it would be expected if it were extracted from a random graph lacking modular structure.

It is clear that such a definition depends on the choice of the null model used, that is, the graph devoid of structure considered, which respects some of the structural characteristics of the graph under study. Given a graph G (N, E) the modularity Q is defined according to

$$Q = \frac{1}{2m} \sum_{j,m \in N}(A_{ij} - P_{ij})\delta(C_i, C_j) \qquad \textbf{(A.14)}$$

here it $m$ represents the number of arcs of the graph G, it $A_{ij}$ is the corresponding element of the adjacency matrix, and it $P_{ij}$ is the probability that nodes i and j are connected in the null model chosen. The module to which node i belongs is denoted by $C_i$ and is $\delta(C_i, C_j)$ only found if nodes i, j belong to the same module (otherwise $\delta(C_i, C_j) = 0$).

The most common null model for the calculation of modularity is the *con fi gurational model* [47.48].

This type of random graph preserves the total number of edges of each node, and therefore preserves not only the total number of edges of the graph but also its degree distribution. In this model, each node can be connected to any other in the graph. One way to think the construction of this model for an undirected graph $G(N, E)$ $N$ nodes and $m$ edges, is that initially each node has available media edge and to form an edge from the node $i$ to the $j$ necessary take one of the edges means of each node. With this, the most usual way to calculate the modularity of a partition is

$$Q = \frac{1}{2m} \sum_{j,m \in N} (A_{ij} - \frac{K_i k_j}{2m}) \delta(C_i, C_j) \qquad \textbf{(A.15)}$$

This partition quality measure can be generalized to heavy and even directed graphs, although the latter case is not the main object of study in this thesis. For heavy graphs, it is enough to consider the *strength* of each node instead of the degree, and modify the normalization factor. Being W the total sum of weights of the graph under consideration, we have

$$Q = \frac{1}{2W} \sum_{j,m \in N} (A_{ij} - \frac{s_i s_j}{2m}) \delta(C_i, C_j) \qquad \textbf{(A.16)}$$

this is the most general way that we will adopt in this thesis to calculate modularity.

### A.3.2 Comparison between partitions

In the same way that there are several quantifiers to evaluate the quality of a clustered partition, there are also different coefficients that can be calculated to evaluate the similarity of two different partitions in communities. In particular in this work two such measures are used, normalized mutual information (NMI) and Rand.

NMI is a measure that comes from information theory, and part of the notion of entropy of a partition $C$ given by

$$H(C) = -\sum_{i=1}^{k} P(i) log_2(P(i))  \quad\quad \textbf{(A.17)}$$

with $C_i$ the community $i$ of the partition $c$ and $P(i)$ the probability of finding a vertex in the community $C_i$, viz $\frac{|C_i|}{N}$.

Let a second partition be $C'$ the mutual information of both partitions is given by

$$I(C, C') = \sum_{i=1}^{k} \sum_{j=1}^{l} P(i,j) log_2 \frac{P(i,j)}{P(i)P(j)} \quad\quad \textbf{(A.18)}$$

with P (i, j) the probability of finding a vertex in the community $i$ of the partition $C$ and also the community $j$ of $C'$. Normalizing by entropy is then defined

$$NMI(C, C') = \frac{I(C,C')}{\sqrt{H(C)H(C')}} \quad\quad \textbf{(A.19)}$$

and assumes values between 0 and 1.

Unlike NMI which compares the node content of each partition, the Rand index compares the pairs of nodes that are preserved in both partitions.

$$R = \frac{a+b}{\binom{N}{2}} \quad\quad \textbf{(TO 20)}$$

with $a$ the number of pairs of nodes that are together in clusters $C$ and $C'$ and $b$ the number of pairs of nodes that are separated.

### A.3.3. Clustering algorithms considered

As mentioned in the previous section, there are multiple methods to extract structure in communities from a graph. In this thesis we will base ourselves mainly on two widely recognized and used methodologies that we will describe below. In the first place, the Clauset-Newman-Moore algorithm [46] was considered, which is a member of a large family of algorithms that, with different heuristics, search for partitions of a network by directly optimizing the quality function Q de fi ned in equation A .16. On the other hand, thealgorithm *Infomap [49] was considered,* which makes use of completely different optimization criteria, based on information theory. In this algorithm the modules are defined in such a way as to minimize the average length of the description of a random walk process that takes place in the graph. The main idea is to describe the random walk with a two-level tag system. Given a partition P, one type of label is used to describe the different communities of the partition and the other class of labels is used to identify nodes within those communities. To efficiently describe a random walk with this code of levels, it is necessary that the partition reflects the flow patterns within the network, so that the different modules correspond to areas of high density of connections where the walker on the random walk spends enough time before moving on to other modules. Given a partition $P$ consisting of $P = P1, P2...Ps$ modules, a random walk of infinite length in the network can be conceptually described by two contributions, one associated with the jumps that occur between different modules. $(P_i, P_j, i \neq j)$ and the other associated with the movements that occur within each of the modules $P_i$. The infomap algorithm quantifies this fact through the cost function described according to

$$L(P) = q_{inter} H(Q) = \sum_i^s p_{intra}^i H(P^i) \tag{A.21}$$

where $q_{inter}$ is the probability of going from one module to another in the walk, $H(Q)$ is the entropy of movements between modules, $p_{intra}$ is the fraction of movements of the walk that have occurred in the module $P_i$ and $H(P_i)$ is the entropy of movements that occur within the module $P_i$. The first term of equation 2.12 gives the average number of bits necessary to describe the movement between modules and the second term gives the average number of bits necessary to describe the movement within the different modules [49] .

# Bibliography

1. Molecular Biology of the Cell. 2002.

2. Csermely P, Korcsmáros T, Kiss HJM, London G, Nussinov R. Structure and dynamics of molecular networks: A novel paradigm of drug discovery. Pharmacol Ther. 2013; 138: 333–408.

3. Watts DJ. Networks, Dynamics, and the Small - World Phenomenon. Am J Sociol. 1999; 105: 493–527.

4. White D, Johansen U. Network Analysis and Ethnographic Problems: Process Models of a Turkish Nomad Clan. Lexington Books; 2005.

5. Robertson SA, Renslo AR. Drug discovery for neglected tropical diseases at the Sandler Center. Future Med Chem. 2011; 3: 1279-1288.

6. Parkkinen JA, Kaski S. Probabilistic drug connectivity mapping. BMC Bioinformatics. 2014; 15: 113.

7. Iskar M, Zeller G, Blattmann P, Campillos M, Kuhn M, Kaminska KH, et al. Characterization of drug-induced transcriptional modules: towards drug repositioning and functional understanding. Mol Syst Biol. 2013; 9: 662.

8. Emig D, Ivliev A, Pustovalova O, Lancashire L, Bureeva S, Nikolsky Y, et al. Drug target prediction and repositioning using an integrated network-based approach. PLoS One. 2013; 8: e60618.

9. Cloete TT, North-West University (south Africa). Eflornithine Derivatives for Enhanced Oral Bioavailability in the Treatment of Human African Trypanosomiasis. 2009.

10. Burri C, Brun R. Eflornithine for the treatment of human African trypanosomiasis. Parasitol Res. 2003; 90 Supp 1: S49–52.

11. Magarinos MP, Carmona SJ, Crowther GJ, Ralph SA, Roos DS, Shanmugam D, et al. TDR Targets: a chemogenomics resource for neglected diseases. Nucleic Acids Res. 2011; 40: D1118-D1127.

12. Gamo FJ, Sanz LM, Vidal J, de Cozar C, Alvarez E, Lavandera JL, et al. Thousands of chemical starting points for antimalarial lead identification. Nature. 2010; 465: 305–310.

13. Crowther GJ, Shanmugam D, Carmona SJ, Doyle MA, Hertz-Fowler C, Berriman M, et al. Identification of Attractive Drug Targets in Neglected-Disease Pathogens Using an In Silico Approach. PLoS Negl Trop Dis. 2010; 4: e804.

14. Guiguemde WA, Shelat AA, Bouck D, Duffy S, Crowther GJ, Davis PH, et al. Chemical genetics of Plasmodium falciparum. Nature. 2010; 465: 311–315.

15. Cheng F, Liu C, Jiang J, Lu W, Li W, Liu G, et al. Prediction of Drug-Target Interactions and Drug Repositioning via Network-Based Inference. PLoS Comput Biol. 2012; 8: e1002503.

16. Alaimo S, Pulvirenti A, Giugno R, Ferro A. Drug – target interaction prediction

through domain-tuned network-based inference. Bioinformatics. 2013; 29: 2004–2008.

17.     Lü L, Linyuan L, Medo M, Yeung CH, Zhang YC, Zhang ZK, et al. Recommender systems. Phys Rep. 2012; 519: 1–49.

18.     Gillis J, Pavlidis P. The impact of multifunctional genes on "guilt by association" analysis. PLoS One. 2011; 6: e17258.

19.     Gillis J, Pavlidis P. The role of indirect connections in gene networks in predicting function. Bioinformatics. 2011; 27: 1860–1866.

20.     McClish DK. Analyzing a portion of the ROC curve. Med Decis Making. 1989; 9: 190-195.

21.     Jeong H, Tombor B, Albert R, Oltvai ZN, Barabási AL. The large-scale organization of metabolic networks. Nature. 2000; 407: 651–654.

22.     Lambiotte R, Ausloos M. Collaborative Tagging as a Tripartite Network. Lecture Notes in Computer Science. 2006. pp. 1114-1117.

23.     Hotho A, Jäschke R, Schmitz C, Stumme G. Information Retrieval in Folksonomies: Search and Ranking. Lecture Notes in Computer Science. 2006. pp. 411-426.

24.     Zhou T, Ren J, Medo M, Zhang YC. Bipartite network projection and personal recommendation. Phys Rev E Stat Nonlin Soft Matter Phys. 2007; 76: 046115.

25.     Tumminello M, Miccichè S, Lillo F, Piilo J, Mantegna RN. Statistically Validated Networks in Bipartite Complex Systems. PLoS One. Public Library of Science; 2011; 6: e17994.

26.     Boccaletti S, Bianconi G, Criado R, del Genio CI, Gómez-Gardeñes J, Romance M, et al. The structure and dynamics of multilayer networks. Phys Rep. 2014; 544: 1–122.

27.     Rogers DJ, Tanimoto TT. A Computer Program for Classifying Plants. Science. 1960; 132: 1115-1118.

28.     Kruger FA, Rostom R, Overington JP. Mapping small molecule binding data to structural domains. BMC Bioinformatics. 2012; 13 Suppl 17: S11.

29.     Flower DR. On the Properties of Bit String-Based Measures of Chemical Similarity. J Chem Inf Comput Sci. 1998; 38: 379–386.

30.     Snarey M, Terrett NK, Willett P, Wilton DJ. Comparison of algorithms for dissimilarity-based compound selection. J Mol Graph Model. 1997; 15: 372–385.

31.     Dixon SL, Koehler RT. The hidden component of size in two-dimensional fragment descriptors: side effects on sampling in bioactive libraries. J Med Chem. 1999; 42: 2887-2900.

32.     Yamanishi Y, Araki M, Gutteridge A, Honda W, Kanehisa M. Prediction of drug-target interaction networks from the integration of chemical and genomic spaces. Bioinformatics. 2008; 24: i232–40.

33.     Berenstein AJ, Magariños MP, Chernomoretz A, Agüero F. A Multilayer Network Approach for Guiding Drug Repositioning in Neglected Diseases. PLoS Negl Trop Dis. 2016; 10: e0004300.

34.     Peters GJ, Backus HHJ, Freemantle S, van Triest B, Codacci-Pisanelli G, van der

Wilt CL, et al. Induction of thymidylate synthase as a 5-fluorouracil resistance mechanism. Biochim Biophys Acta. 2002; 1587: 194–205.

35. Filatov D, Ingemarson R, Gräslund A, Thelander L. The role of herpes simplex virus ribonucleotide reductase small subunit carboxyl terminus in subunit interaction and formation of iron-tyrosyl center structure. J Biol Chem. 1992; 267: 15816-15822.

36. Serrano MA, Boguñá M, Vespignani A. Extracting the multiscale backbone of complex weighted networks. Proc Natl Acad Sci US A. 2009; 106: 6483-6488.

37. Bruncko M, McClellan WJ, Wendt MD, Sauer DR, Geyer A, Dalton CR, et al. Naphthamidine urokinase plasminogen activator inhibitors with improved pharmacokinetic properties. Bioorg Med Chem Lett. 2005; 15: 93–98.

38. Yıldırım MA, Goh KI, Cusick ME, Barabási AL, Vidal M. Drug — target network. Nat Biotechnol. 2007; 25: 1119-1126.

39. Wasserman S, Faust K. Social Network Analysis in the Social and Behavioral Sciences. Social Network Analysis. pp. 3–27.

40. Boccaletti S, Latora V, Moreno Y, Chavez M, Hwang D. Complex networks: Structure and dynamics. Phys Rep. 2006; 424: 175–308.

41. Newman MEJ. The Structure and Function of Complex Networks. SIAM Rev. 2003; 45: 167–256.

42. Barrat A, Barthelemy M, Pastor-Satorras R, Vespignani A. The architecture of complex weighted networks. Proceedings of the National Academy of Sciences. 2004; 101: 3747-3752.

43. Bonacich P. Power and Centrality: A Family of Measures. Am J Sociol. 1987; 92: 1170-1182.

44. Kleinberg JM. Authoritative sources in a hyperlinked environment. J ACM. 1999; 46: 604–632.

45. Fortunato S. Community detection in graphs. Phys Rep. 2010; 486: 75–174.

46. Newman MEJ, Girvan M. Finding and evaluating community structure in networks. Physical Review E. 2004; 69. doi:10.1103 / physreve.69.026113

47. Molloy M, Reed B. The Size of the Giant Component of a Random Graph with a Given Degree Sequence. Comb Probab Comput. 1998; 7: 295-305.

48. Molloy M, Reed B. A critical point for random graphs with a given degree sequence. Random Struct Algorithms. 1995; 6: 161-180.

49. Rosvall M, Bergstrom CT. Maps of random walks on complex networks reveal community structure. Proceedings of the National Academy of Sciences. 2008; 105: 1118-1123.